\documentclass[11pt]{article}
\usepackage{amsmath, amsthm, amssymb}
\usepackage{float,epsfig}
\usepackage{mathdots}
\usepackage{rotating}
\usepackage{subfigure}
\usepackage{caption}
\usepackage{tikz}
\usepackage{color}
\usepackage{algorithm}
\usepackage{multicol}
\usepackage[utf8]{inputenc}
\usepackage{braket}
\usepackage{multirow}
\usepackage{qcircuit}
\usepackage[overload]{empheq}
\usepackage{hyperref}
\usepackage{xcolor}

\usepackage[noend]{algpseudocode}
\usepackage{mathtools}

\usepackage{xparse}

\NewDocumentCommand{\INTERVALINNARDS}{ m m }{
    #1 {,} #2
}
\NewDocumentCommand{\interval}{ s m >{\SplitArgument{1}{,}}m m o }{
    \IfBooleanTF{#1}{
        \left#2 \INTERVALINNARDS #3 \right#4
    }{
        \IfValueTF{#5}{
            #5{#2} \INTERVALINNARDS #3 #5{#4}
        }{
            #2 \INTERVALINNARDS #3 #4
        }
    }
}

\usepackage{lipsum}



\begin{document}

\newtheorem{theorem}{\bf Theorem}[section]
\newtheorem{proposition}[theorem]{\bf Proposition}
\newtheorem{definition}[theorem]{\bf Definition}
\newtheorem{corollary}[theorem]{\bf Corollary}
\newtheorem{example}[theorem]{\bf Example}
\newtheorem{exam}[theorem]{\bf Example}
\newtheorem{remark}[theorem]{\bf Remark}
\newtheorem{lemma}[theorem]{\bf Lemma}
\newcommand{\nrm}[1]{|\!|\!| {#1} |\!|\!|}

\newcommand{\calL}{{\mathcal L}}
\newcommand{\calX}{{\mathcal X}}
\newcommand{\calY}{{\mathcal Y}}
\newcommand{\calZ}{{\mathcal Z}}
\newcommand{\calW}{{\mathcal W}}
\newcommand{\calA}{{\mathcal A}}
\newcommand{\calB}{{\mathcal B}}
\newcommand{\calC}{{\mathcal C}}
\newcommand{\calK}{{\mathcal K}}
\newcommand{\C}{{\mathbb C}}
\newcommand{\Z}{{\mathbb Z}}
\newcommand{\R}{{\mathbb R}}
\renewcommand{\SS}{{\mathbb S}}
\newcommand{\LL}{{\mathbb L}}
\newcommand{\st}{{\star}}

\def\kernel{\mathop{\rm kernel}\nolimits}
\def\sigan{\mathop{\rm span}\nolimits}

\newcommand{\klasse}{{\boldsymbol \Delta}}

\newcommand{\ba}{\begin{array}}
\newcommand{\ea}{\end{array}}
\newcommand{\von}{\vskip 1ex}
\newcommand{\vone}{\vskip 2ex}
\newcommand{\vtwo}{\vskip 4ex}
\newcommand{\dm}[1]{ {\displaystyle{#1} } }

\newcommand{\be}{\begin{equation}}
\newcommand{\ee}{\end{equation}}
\newcommand{\beano}{\begin{eqnarray*}}
\newcommand{\eeano}{\end{eqnarray*}}
\newcommand{\inp}[2]{\langle {#1} ,\,{#2} \rangle}
\def\bmatrix#1{\left[ \begin{matrix} #1 \end{matrix} \right]}
\def \noin{\noindent}
\newcommand{\evenindex}{\Pi_e}



\def \R{{\mathbb R}}
\def \C{{\mathbb C}}
\def \K{{\mathbb K}}
\def \H{{\mathbb H}}

\def \T{{\mathbb T}}
\def \Pb{\mathrm{P}}
\def \N{{\mathbb N}}
\def \Ib{\mathrm{I}}
\def \Ls{{\Lambda}_{m-1}}
\def \Gb{\mathrm{G}}
\def \Hb{\mathrm{H}}
\def \Lam{{\Lambda}}

\def \Qb{\mathrm{Q}}
\def \Rb{\mathrm{R}}
\def \Mb{\mathrm{M}}
\def \norm{\nrm{\cdot}\equiv \nrm{\cdot}}

\def \P{{\mathbb{P}}_m(\C^{n\times n})}
\def \A{{{\mathbb P}_1(\C^{n\times n})}}
\def \H{{\mathbb H}}
\def \L{{\mathbb L}}
\def \G{{\F_{\tt{H}}}}
\def \S{\mathbb{S}}
\def \s{\mathbb{s}}
\def \sigmin{\sigma_{\min}}
\def \elam{\Lambda_{\epsilon}}
\def \slam{\Lambda^{\S}_{\epsilon}}
\def \Ib{\mathrm{I}}
\def \Tb{\mathrm{T}}
\def \d{{\delta}}

\def \Lb{\mathrm{L}}
\def \N{{\mathbb N}}
\def \Ls{{\Lambda}_{m-1}}
\def \Gb{\mathrm{G}}
\def \Hb{\mathrm{H}}
\def \Delta{\triangle}
\def \Rar{\Rightarrow}
\def \p{{\mathsf{p}(\lam; v)}}

\def \D{{\mathbb D}}

\def \tr{\mathrm{Tr}}
\def \cond{\mathrm{cond}}
\def \lam{\lambda}
\def \sig{\sigma}
\def \sign{\mathrm{sign}}

\def \ep{\epsilon}
\def \diag{\mathrm{diag}}
\def \rev{\mathrm{rev}}
\def \vec{\mathrm{vec}}

\def \ham{\mathsf{Ham}}
\def \herm{\mathsf{Herm}}
\def \sym{\mathsf{sym}}
\def \odd{\mathsf{sym}}
\def \en{\mathrm{even}}
\def \rank{\mathrm{rank}}
\def \pf{{\bf Proof: }}
\def \dist{\mathrm{dist}}
\def \rar{\rightarrow}

\def \rank{\mathrm{rank}}
\def \pf{{\bf Proof: }}
\def \dist{\mathrm{dist}}
\def \Re{\mathsf{Re}}
\def \Im{\mathsf{Im}}
\def \re{\mathsf{re}}
\def \im{\mathsf{im}}

\def \sym{\mathsf{sym}}
\def \sksym{\mathsf{skew\mbox{-}sym}}
\def \odd{\mathrm{odd}}
\def \even{\mathrm{even}}
\def \herm{\mathsf{Herm}}
\def \skherm{\mathsf{skew\mbox{-}Herm}}
\def \str{\mathrm{ Struct}}
\def \eproof{$\blacksquare$}

\def \bS{{\bf S}}
\def \cA{{\cal A}}
\def \E{{\mathcal E}}
\def \X{{\mathcal X}}
\def \F{{\mathcal F}}
\def \cH{\mathcal{H}}
\def \cJ{\mathcal{J}}
\def \tr{\mathrm{Tr}}
\def \range{\mathrm{Range}}
\def \adj{\star}

\def \pal{\mathrm{palindromic}}
\def \palpen{\mathrm{palindromic~~ pencil}}
\def \palpoly{\mathrm{palindromic~~ polynomial}}
\def \odd{\mathrm{odd}}
\def \even{\mathrm{even}}

\newcommand{\tm}[1]{\textcolor{magenta}{ #1}}
\newcommand{\tre}[1]{\textcolor{red}{ #1}}
\newcommand{\tb}[1]{\textcolor{blue}{ #1}}
\newcommand{\tg}[1]{\textcolor{green}{ #1}}

\title{Quantum circuit model for discrete-time three-state quantum walks on Cayley graphs}
\author{ Rohit Sarma Sarkar\thanks{Department of Mathematics,
IIT Kharagpur, India, E-mail: {rohit15sarkar@yahoo.com}
 } \, and \, Bibhas Adhikari\thanks{Corresponding author, Fujitsu Research of America, Inc.,
Santa Clara, USA, E-mail: badhikari@fujitsu.com and
bibhas.adhikari@gmail.com } 
}

\date{}

\maketitle
\thispagestyle{empty}

\noindent{\bf Abstract.} We develop qutrit circuit models for discrete-time  three-state quantum walks on Cayley graphs corresponding to Dihedral groups $D_N$ and the additive groups of integers modulo any positive integer $N$. The proposed circuits comprise of elementary qutrit gates such as qutrit rotation gates, qutrit-$X$ gates and two-qutrit controlled-$X$ gates. First, we propose qutrit circuit representation of special unitary matrices of order three, and the block diagonal special unitary matrices with $3\times 3$ diagonal blocks, which correspond to multi-controlled $X$ gates and permutations of qutrit Toffoli gates. We show that one-layer qutrit circuit model need $O(3nN)$ two-qutrit control gates and $O(3N)$ one-qutrit rotation gates for these quantum walks when $N=3^n$. Finally we numerically simulate these circuits to mimic its performance such as time-averaged probability of finding the walker at any vertex on noisy quantum computers. The simulated results for the time-averaged probability distributions for noisy and noiseless walks are further compared using KL-divergence and total variation distance. These results show that noise in gates in the circuits significantly impacts the distributions than amplitude damping or phase damping errors. \\



\section{Introduction}

Quantum computing has acquired stellar progress over recent years showcasing quantum algorithms can have more than polynomial speed-ups compared to their classical counterparts. Hence, developing new efficient quantum algorithms has been a perceptible goal for researchers over the years. Quantum walks, a quantum analogue of classical random walks, represent a universal model for quantum computation \cite{Aharonov2001,Childs2009,Lovett2010,Childs2013} and act as a  great platform for designing fast quantum algorithms. 
Similar to its classical counterpart, quantum walks on graphs are divided into two models based on nature of time evolution viz. Discrete-Time Quantum Walks (DTQWs) and Continuous-Time Quantum Walks (CTQWs) \cite{Childs2002},\cite{Andraca2012}. Apart from the difference in time evolution, DTQWs act on a larger Hilbert space on account of requiring a quantum coin operator which defines the evolution dynamics of the walker. Several works analyzing fundamental properties of DTQWs (viz. periodicity and localization) exist for three state DTQWs on several graphs like lines, cycles, mixed paths/cycles and Cayley graphs of non-abelian groups viz. symmetric and Dihedral groups \cite{Inui2004,Banerjee2022,SarmaSarkar2023,Liu2021,Kubota2021,Kajiwara2019}. In \cite{Bisio2016},\cite{D'Ariano2019}, the authors study DTQWs on Cayley graphs of Dihedral groups along with their variants. For a detailed review on quantum walks, see \cite{Andraca2012}. It is also of note that the discrete-time quantum walks spreads quadratically faster in position space in comparison to classical random walks \cite{Kempe2003,Andraca2012}.

In order to study quantum walks, we approach it through quantum circuit model of quantum computation. In particular, we shall consider qutrit circuits as opposed to the primarily used qubit circuits to incorporate the degree of freedom provided by the three dimensional quantum coin in our model. Over the years, one of the primary focus in research has been directed towards construction of efficient circuits for the quantum walk model. It is of note that unlike binary system of classical computers, superconducting \cite{Yan2019} and trapped ion computers\cite{Alderete2020} theoretically posses discrete energy levels of an infinite spectrum, which in turn, makes them eligible to work on qudit systems. Many quantum walks related experiments have been carried out on real quantum hardware using qubits \cite{Balu2017,Yan2019,Alderete2020,Acasiete2020,Singh2020}. However, it is still a challenge to carry out the same experiments efficiently on qutrits and in turn, qudits \cite{Zhou2019}. Qutrit circuit model has been the object of interest for researchers worldwide and treated as an alternative to qubit circuits \cite{Gokhale2019}. Works pertaining to constructing circuits of quantum walks exist in literature\cite{Douglas2009} and this has been extended to developing qutrit circuits for three state lazy quantum walk on the line \cite{Saha2021} and qudit circuits on a lattice for $d$-state DTQWs\cite{Saha2022}. Works also exist in literature showcasing generalizations of certain quantum algorithms (viz. Shor's Alogrithm, Deustch Josza Algorithm) using qutrits \cite{Fan2007,Bocharov2016,Bocharov2017}. Moreover, synthesis of certain qudit gates can be found in \cite{Di2013}. 



 Two-state DTQWs on cycle graphs using Hadamard coin was first introduced in \cite{Aharonov2001}, where the authors studied the limiting behaviour of the walk. A three-state DTQW, known as lively quantum walk on cycle graphs was introduced by Kajiwara et al. in \cite{Kajiwara2019,Sadowski2016}. Periodicity property of these walks is studied in \cite{Sarkar2020}. One-dimensional DTQWs on Cayley graphs of Dihedral groups using the Hadamard coin was first proposed by Dai et al. \cite{Dai2018}. This was further extended to three-state DTQW on Cayley graphs of Dihedral groups by Liu et al. \cite{Liu2021} using Grover coin, in which they studied the time-averaged probability of the walk. Recently, we have explored the periodicity and localization properties of DTQWs on Cayley graphs corresponding to Diheral groups for generalized Grover coins in \cite{SarmaSarkar2023}. Another model of DTQW that is studied recently include DTQWs for Cayley graphs of symmetric groups \cite{Banerjee2022}. Other similar models of quantum walks like quantum walks on graphs that are generated by free groups and virutally abelian quantum walks are analyzed in \cite{Acevedo2005,D'Ariano2016}. Works involving study of periodicity and localiation of DTQWs in line graphs, mixed paths and cycles can be found in \cite{Kubota2021}. For we detailed surveys on DTQWs and DTQWs on Cayley graphs, see \cite{Andraca2012}.

In this paper, we develop qutrit quantum circuit models for three-state DTQWs on Cayley graphs corresponding to Dihedral group $D_N$ and additive groups of integers modulo a positive integer $\Z_N$, where $3^{n-1}< N\leq 3^n$ for some positive integer $n\geq 1$. For both the graphs  we consider the quantum coins as special unitary matrices of order $3$. The quantum circuits are defined through some elementary qutrit gates which include one-qutrit rotation gates and two-qutrit controlled gates. A detailed analysis of circuit complexity of the models is also given. Finally, the models are numerically simulated based on well known noise models in order to investigate the performance of the circuit models in noisy quantum computers by setting the coin operators as  generalized Grover coins which are one-parameter one-qutrit gates and can be expressed as linear sum of permutation matrices. We reproduce the localization property of these walks that exist in literature through the developed circuit models and also review the changes in the time-averaged probability of finding the walker on the vertices of the graphs due to the noises. 

The developed circuit models require quantum circuit representation of generic one-qutrit gates that are special unitary matrices, block unitary matrices with $3\times 3$ diagonal blocks that are special unitary matrices, and finally decomposing multi-controlled gates such as generalized Toffoli gates into elementary qutrit gates. We provide a decomposition of generic one-qutrit gates into sequence of qutrit gates that can be viewed as an analogue of $ZYZ$ decomposition of special unitary matrices of order $2.$ We further develop a scalable quantum circuit implementation of block diagonal unitary matrices which correspond to multi-controlled qutrit gates. Finally we provide a mechanism for compiling multi-controlled qutrit gates into elementary qutrit gates.



The rest of the paper is organized as follows. In Section \ref{sec:2}, we review the definitions and related mathematical details of DTQWs on the Cayley graphs corresponding to $D_N$ and $\Z_N,$ and on elementary qutrit gates. Section \ref{QutritCircuit} addresses the quantum circuit implementation of generic special unitary matrices of order $3,$ qutrit circuit models for the DTQWs. The scalable quantum circuit representation of block diagonal unitary matrices and representing multi-controlled qutrit gates into elementary qutrit gates are given in Section \ref{Sec:4}. The circuit complexity of the developed circuit model for the DTQWs is also given in this section. Finally, we report the numerical simulation results in Section \ref{sec:5}.

\section{Preliminaries}\label{sec:2}

In this section, we briefly discuss Cayley graphs and the DTQW models on these graphs from \cite{Dummit1991,Dai2018,Liu2021,SarmaSarkar2023}. Then we provide an overview of qutrit gates that are building blocks in the construction of quantum qutrit circuit models for the DTQWs. 

\subsection{Cayley graphs} 

Given a group $(G,\circ)$ and a generating set $H\subseteq G$ of $G,$ the Cayley graph corresponding to the pair $(G,H)$ is defined as $\mathrm{Cay(G,H)}=(V,E)$ where $V$ is the set of vertices in $G$ and two elements $a,b\in V$ are linked by a directed edge from $a$ to $b$ if $b=a\circ c$ for some $c\in H$ and this edge is denoted by $(a,b)$ \cite{Dummit1991}. In particular, if $c=c^{-1}$ then the edge is both way directed and hence we call it an undirected edge. Hence, for underlying commutative groups, the corresponding Cayley graphs are undirected.

For instance, if $G=\Z_n,$ the additive group of integers modulo $N$ and $H=\{1,-1\}$ then $\mathrm{Cay(\Z_N,\{1,-1\})}$ is the undirected cycle graph on $N$ vertices. As an example we exhibit a cycle graph with $4$ vertices or $\mathrm{Cay(\Z_4,\{1,-1\})}$ in Figure \ref{CayGs}.

\begin{figure}[H]
\centering
\begin{subfigure}
\centering
\begin{tikzpicture} 
\tikzset{vertex/.style = {shape=circle,draw,minimum size=1.0em}}
\tikzset{edge/.style = {->,> = latex}}
\node[vertex,label={$0 \mbox{ mod } 4$}] (a) at  (0,0) {};
\node[vertex,label={$1 \mbox{ mod } 4$}] (b) at  (0,2) {};
\node[vertex,label={$2 \mbox{ mod } 4$}] (c) at  (2,2) {};
\node[vertex,label={$3 \mbox{ mod } 4$}] (d) at  (2,0) {};
\draw[blue](a)--(b);
\draw[blue] (b) -- (c);
\draw[blue] (c) to (d);
\draw[blue] (d) to (a);
\end{tikzpicture} 
\label{fig1CayleyCirc}
\end{subfigure}
\hspace{1.0cm} 
\begin{subfigure}
\centering
 \begin{tikzpicture} 
\tikzset{vertex/.style = {shape=circle,draw,minimum size=1.0em}}
\tikzset{edge/.style = {->,> = latex}}
\node[vertex,label={$(0,0)$}] (a) at  (0,0) {$e$};
\node[vertex,label={$(0,1)$}] (b) at  (0,2) {$a$};
\node[vertex,label={$(0,2)$}] (c) at  (2,2) {$a^2$};
\node[vertex,label={$(0,3)$}] (d) at  (2,0) {$a^3$};
\node[vertex,label={$(1,0)$}] (a1) at  (-1,-1) {$b$};
\node[vertex,label={$(1,1)$}] (b1) at  (-1,3) {$ba$};
\node[vertex,label={$(1,3)$}] (c1) at  (3,-1) {$ba^3$};
\node[vertex,label={$(1,2)$}] (d1) at  (3,3) {$ba^2$};
\draw[edge,red](a)--(b);
\draw[edge,red] (b) -- (c);
\draw[edge,red] (c) to (d);
\draw[edge,red] (d) to (a);
\draw[blue](a)--(a1);
\draw[blue] (d) -- (c1);
\draw[blue] (c) to (d1);
\draw[blue] (b) to (b1);
\draw[edge,red](a1)--(c1);
\draw[edge,red] (c1) -- (d1);
\draw[edge,red] (d1) to (b1);
\draw[edge,red] (b1) to (a1);
\end{tikzpicture} 
\end{subfigure}
\caption{ $\mathrm{Cay(\mathbb{Z}_4,\{1,-1\})}$ (left) and $\mathrm{Cay(D_4,\{a,b\})}$ (right)}
\label{CayGs}
\end{figure}

 
 For a positive integer $N,$ a Dihedral group $G$ is defined by two elements, say $a,b,$ called the generators of $G$ such that $b^2=a^N=e$, the identity element of the group and $aba=b$. From now onward, we denote this Dihedral group as $D_N=\langle \{a,b\}\rangle.$ Geometrically, $D_N$ is the group of symmetries of the regular $N$-gon \cite{Dummit1991}. The elements of $D_N$ is denoted as $b^sa^r,$ where $s\in \{0,1\},r\in \{0,1,\hdots,N-1\}$, and $b$ represents reflection about an axis of symmetry and the element $a$ represents a rotation by an angle of $\frac{2\pi}{N}$ about the center. If $s=0$, the $N$-gon admits rotation and the $N$-gon admits a reflection if $s=1$. Clearly, $D_N$ has $2N$ elements which can be labelled as the ordered pairs $(s,r),$ $0\leq s\leq 1,$ $0\leq r\leq N-1$, with $\{a,b\}$ a generating subset of $D_N.$ The Cayley graph $\mathrm{Cay(D_N,\{a,b\})}$ is a mixed graph meaning there exist directed edges between vertices $a^r$ and $a^{r+1}$ i.e. $a^r\rightarrow a^{r+1}$ along with $ba^{m+1}\rightarrow ba^{m}$. Further the undirected edges exist between the vertices $a^r$ and $ba^{r},$ for all $0\leq r\leq N-1$. We provide a labelling of the vertices which is $(s,r)$,  $s\in \{0,1\},r\in \{0,1,\hdots,N-1\}$ as described above. Then $\mathrm{Cay(D_N,\{a,b\})}$ can be depicted as two concentric directed cycle graphs possessing opposite orientations, where the vertices of the inner and outer cycle graphs are given by $(0,r)$ and $(1,r)$ respectively, and the undirected edges link the vertices $(0,r)$ with $(1,r).$ For example, in Figure \ref{CayGs}, we exhibit $\mathrm{Cay(D_4,\{a,b\})}.$




\subsection{Discrete-time quantum walks on $\mathrm{Cay(\Z_N,\{1,-1\})}$ and $\mathrm{Cay(D_N,\{a,b\})}$}  \label{sec:dtqws}

A DTQW on a graph is governed by a unitary operator $U=S(C\otimes I),$ which is applied repeatedly over time to the initial state $\ket{\psi(0)}$ of the walker. The operator $S$ is called the shift operator and $C$ is called the coin operator. Hence $U$ acts on the space $\mathcal{H}_C\otimes \mathcal{H}_p$ where $\mathcal{H}_C$ is the coin space whose dimension gives the internal degree of freedom of the quantum coin associated with the quantum walk and $\mathcal{H}_p$ is the position space spanned by the quantum states localized at the vertices of the graph. Here $\otimes$ denotes the tensor product between vector spaces. The state of the quantum walk at time $t$ for an initial state $\ket{\psi(0)}$ is given by $\ket{\psi(t)}=U^t\ket{\psi(0)}$.

Below we review the DTQWs on $\mathrm{Cay(\Z_N,\{1,-1\})}$ and $\mathrm{Cay(D_N,\{a,b\})}$ from \cite{Liu2021,SarmaSarkar2023} and \cite{Kajiwara2019,Sarkar2020}, respectively.

\subsubsection{Three-state lively quantum walk model on $\mathrm{Cay(\mathbb{Z}_N,\{1,-1\})}$}

The vertices of the cycle graph  $\mathrm{Cay(\mathbb{Z}_N,\{1,-1\})}$ can be represented as $m$, where $0\leq m\leq N-1$ and there are undirected edges between the vertices labeled $m$ and $m\pm 1$. The lively DTQW was first introduced in \cite{Sadowski2016} using the Grover coin, later We explored the periodicity properties of these walks in \cite{Sarkar2020} by considering the coin operators as generalized Grover coins.

The lively DTQW on cycle graph \cite{Sarkar2020} is defined on the Hilbert space $\mathcal{H}=\mathcal{H}_C\otimes \mathcal{H}_{V}\in \mathbb{C}^3\otimes \mathbb{C}^N$ where $\mathcal{H}_C=\mbox{span}\{\ket{0}_3,\ket{1}_3,\ket{2}_3\}$ such that $\{\ket{l}_3|l=0,1,2\}$ is the canonical basis of $\mathbb{C}^3,$ the coin space and $\mathcal{H}_V$ is the Hilbert space spanned by the vertices $\ket{m}_N\in V,\mbox{ }0\leq m\leq N-1$ of the cycle graph $\mathrm{Cay(\mathbb{Z}_n,\{1,-1\})}$ which are, in turn, the canonical basis of $\mathbb{C}^N$. Hence, for lively quantum walks, the proposed quantum walk corresponds to the unitary matrix $U=S(C\otimes I_N)$ which evolves in time where $C=[c_{ij}]\in U(3)$ and $S$ is the shift operator, which is as follows. \begin{eqnarray}\label{shift2}
    S &=& \ket{0}_3\bra{0}_3\otimes\sum_{m=0}^{N-1}\ket{m}_N\bra{m+1(\mod N)}_N \nonumber \\ 
    && +\ket{1}_3\bra{1}_3\otimes\sum_{m=0}^{N-1}\ket{m}_N\bra{m-1(\mod N)}_N \nonumber \\ 
    && +\ket{2}_3\bra{2}_3\otimes \sum_{m=0}^{N-1} \ket{m+a \mbox( mod N )}_N\bra{m}_N
\end{eqnarray}  for some liveliness factor $a\leq \lfloor \frac{N}{2}\rfloor$ i.e. the walker moves right,left or can jump to another vertex depending on the coin states. When $a=0$, the walk takes the form of the standard lazy three state quantum walk i.e. the walker may also stay put instead of jumping off to another vertex.\par
Hence, the discrete-time evolution of the walk is defined by $\ket{\psi(t)}=U^t\ket{\psi(0)}$ for an initial state $\ket{\psi(0)}\in \mathcal{H}$. Consequently, $$\ket{\psi(t)}=U^t\ket{\psi(0)}=\sum_{m=0}^{N-1}\sum_{l\in \{0,1,2\}}\psi(l,m,t) \ket{l}_3\otimes \ket{m}_N=\sum_{m=0}^{N-1}\ket{\psi(m,t)}\otimes \ket{m}_N,$$  where \begin{eqnarray*}
    \sum_{l=0}^2\sum_{r=0}^{N-1}|\psi(l,m,t)|^2=1.
\end{eqnarray*} 



\subsubsection{Three-state quantum walk model on  $\mathrm{Cay(D_N,\{a,b\})}$}\label{sec:cd}


Similar to lively quantum walk in cycles, the three-state DTQW on $\mathrm{Cay(D_N,\{a,b\})}$ is defined on the Hilbert space $\mathcal{H}=\mathcal{H}_C\otimes \mathcal{H}_{V}$ where $\mathcal{H}_C=\mbox{span}\{\ket{0}_3,\ket{1}_3,\ket{2}_3\}$ is the coin space and $\mathcal{H}_V$ is the Hilbert space spanned by the vertices of $\ket{(s,r)}\in V$ of $\mathrm{Cay(D_N,\{a,b\})}$. 

Hence, $\mathcal{H}_V=\mbox{span}\{\ket{s}_2\ket{r}_N| s\in \{0,1\},r\in \{0,1,\hdots,N-1\}\}\cong \C^2\otimes\C^N$. Thus the vertex set is given by $$\{\underbrace{\ket{0}_2\ket{0}_N,\ket{0}_2\ket{1}_N,\hdots \ket{0}_2\ket{N-1}_N}_{\mbox{rotation without reflection}},\overbrace{\ket{1}_2\ket{0}_N,\ket{1}_2\ket{1}_N,\hdots \ket{1}_2\ket{N-1}_N}^{\mbox{rotation after reflection}}\}$$ where the first qubit $\ket{.}_2$ represents the reflection qubit such that $\{\ket{s}_2| \, s=0,1\}$ is the canonical basis of $\mathbb{C}^2$ and $\{\ket{r}_N|0\leq r\leq N-1\}$ is the canonical basis of  $\mathbb{C}^N$. It is evident that the vertices labeled $(0,r)$ are the $r$-th vertices and the vertices labeled $(1,r)$ are the $(N+r)$-th vertices, $0\leq r\leq N-1.$ The proposed quantum walk evolves via the following unitary matrix $U=S(C\otimes I_2 \otimes I_N),$ where $C=[c_{ij}]\in U(3)$ and $S$ is the shift operator defined in the following way.  We follow the walk described in \cite{Liu2021,SarmaSarkar2023}, i.e. the walker performs one rotation in the direction of the edges of the cycle on which the walker resides if the coin state is $\ket{0}_3$. If the reflection state is $\ket{0}_2$, the walker moves along the inner directed cycle and if the reflection state is $\ket{1}_2$, the walker moves along the outer directed cycle. The walker remains at the same position if the coin state is $\ket{1}_3$, and walker jumps cycles via one reflection if the coin state is $\ket{2}_3$.

The the shift operator is defined as 
\begin{eqnarray}\label{shift3}
    S &=& \ket{0}_3\bra{0}_3\otimes\ket{0}_2\bra{0}_2\otimes\sum_{r=0}^{N-1}\ket{r}_N\bra{r-1(\mod N)}_N \nonumber \\ 
    && +\ket{0}_3\bra{0}_3\otimes\ket{1}_2\bra{1}_2\otimes\sum_{r=0}^{N-1}\ket{r}_N\bra{r+1(\mod N)}_N \nonumber \\ 
    && +\ket{1}_3\bra{1}_3\otimes\ket{0}_2\bra{0}_2\otimes \sum_{r=0}^{N-1} \ket{r}_N\bra{r}_N+\ket{1}_3\bra{1}_3\otimes\ket{1}_2\bra{1}_2\otimes  \sum_{r=0}^{N-1}\ket{r}_N\bra{r}_N \nonumber \\ && + \ket{2}_3\bra{2}_3\otimes\ket{0}_2\bra{1}_2\otimes \sum_{r=0}^{N-1}\ket{r}_N\bra{r}_N+\ket{2}_3\bra{2}_3\otimes\ket{1}_2\bra{0}_2\otimes  \sum_{r=0}^{N-1}\ket{r}_N\bra{r}_N. 
\end{eqnarray}

The  discrete-time evolution of the walk is defined by $\ket{\psi(t)}=U^t\ket{\psi(0)}$ for an initial state $\ket{\psi(0)}\in \mathcal{H}$ where $U$ is a $6N\times 6N$ matrix. Consequently, $$\ket{\psi(t)}=U^t\ket{\psi(0)}=\sum_{s=0,1}\sum_{r=0}^{N-1}\sum_{l\in \{0,1,2\}}\psi(l,s,r,t) \ket{l}_3\otimes \ket{s}_2\ket{m}_N=\sum_{m=0}^{N-1}\ket{\psi(m,t)}\otimes \ket{m}_N,$$  where \begin{eqnarray*}
    \sum_{l=0}^2\sum_{s=0}^1\sum_{r=0}^{N-1}|\psi(l,s,r,t)|^2=1.
\end{eqnarray*} 



\begin{remark}
Note that, constructing a qutrit circuit for this walk is  challenging since the reflection state is two-dimensional i.e. it represents a qubit. However, we shall see later in Section \ref{QutritCircuit} that constructing a circuit by taking the reflection state as a qutrit instead of qubit preserves the walk. 
\end{remark} 

\subsection{Qutrit rotation gates, qutrit $X$ gates and two-qutrit controlled gates}\label{uniq}

Now we recall some elementary quantum gates for qutrits such as qutrit rotation gates, qutrit $X$ gates along with the two-qutrit controlled gates which will play the pivotal role in developing a circuit model of the quantum walks discussed above. Obviously, the qubit rotation gates are given by $$R_X(\theta)=\bmatrix{\cos{\theta}&\iota \sin{\theta}\\ \iota \sin{\theta}&\cos{\theta}},R_Y(\theta)=\bmatrix{\cos{\theta}&\sin{\theta}\\-\sin{\theta}&\cos{\theta}},R_Z(\theta)=\bmatrix{e^{\iota \theta}&0\\0&e^{-\iota \theta}},$$ which represent the rotation of a qubit on the Bloch sphere around an angle $\theta$ with respect to the $X$, $Y$ and $Z$ axes respectively. Here $\iota=\sqrt{-1}.$

Now since a qutrit has three dimensions, we use the following elementary qutrit gates which were first used in \cite{Di2013,Wang2020}. These gates are formed by applying the qubit rotation gates on the subspace of dimension $2$ for a qutrit keeping the remaining dimension unchanged. Thus we have the following \textit{qutrit rotation gates}. 

\begin{eqnarray*}
    && R_{X01}(\theta)=\bmatrix{\cos{\theta} & \iota\sin{\theta}&0\\\iota\sin{\theta}&\cos{\theta}&0\\0&0&1}, \,  R_{X12}(\theta) =\bmatrix{1&0&0\\0&\cos{\theta} & \iota\sin{\theta}\\0&\iota\sin{\theta}&\cos{\theta}}, \, R_{X02}(\theta) = \bmatrix{\cos{\theta} & 0&\iota\sin{\theta}\\0&1&0\\\iota\sin{\theta}&0&\cos{\theta}} \\
   && R_{Y01}(\theta)=\bmatrix{\cos{\theta} & \sin{\theta}&0\\-\sin{\theta}&\cos{\theta}&0\\0&0&1}, \, 
   R_{Y12}(\theta) =\bmatrix{1 & 0&0\\0&\cos{\theta} & \sin{\theta}\\0&-\sin{\theta}&\cos{\theta}}, \,
   R_{Y02}(\theta) =\bmatrix{\cos{\theta} & 0&\sin{\theta}\\0&1&0\\-\sin{\theta}&0&\cos{\theta}} \\
   &&  R_{Z01}(\theta) =\bmatrix{e^{\iota \theta}&0&0\\0&e^{-\iota \theta}&0\\0&0&1}, \, 
    R_{Z12}(\theta) =\bmatrix{1&0&0\\0&e^{\iota \theta}&0\\0&0&e^{-\iota \theta}}, \, 
    R_{Z02}(\theta)=\bmatrix{e^{\iota \theta}&0&0\\0&1&0\\0&0&e^{-\iota \theta}}.
\end{eqnarray*} Thus if $i,j, k$ denote the three dimensions of a qubit then $R_{Sij}(\theta)$ denotes the $S$ qubit (subspace generated by $i,j$) rotation of the qutrit keeping the dimension $k$ fixed. We also use the following gate in sequel. $$S(\theta) = \bmatrix{e^{\iota \theta}&0&0\\0&e^{\iota \theta}&0\\0&0&e^{\iota \theta}}.$$

We also recall the following single qutrit gates considered in \cite{Saha2021} as follows.  \begin{eqnarray}
&& X_{0,1}=\bmatrix{0&1&0\\1&0&0\\0&0&1}, X_{1,2}=\bmatrix{1&0&0\\0&0&1\\0&1&0}, X_{0,2}=\bmatrix{0&0&1\\0&1&0\\1&0&0}, X_{+1}=\bmatrix{0&0&1\\1&0&0\\0&1&0},X_{+2}=\bmatrix{0&1&0\\0&0&1\\1&0&0}.\nonumber
\end{eqnarray} 
Obviously, $X_{p,q}$ gate maps $\ket{p}_3$ (respectively $\ket{q}_3$) to $\ket{q}_3$ ($\ket{p}_3$) where for $p,q\in \{0,1,2\}$ and $X_{+a}$ gate is a linear transformation from $\ket{p}_3$ to $\ket{a+p\mbox{ mod 3}}_3$. We shall call these single qutrit gates as \textit{qutrit-$X$ gates} where $X\in \{X_{0,1},X_{1,2},X_{0,2},X_{+1},X_{+2}\}$. 

Finally, we discuss \textit{two-qutrit controlled-$X$ gates} represented by a quantum circuit given by \vspace{0.25cm}

\centerline{\Qcircuit @C=1em @R=.7em {
 &\lstick{A}&\qw& \gate{\raisebox{.5pt}{\textcircled{\raisebox{-.9pt} {a}}} }& \qw\\
 &\lstick{B}&\qw& \gate{X} \qwx[-1]& \qw\\}} \vspace{0.25cm} 
 Obviously, this gate applies one-qutrit $X$ gate on the target (second) qutrit when the control (first) qutrit is $a\in \{0,1,2\}$. When $a=2$, such a gate is called Muthukrishnan-Stroud gate {\cite{Muthukrishnan2000}}. It is also of note that the circuit\\
\centerline{\Qcircuit @C=1em @R=.7em {
 &\lstick{}&\qw&\gate{\raisebox{.5pt}{\textcircled{\raisebox{-.9pt} {0}}} }& \qw\\
 &\lstick{}&\qw& \gate{X} \qwx[-1]& \qw\\}} is equivalent to \\\centerline{\Qcircuit @C=1em @R=.7em {
 &\lstick{}&\qw&\gate{X_{+2}} &\gate{\raisebox{.5pt}{\textcircled{\raisebox{-.9pt} {2}}} }&\gate{X_{+1}}  &\qw\\
 &\lstick{}&\qw&\qw &\gate{X} \qwx[-1]& \qw&\qw\\}} \\Similarly, the circuit\\ \centerline{\Qcircuit @C=1em @R=.7em {
 &\lstick{}&\qw& \gate{\raisebox{.5pt}{\textcircled{\raisebox{-.9pt} {1}}} }& \qw\\
 &\lstick{}&\qw& \gate{X} \qwx[-1]& \qw\\}} is equivalent to \\\centerline{\Qcircuit @C=1em @R=.7em {
 &\lstick{}&\qw&\gate{X_{+1}} &\gate{\raisebox{.5pt}{\textcircled{\raisebox{-.9pt} {2}}} }&\gate{X_{+2}}  &\qw\\
 &\lstick{}&\qw&\qw &\gate{X} \qwx[-1]& \qw&\qw\\}} 
\vspace{0.25cm}

We denote a $n$-qutrit gate, also known as \textit{$n$-qutrit controlled-$U$} gate where $U$ is a $(n-1)$-qutrit gate as follows. 

\centerline{\Qcircuit @C=1em @R=.7em {
 &\lstick{1}& \qw&\gate{\raisebox{.5pt}{\textcircled{\raisebox{-.9pt} {a}}} }&\qw \\&\lstick{2}& \qw&\multigate{2}{U}\qwx[-1]&\qw\\&\lstick{\vdots}& \qw&\ghost{U}&\qw\\&\lstick{n}& \qw&\ghost{U}&\qw}}.\vspace{0.25cm}\\ The circuit above means that $U$ acts on the $2^{nd}$ to $n^{th}$ qutrits if and only if the first qutrit is $\ket{a}_3,a\in \{0,1,2\}$.
 
 Another gate, we shall introduce in this work is the qudit SWAP gate i.e. given $0\leq a\leq b\leq d-1$, $\ket{a}_d\ket{b}_d\underbrace{\rightarrow}_{{SWAP^{(d)}_{a,b}}}\ket{b}_d\ket{a}_d$. The $SWAP^{(d)}_{a,b}$ is denoted by the following circuit.
    \begin{eqnarray}\label{Swap1}
\Qcircuit @C=0.25cm @R=.25cm {
\lstick{\ket{a}_d} & \qw \link{1}{1} & & \link{1}{-1} & \rstick{\ket{b}_d} \qw \\
& & \\
\lstick{\ket{b}_d} & \qw \link{-1}{1} & & \link{-1}{-1} & \rstick{\ket{a}_d} \qw}\end{eqnarray} 
For $d=3$, we get the qutrit version of the SWAP gate. It is of note that the ${SWAP^{(d)}_{a,b}}$ gate can be constructed using three controlled qudit-X gates in the following way 
\begin{eqnarray}\label{Swap2}
\Qcircuit @C=0.25cm @R=.25cm {
\lstick{}&\qw&\gate{\raisebox{.5pt}{\textcircled{\raisebox{-.9pt} {b}}}}&\gate{X_{a,b}}&\gate{\raisebox{.5pt}{\textcircled{\raisebox{-.9pt} {b}}} }&\qw\\
\lstick{}&\qw&\gate{X_{a,b}}\qwx[-1]&\gate{\raisebox{.5pt}{\textcircled{\raisebox{-.9pt} {b}}} }\qwx[-1]&\gate{X_{a,b}}\qwx[-1]&\qw\\}\end{eqnarray} where $X_{a,b}$ gate maps $\ket{a}_d$ to $\ket{b}_d$ and $\ket{b}_d$ to $\ket{a}_d$.\\

\section{Qutrit circuit models for three-state quantum walks}\label{QutritCircuit}

It is of note that, the coin operator in a three-state quantum walk or a lively quantum walk is a single qutrit gate. Hence, it is a perceptible goal to provide a qutrit circuit model in order to realize the three-state DTQWs defined in Section \ref{sec:dtqws} on $\mathrm{Cay}(D_N,\{a,b\})$) and $\mathrm{Cay}(\mathbb{Z}_N,\{1,-1\})$, where $3^{n-1}\leq N\leq 3^n$ for some positive integer $n$. In this section, we shall construct qutrit quantum circuit for the said quantum walks using single qutrit rotation gates and qutrit controlled-X gates/M-S gates. Our circuit construction is similar to the model proposed for two-state DTQWs on a cycle using Hadamard coins in \cite{Douglas2009}. 

First we provide a technique to decompose a generic single qutrit gate into single qutrit rotation gates in the following section. A similar technique is used in \cite{Fedullo1992} and a decomposition approach using Cartan algebra is carried in \cite{Di2013}.

\subsection{Decomposition of generic qutrit gates into qutrit-rotation gates}
 
  It is well known that for a vector $\bmatrix{a\\b}\in \mathbb{C}^2$, there exists a $2\times 2$ special unitary matrix $\bmatrix{\overline{a}&\overline{b}\\-b & a}$ such that $\frac{1}{\sqrt{|a|^2+|b^2|}}\bmatrix{\overline{a}&\overline{b}\\-b & a}\bmatrix{a\\b}=\bmatrix{\sqrt{|a|^2+|b|^2}\\0}$.\\

  It is also obvious that any $3\times 3$ unitary matrix is a single qutrit gate. Let the matrix be $C=\bmatrix{c_{11}&c_{12}&c_{13} \\ c_{21}&c_{22}&c_{23} \\ c_{31}&c_{32}&c_{33}}\in SU(3)$ such that each $c_{ij}$ is of the form $r_{ij}e^{-i\alpha_{ij}}$ where  $r_{ij}\geq 0$ and $\alpha_{ij}\in \mathbb{R}$. Hence, by left multiplying the matrix $M_1=\frac{1}{\sqrt{r_{11}^2+r_{31}^2}}\bmatrix{r_{11}e^{i\alpha_{11}}&0&r_{31}e^{i\alpha_{31}}\\0&1&0\\-r_{31}e^{-i\alpha_{31}}&0&r_{11}e^{-i\alpha_{11}}}$ to $C$, we get \begin{eqnarray*}
      M_1C=\bmatrix{\sqrt{r_{11}^2+r_{31}^2}&\times &\times\\c_{21}&\times&\times\\0&\times &\times}
  \end{eqnarray*} It is of note that $M_1$ can be written as \begin{eqnarray*}
      M_1:=M_1(\theta_1,\phi_1,\psi_1)= \bmatrix{\cos{\theta_1}e^{i\phi_1}&0&\sin{\theta_1}e^{i\psi_1}\\0&1&0\\-\sin{\theta_1}e^{-i\psi_1}&0&\cos{\theta_1}e^{-i\phi_1}}
  \end{eqnarray*} where $\theta_1=\arctan{\frac{r_{31}}{r_{11}}},\phi_1=\alpha_{11},\psi_1=\alpha_{31}$. Also, from Z-Y-Z decomposition \cite{NielsenChuang2002}, \begin{eqnarray}\label{M02}
      M_1(\theta_1,\phi_1,\psi_1)=R_{Z02}\left(\frac{\phi_1+\psi_1}{2}\right)R_{Y02}\left(\theta_1\right)R_{Z02}\left(\frac{\phi_1-\psi_1}{2}\right)
  \end{eqnarray}
Similarly, left multiplying the matrix $M_1C$ by the matrix $M_2=\frac{1}{\sqrt{r_{11}^2+r_{21}^2+r_{31}^2}}\bmatrix{\sqrt{r_{11}^2+r_{21}^2}& r_{21}e^{i\alpha_{21}}&0\\-r_{21}e^{-i\alpha_{21}}&\sqrt{r_{11}^2+r_{21}^2}&0\\0&0&1}$, we obtain \begin{eqnarray*}
M_2M_1C=\bmatrix{\sqrt{r_{11}^2+r_{21}^2+r_{31}^2}&\times&\times\\0&\times&\times\\0&\times&\times}=\bmatrix{1&\times&\times\\0&\times&\times\\0&\times&\times}
\end{eqnarray*} since $|c_{11}|^2+|c_{21}|^2+|c_{31}|^2=1$ and similarly, \begin{eqnarray*}
M_2:=M_2(\theta_2,\phi_2,\psi_2)=\bmatrix{\cos{\theta_2}e^{i\phi_2}&\sin{\theta_2}e^{i\psi_2}&0\\-\sin{\theta_2}e^{-i\psi_2}&\cos{\theta_2}e^{-i\phi_2}&0\\0&0&1}
\end{eqnarray*}  where $\theta_2=\arctan{\frac{\sqrt{r_{11}^2+r_{31}^2}}{r_{21}}},\phi_2=0,\psi_2=\alpha_{21}$ i.e. we obtain $M_2(\theta_2,0,\psi_2)$. Further from, Z-Y-Z decomposition again, we have 
  \begin{eqnarray}\label{M01}
      M_2(\theta_2,0,\psi_2)&=& R_{Z01}\left(\frac{\psi_2}{2}\right)R_{Y01}\left(\theta_2\right)R_{Z01}\left(\frac{-\psi_2}{2}\right).
  \end{eqnarray}
 
Continuing in this way, we see that there exists \begin{eqnarray*}
M_3=M_3(\theta_3,\phi_3,\psi_3)=\bmatrix{1&0&0\\0&\cos{\theta_3}e^{i\phi_3}&\sin{\theta_3}e^{i\psi_3}\\0&-\sin{\theta_3}e^{-i\psi_3}&\cos{\theta_3}e^{-i\phi_3}}
\end{eqnarray*}
such that $M_3M_2M_1C=\underbrace{\bmatrix{1 &\times&\times\\0&r&\times\\0&0&\times}}_R$ where $r>0.$

Clearly, $R$ is an upper triangular matrix which is also a special unitary matrix. Hence, $R$ is a special unitary diagonal matrix. Further, since two diagonal entries of $R$ are real and positive, hence entries in $R$ cannot be complex on account of the determinant being $1$. Hence $R=I$ i.e. the identity matrix. We also have \begin{eqnarray}\label{M12}
    M_3(\theta_3,\phi_3,\psi_3)&=& R_{Z12}\left(\frac{\phi_3+\psi_3}{2}\right) R_{Y12}\left(\theta_3\right)R_{Z12}\left(\frac{\phi_3-\psi_3}{2}\right).
\end{eqnarray}
Hence \begin{eqnarray*}
M_3(\theta_3,\phi_3,\psi_3)M_2(\theta_2,0,\psi_2)M_1(\theta_1,\phi_1,\psi_1)C&=&I.
\end{eqnarray*} Thus we obtain
   \begin{eqnarray*}
C&=&M^*_1(\theta_1,\phi_1,\psi_1)M^*_2(\theta_2,0,\psi_2)M_1(\theta_1,\phi_1,\psi_1) M^*_3(\theta_3,\phi_3,\psi_3)\\
    &=&M_1(-\theta_1,-\phi_1,\psi_1)M_2(-\theta_2,0,\psi_2)M_3(-\theta_3,-\phi_3,\psi_3)
\end{eqnarray*} for any arbitrary $C\in SU(3)$. Thus, from the above discussion, we have the following theorem. 

\begin{theorem}\label{qutritZYZ}
Any $3\times 3$ special unitary matrix $U=[u_{ij}]_{3\times 3}$ has the following parametric representation 
    \begin{eqnarray*}
        u_{11}&=&\cos{\theta_1}\cos{\theta_2}e^{-i\phi_1}\\
        u_{12}&=&-(\sin{\theta_1}\sin{\theta_3}e^{i(\psi_1-\psi_3)}+\cos{\theta_1}\sin{\theta_2}\cos{\theta_3}e^{i(\psi_2-\phi_1-\phi_3)})\\
        u_{13}&=&(-\sin{\theta_1}\cos{\theta_3}e^{i(\phi_3+\psi_1)}+\cos{\theta_1}\sin{\theta_2}\sin{\theta_3}e^{i(\psi_2-\phi_1+\psi_3)})\\
        u_{21}&=&\sin{\theta_2}e^{-i\psi_2}\\
        u_{22}&=&\cos{\theta_2}\cos{\theta_3}e^{-i\phi_3}\\
        u_{23}&=&\cos{\theta_2}\sin{\theta_3}e^{-i(\psi_3-\pi)}\\
        u_{31}&=&\sin{\theta_1}\cos{\theta_2}e^{-i\psi_1}\\
        u_{32}&=&(\cos{\theta_1}\sin{\theta_3}e^{i(\phi_1-\psi_3)}-\sin{\theta_1}\sin{\theta_2}\cos{\theta_3}e^{i(\psi_2-\psi_1-\phi_3)})\\
        u_{33}&=&(\cos{\theta_1}\cos{\theta_3}e^{i(\phi_1+\phi_3)}+\sin{\theta_1}\sin{\theta_2}\sin{\theta_3}e^{i(-\psi_1+\psi_2+\psi_3)}),
    \end{eqnarray*} where $\theta_1,\theta_2,\theta_3,\phi_1,\phi_3,\psi_1,\psi_2,\psi_3\in \mathbb{R}.$
Besides, $U$ has the following factorization through qutrit-rotation gates. \begin{eqnarray*}
    U&=&R_{Z02}\left(\frac{-\phi_1+\psi_1}{2}\right) R_{Y02}\left(-\theta_1\right) R_{Z02}\left(\frac{-\phi_1-\psi_1}{2}\right) \\ && R_{Z01}\left(\frac{\psi_2}{2}\right) R_{Y01}\left(-\theta_2\right) R_{Z01}\left(\frac{-\psi_2}{2}\right) \\ && R_{Z12}\left(\frac{-\phi_3+\psi_3}{2}\right) R_{Y12}\left(-\theta_3\right) R_{Z12}\left(\frac{-\phi_3-\psi_3}{2}\right).
\end{eqnarray*} 
\end{theorem}

\begin{remark}
\begin{enumerate}
    \item The Theorem \ref{qutritZYZ}, can be considered as qutrit analogue of the qubit Z-Y-Z decomposition.
    \item It is to be noted that the matrix $U$ in theorem \ref{qutritZYZ}, requires $8$-parameters which is precisely the dimension of the manifold $SU(3)$.
    
    \item Given any matrix from $SU(3)$, the $9$ equations are easily solvable. For example, let us take $V=[v_{ij}]_{3\times 3}\in SU(3)$. Then through some computation, we obtain \begin{eqnarray*}
        \theta_2&=&\arcsin{|v_{21}|}\\
        \psi_2&=&\arctan{\arg(v_{21})}\\
        \theta_1&=&\arccos{\frac{|v_{11}|}{\cos{\theta_2}}},\mbox{where }\theta_2 \neq \frac{\pi}{2}\\
        \phi_1&=&-\arctan{\arg(v_{11})}\\
        \psi_1&=&-\arctan{\arg(v_{31})}\\
        \theta_3&=&\arccos{\frac{|v_{22}|}{\cos{\theta_2}}},\mbox{where }\theta_2\neq \frac{\pi}{2}\\
        \phi_3&=&-\arctan{\arg(v_{22})}\\
        \psi_3&=&-\arctan{\arg(v_{23})}+\pi.\\
    \end{eqnarray*} We shall take $\psi_3=0,\theta_1=\frac{\pi}{2}=\theta_3,\mbox{when }\theta_2=\frac{\pi}{2}.$
    
    \item Further any $3\times 3$ unitary matrix can be written as $e^{i\alpha}U$ for some real $\alpha$ where $U\in SU(3)$. Thus any unitary matrix has a similar decomposition. 
\end{enumerate}
\end{remark}


Below We provide a factorization of $3\times 3$ unitary diagonal matrices through qutrit rotation gates that will be used in sequel.

\begin{lemma}\label{diagonal}
Given a $3\times 3$ diagonal matrix $D=\mbox{diag}(e^{\iota\alpha},e^{\iota\beta},e^{\iota\zeta})\in U(3), \alpha,\beta,\zeta\in \R$. Then \begin{eqnarray*}
    D&=&\exp{\left(\frac{\alpha+\beta+\zeta}{3}I\right)}R_{Z01}\left(\frac{2\alpha-\beta-\zeta}{3}\right) R_{Z12}\left(\frac{\alpha+\beta-2\zeta}{3}\right).
\end{eqnarray*}    
\end{lemma}
\pf Follows directly from the definition of $R_{Z01},R_{Z12}$. \hfill{$\square$}

\begin{corollary}\label{spdiagonal}
A $3\times 3$ diagonal unitary matrix of the form $\Tilde{D}=\mbox{diag}(e^{\iota\alpha},e^{\iota\beta},e^{-\iota(\alpha+\beta)})\in SU(3),\alpha,\beta\in \R$ can be decomposed as \begin{eqnarray*}
    \Tilde{D}=R_{Z01}(\alpha)R_{Z12}(\alpha+\beta)
    = R_{Z02}(\alpha)R_{Z12}(\beta)
    = R_{Z02}(\alpha+\beta)R_{Z01}(-\beta).
\end{eqnarray*}        
\end{corollary}
\pf The proof directly follows by comparing RHS and LHS. \hfill{$\square$}

\subsection{Qutrit circuit model for DTQW on $\mathrm{Cay(D_N,\{a,b\})}$} 

Recall that the vertices of $\mathrm{Cay(D_N,\{a,b\})}$ are labelled as $(s,r),s\in \{0,1\},r\in\{0,1\hdots,N-1\}$ where $s$ denotes the reflection and $r$ denotes rotation. Thus the position space is given by the Hilbert space $\mathcal{H}_V=\mbox{span}\{\ket{s}_2\ket{r}_N \, : \, s\in \{0,1\}, r\in \{0,1,\hdots,N-1\}\}\cong \mathbb{C}^2\otimes \mathbb{C}^N$. Now in order to incorporate this labelling into the proposed qutrit circuit model, we consider the quantum states corresponding to the vertex set as $\{\ket{s}_3\ket{r_n}_3\ket{r_{n-1}}_3\hdots \ket{r_1}_3 \,:\, s\in \{0,1\},r_j\in \{0,1,2\}, j\in\{1,2,\hdots,n-1\}\}$ where the ternary representation of $r$ is given by $r=\sum_{j=1}^{n}r_j3^{j-1}$ i.e. $\ket{r}_N=\ket{r_n}_3\ket{r_{n-1}}_3\hdots \ket{r_1}_3$. It is also of note that the reflection state $\ket{s}_2$ belongs to $\mathbb{C}^2$, which we consider as a qutrit i.e. $\ket{s}_3$. Hence, $\ket{s}_3\in \{\ket{0}_3,\ket{1}_3,\ket{2}_3\},$ the canonical basis sates of $\mathbb{C}^3$. Thus from our construction, we label the $r$-th vertex of the inner cycle of $\mathrm{Cay(D_N,\{a,b\})}$ labelled by $(0,r)$ as $\ket{0}_3\ket{r_n}_3\ket{r_{n-1}}_3\hdots \ket{r_1}_3$ where $0\leq r_j\leq 2;$ and the state $\ket{1}_3\ket{r_n}_3\ket{r_{n-1}}_3\hdots \ket{r_1}_3$ represents the $N+r$-th vertex labeled by $(1,r)$ i.e. the $r$-th vertex in the outer cycle of $\mathrm{Cay(D_N,\{a,b\})}$.  Because of such qutrit circuit representation of vertices, it is evident that $n+1$ qutrits are required to describe the position space, in which the first qutrit represents the reflection state and the rest $n$-qutrits for the reflection state. We shall use another qutrit for the quantum coin state from $\C^3$. Hence, for the quantum circuit model of the quantum walk, $n+2$ qutrits are required i.e. the unitary matrix corresponding to the qutrit quantum circuit is of order $3^{n+2}$.

Moreover, the circuit should be constructed in such a way that the input states of the form $\ket{c}_3\ket{2}_3\ket{r_n}_3\ket{r_{n-1}}_3\hdots \ket{r_1}_3$ remain invariant where $\ket{c}_3\in \mathcal{H}_C$ denotes the coin state such that $c\in \{0,1,2\},r_j\in \{0,1,2\},j\in \{1,2,\hdots,n\}$. By constructing the circuit in such a way, we shall prove that the quantum circuit thus formed does not impede on the structure of the walk. First, we construct the following qutrit quantum gates using a similar approach used in \cite{Douglas2009}.


\vspace{0.25cm}

 \textsc{Increment}: \\\centerline{\Qcircuit @C=1em @R=.7em {
 &\lstick{1}&\qw&\gate{X_{+1}} &\qw & \qw &\qw &\qw&\qw\\
 &\lstick{2}&\qw&\gate{\raisebox{.5pt}{\textcircled{\raisebox{-.9pt} {2}}} }\qwx[-1] &\gate{X_{+1}}  & \qw &\hdots &\hdots&\qw\\
 &\lstick{3}&\qw&\gate{\raisebox{.5pt}{\textcircled{\raisebox{-.9pt} {2}}} } \qwx[-1] &\gate{\raisebox{.5pt}{\textcircled{\raisebox{-.9pt} {2}}} }\qwx[-1] &\gate{X_{+1}} &\hdots &\hdots&\qw\\
 &\lstick{\vdots}&\qw&\vdots &\vdots &\vdots &\vdots &\vdots&\qw\\
 &\lstick{n-1}&\qw&\gate{\raisebox{.5pt}{\textcircled{\raisebox{-.9pt} {2}}} }  &\gate{\raisebox{.5pt}{\textcircled{\raisebox{-.9pt} {2}}} }  &\gate{\raisebox{.5pt}{\textcircled{\raisebox{-.9pt} {2}}} }  &\gate{X_{+1}}  &\qw&\qw\\
 &\lstick{n}&\qw&\gate{\raisebox{.5pt}{\textcircled{\raisebox{-.9pt} {2}}} } \qwx[-1]&\gate{\raisebox{.5pt}{\textcircled{\raisebox{-.9pt} {2}}} } \qwx[-1]& \gate{\raisebox{.5pt}{\textcircled{\raisebox{-.9pt} {2}}} } \qwx[-1] &\gate{\raisebox{.5pt}{\textcircled{\raisebox{-.9pt} {2}}} }\qwx[-1] &\gate{X_{+1}}&\qw\\}} 

\vspace{0.25cm}

and

\vspace{0.25cm}
 
  \textsc{Decrement}:\\\centerline{\Qcircuit @C=1em @R=.7em {
 &\lstick{1}&\qw&\gate{X_{+2}} &\qw & \qw &\qw &\qw&\qw\\
 &\lstick{2}&\qw&\gate{\raisebox{.5pt}{\textcircled{\raisebox{-.9pt} {0}}} }\qwx[-1] &\gate{X_{+2}}  & \qw &\hdots &\hdots&\qw\\
 &\lstick{3}&\qw&\gate{\raisebox{.5pt}{\textcircled{\raisebox{-.9pt} {0}}} } \qwx[-1] &\gate{\raisebox{.5pt}{\textcircled{\raisebox{-.9pt} {0}}} }\qwx[-1] &\gate{X_{+2}} &\hdots &\hdots&\qw\\
 &\lstick{\vdots}&\qw&\vdots &\vdots &\vdots &\vdots &\vdots&\qw\\
 &\lstick{n-1}&\qw&\gate{\raisebox{.5pt}{\textcircled{\raisebox{-.9pt} {0}}} }  &\gate{\raisebox{.5pt}{\textcircled{\raisebox{-.9pt} {0}}} }  &\gate{\raisebox{.5pt}{\textcircled{\raisebox{-.9pt} {2}}} }  &\gate{X_{+2}}  &\qw&\qw\\
 &\lstick{n}&\qw&\gate{\raisebox{.5pt}{\textcircled{\raisebox{-.9pt} {0}}} } \qwx[-1]&\gate{\raisebox{.5pt}{\textcircled{\raisebox{-.9pt} {0}}} } \qwx[-1]& \gate{\raisebox{.5pt}{\textcircled{\raisebox{-.9pt} {0}}} } \qwx[-1] &\gate{\raisebox{.5pt}{\textcircled{\raisebox{-.9pt} {0}}} }\qwx[-1] &\gate{X_{+2}}&\qw\\}} 
 
 \vspace{0.25cm} 
 
 Clearly, these gates comprise of multi-controlled qutrit gates. Each string of qutrits representing the rotation state of the vertex after passing through increment or decrement will always give us the immediate adjacent rotation state i.e.  
 \begin{eqnarray*}
\ket{r_n}_3\ket{r_{n-1}}_3\hdots \ket{r_1}_3 \overbrace{\rightarrow}_{\mbox{increment}}\ket{r'_n}_3\ket{r'_{n-1}}_3\hdots \ket{r_1}_3\\ 
 \ket{r_n}_3\ket{r_{n-1}}_3\hdots \ket{r_1}_3\underbrace{\rightarrow}_{\mbox{decrement}}\ket{\Tilde{r}_n}_3\ket{\Tilde{r}_{n-1}}_3\hdots \ket{\Tilde{r}_1}_3
 \end{eqnarray*} where $\sum_{j=1}^nr'_j3^{j-1}=(r+1)\mod N$ and $\sum_{j=1}^n\Tilde{r}_j3^{j-1}=(r-1)\mod N$.  It is also evident that  when the coin state is $\ket{0}_3$ and reflection state is $\ket{0}_3$ then using the \textsc{Increment} gates on the rotation states, the state $\ket{0}_3\ket{0}_3\ket{r_n}_3\ket{r_{n-1}}_3\hdots \ket{r_1}_3$ is mapped to $\ket{0}_3\ket{0}_3\ket{r'_n}_3\ket{r'_{n-1}}_3\hdots \ket{r'_1}_3$. Similarly, when the coin state is $\ket{0}_3$ and reflection state is $\ket{1}_3$, then using the \textsc{Decrement} gates on the rotation states, the state $\ket{0}_3\ket{1}_3\ket{r_n}_3\ket{r_{n-1}}_3\hdots \ket{r_1}_3$ is mapped to $\ket{0}_3\ket{1}_3\ket{\Tilde{r}_n}_3\ket{\Tilde{r}_{n-1}}_3\hdots \ket{\Tilde{r}_1}_3$. Besides, the \textsc{Increment} and \textsc{Decrement} gates also maintain a periodic condition of the walk, which means $\ket{0}_3\ket{0}_3\hdots\ket{0}_3$ and $\ket{2}_3\ket{2}_3\hdots\ket{2}_3$ are adjacent rotation states. Hence these gates map $r$-th (resp. $(N+r)$-th) vertex to $r\pm 1$-th ( resp. $(N+r\pm 1)$ -th ) vertex.

 Now, using all previous circuits discuss above, we present the circuit model for three-state quantum walks on Cayley graph of Dihedral group $\mathrm{Cay}(D_N,\{a,b\})$ where $N=3^n$ as follows. \vspace{0.25cm}
\\\centerline{\Qcircuit @C=1em @R=.7em {
 &\lstick{1}&\qw&\gate{C}& \gate{C^\star}&\qw&\gate{\raisebox{.5pt}{\textcircled{\raisebox{-.9pt} {2}}} }&\qw &\gate{\raisebox{.5pt}{\textcircled{\raisebox{-.9pt} {0}}} }&\qw&\gate{\raisebox{.5pt}{\textcircled{\raisebox{-.9pt} {0}}} }&\qw\\ 
 &\lstick{2}&\qw&\qw&\gate{\raisebox{.5pt}{\textcircled{\raisebox{-.9pt} {2}}} }\qwx[-1]&\qw&\gate{X_{0,1}}\qwx[-1]&\qw&\gate{\raisebox{.5pt}{\textcircled{\raisebox{-.9pt} {0}}} }\qwx[-1]&\qw&\gate{\raisebox{.5pt}{\textcircled{\raisebox{-.9pt} {1}}} }\qwx[-1]&\qw\\
&\lstick{3}&\qw&\qw& \qw&\qw&\qw&\qw&\multigate{4}{\textsc{Increment}}\qwx[-1]&\qw&\multigate{4}{ \textsc{Decrement}}\qwx[-1]&\qw\\
&\lstick{4}&\qw&\qw& \qw&\qw&\qw&\qw&\ghost{ \textsc{Increment}}&\qw&\ghost{ \textsc{Decrement}}&\qw\\
&\lstick{\vdots}&\qw&\qw& \qw&\qw&\qw&\qw&\ghost{ \textsc{Increment}}&\qw&\ghost{\textsc{Decrement}}&\qw\\
&\lstick{n+1}&\qw&\qw& \qw&\qw&\qw&\qw&\ghost{ \textsc{Increment}}&\qw&\ghost{ \textsc{Decrement}}&\qw\\
&\lstick{n+2}&\qw&\qw& \qw&\qw&\qw&\qw&\ghost{ \textsc{Increment}}&\qw&\ghost{ \textsc{Decrement}}&\qw\\}}  

\vspace{0.25cm}  

where $C$ and $C^\star$ are the coin operator and its inverse respectively. 

From our proposed circuit, we observe the following. 
\begin{itemize}
    \item Let $N=27$ and the walker starts with the state $\ket{0}_3^{\otimes 5}$ at $t=0$. Let the coin $C=(c_{ij})_{3\times 3}$. Then at $t=1$, the walker has the state $\ket{\psi(1)}=c_{11}\ket{0}_3\ket{0}_3\ket{001}_3+c_{21}\ket{1}_3\ket{0}_3\ket{000}_3+c_{31}\ket{2}_3\ket{1}_3\ket{000}_3$. This is due to the fact that $C\ket{0}_3=c_{11}\ket{0}_3+c_{21}\ket{1}_3+c_{31}\ket{2}_3$. Now since the second qutrit is not $\ket{2}_3$, it remains unaffected through the second gate in the circuit and then the $X_{0,1}$ gate applies when the first qutrit is $\ket{2}_3$. The final two gates act as increment/decrement operators when the first two qutrits of the state of the walker are $\ket{0}_3\ket{0}_3$ and $\ket{0}\ket{1}_3$ respectively and we get our result. Similarly for $t=2$, we get the following state \begin{eqnarray*}
    \ket{\psi(2)}=c^2_{11}\ket{0}_3\ket{0}_3\ket{002}_3+c_{11}c_{21}\ket{1}_3\ket{0}_3\ket{001}_3+c_{11}c_{31}\ket{2}_3\ket{1}_3\ket{001}_3\\+c_{12}c_{21}\ket{0}_3\ket{0}_3\ket{001}_3+c_{22}c_{21}\ket{1}_3\ket{0}_3\ket{000}_3+c_{32}c_{21}\ket{2}_3\ket{1}_3\ket{000}_3\\+c_{31}c_{13}\ket{0}_3\ket{1}_3\ket{222}_3+c_{31}c_{23}\ket{1}_3\ket{1}_3\ket{000}_3+c_{31}c_{33}\ket{2}_3\ket{0}_3\ket{000}_3
\end{eqnarray*} 
The probability of finding the walker at the starting point $(0,0)$ i.e. $\ket{0}_3\ket{000}_3$ at $t=2$ is $|c_{22}c_{21}|^2+|c_{31}c_{33}|^2$.\par
When $N=25$, for $t=2$, the state becomes \begin{eqnarray*}
    \ket{\psi(2)}=c^2_{11}\ket{0}_3\ket{0}_3\ket{002}_3+c_{11}c_{21}\ket{1}_3\ket{0}_3\ket{001}_3+c_{11}c_{31}\ket{2}_3\ket{1}_3\ket{001}_3\\+c_{12}c_{21}\ket{0}_3\ket{0}_3\ket{001}_3+c_{22}c_{21}\ket{1}_3\ket{0}_3\ket{000}_3+c_{32}c_{21}\ket{2}_3\ket{1}_3\ket{000}_3\\+c_{31}c_{13}\ket{0}_3\ket{1}_3\ket{221}_3+c_{31}c_{23}\ket{1}_3\ket{1}_3\ket{000}_3+c_{31}c_{33}\ket{2}_3\ket{0}_3\ket{000}_3
\end{eqnarray*} It is easy to see that for $t=1$ and $t=2$, the final state becomes a linear combination of $3$ and $9$ quantum states respectively and the walker passes over $3$ and $6$ distinct vertices respectively.
    \item The input state $\ket{l}_3\ket{2}_3\ket{r_n}_3\hdots\ket{r_1}_3$ is invariant for all $l\in \{0,1,2\},r_j\in\{0,1,2\},j\in \{1,\hdots,n\}$.
    \item The quantum circuit is scalable i.e. using the circuit for DTQW on $\mathrm{Cay}(D_{3^n},\{a,b\})$, one can construct the circuit for  DTQW on $\mathrm{Cay}(D_{3^{n+1}},\{1,-1\})$, by addition of multi-controlled qutrit-X gates to construct \textsc{Increment} and \textsc{Decrement} gates.
\end{itemize}

The quantum circuit proposed so far can also be modified for $\mathrm{Cay}(D_N,\{a,b\}),$ $3^{n-1}< N<3^n$. The main idea is to incorporate necessary qutrit controlled gates in between the \textsc{Increment} and \textsc{Decrement} gates so that the periodic conditions are suitably modified. Let, $N=k < 3^n$ and hence, $k=\sum_{j=1}^{n}3^{j-1}k_j$ where $k_j\in \{0,1,2\},0\leq j\leq n$ and not all $k_j$'s are equal to $2$. Then we assign necessary multi-controlled qutrit $X$ gates in order to modify the periodic conditions so that the $\ket{s}_3\ket{k_{n}}_3\hdots \ket{k_1}_3$ and $\ket{s}_3\ket{0}_3\hdots \ket{0}_3$ are adjacent vertices for $s\in \{0,1\}$. Let us consider $\ket{k'_n}_3\hdots \ket{k'_1}_3$ such that $\ket{k'_n}_3\hdots \ket{k'_1}_3$ and $\ket{k_{n}}_3\hdots \ket{k_1}_3$ differ in exactly one qutrit. We call these vertices as \textit{Hamming-1} vertices. Hence, the main idea is to find suitable set of multi-controlled qutrit $X$ gates such that we can construct a mapping between the vertices in the following way \begin{eqnarray*}
    \ket{0}_3\ket{k_n}_3\hdots\ket{k_1}_3\underbrace{\rightarrow} \ket{0}_3\ket{k'_n}_3\hdots\ket{k'_1}_3\underbrace{\rightarrow\hdots}_{\mbox{Hamming-1 vertices}} \ket{0}_3\ket{2}_3\hdots\ket{2}_3\rightarrow \ket{0}_3\ket{0}_3\hdots\ket{0}_3\\
    \ket{1}_3\ket{0}_3\hdots\ket{0}_3\underbrace{\rightarrow} \ket{1}_3\ket{2}_3\hdots\ket{2}_3\underbrace{\rightarrow\hdots}_{\mbox{Hamming-1 vertices}} \ket{1}_3\ket{k'_n}_3\hdots\ket{k'_1}_3\rightarrow \ket{1}_3\ket{k_n}_3\hdots\ket{k_1}_3   
\end{eqnarray*}The mapping from one state $\ket{k_n}_3\hdots \ket{k_1}_3$ to another state $\ket{0}_3\hdots \ket{0}_3$ by creating a sequence of Hamming-1 vertices, is similar to the construction of qubit circuits that maps one state to another through multi-controlled CNOT gates using Gray codes \cite{NielsenChuang2002}.

For any $N$, the basic idea is to incorporate a multi-controlled qutrit-$X$ gate with the target qutrit located at the single position where the two adjacent Hamming-1 vertices differ, and then undoing the process. To be more descriptive, suppose we are given a starting state $a=\ket{a_n}_3\hdots\ket{a_1}_3 $ and ending state $b=\ket{b_n}_3\hdots\ket{b_1}_3$ and our aim is to construct a quantum gate that maps $a$ to $b$ and $b$ to $a$. Then we want to find a sequence of Hamming-1 vertices such that \begin{eqnarray*}
    a\underbrace{\rightarrow} a^{(1)},\underbrace{\hdots\hdots}_{\mbox{Hamming-1 vertices}}, a^{(n-1)}\rightarrow a^{(n)}= b.
\end{eqnarray*}

Thus, firstly we shall find a multi-controlled qutrit-$X$ gate that swaps $a$ and $a^{(1)}$. Suppose $a$ and $a^{(1)}$ differ at the $i$-th digit. Then a multi-controlled $X$ gate can be used to flip the qutrit on the $i$-th position. The multi-controlled qutrit $X$ gate is constructed keeping in mind the values of the other qutrits being same to those in both $a$ and $a^{(1)}$. This process can be repeated for $a^{(1)}$ and $a^{(2)}$ until a multi-controlled qutrit $X$ gate is obtained that maps $a^{(n-1)}$ to $b$ and vice versa. Hence, using this construction, one can map from $a$ to $b$. In order to complete the circuit however, the circuit must also map from $b$ to $a$. A similar construction can be designed as follows. Suppose $b$ and $a^{(n-1)}$ differ in the $j^{th}$ qutrit. Then apply a multi-controlled qutrit $X$ gate with the $j^{th}$ qutrit as target and the gate being conditional on the other qutrits having the same values as in both $b$ and $a^{(n-1)}$. Hence, continuing this way, a map can be  constructed in order to obtain the sequence \begin{eqnarray*}
    b\underbrace{\rightarrow} a^{(n-1)}\underbrace{\rightarrow\hdots}_{\mbox{Hamming-1 vertices}} a^{(1)}\rightarrow a.
\end{eqnarray*}

Let us look at an example. Consider three-state DTQW on $\mathrm{Cay}(D_{25},\{a,b\})$. In this case $N=25$ which is less than $27=3^3$. Hence in this case, the last vertex of both directed cycles in the graph is represented  by $\ket{s}_3\ket{2}_3\ket{2}_3\ket{0}_3, s\in \{0,1\}$. Hence, we shall incorporate multi-controlled qutrit gates comprising of generalized multi-controlled qutrit $X$ gates in between the \textsc{Increment} and \textsc{Decrement} gates such that $\ket{s}_3\ket{0}_3\ket{0}_3\ket{0}_3$ and $\ket{s}_3\ket{2}_3\ket{2}_3\ket{0}_3$ are adjacent vertices and the input states $\ket{l}_3\ket{s}_3\ket{2}_3\ket{2}_3\ket{2}_3$ and $\ket{c}_3\ket{s}_3\ket{2}_3\ket{2}_3\ket{1}_3$ remains invariant in the circuit where $\ket{c}_3$ is the coin state such that $c\in \{0,1,2\}$. Thus we construct the following two unitaries.
 
\textsc{Instop}: \\ \centerline{\Qcircuit @C=1em @R=.7em {
 &\lstick{}&\qw&\gate{X_{0,2}} &\gate{\raisebox{.5pt}{\textcircled{\raisebox{-.9pt} {0}}} } & \gate{\raisebox{.5pt}{\textcircled{\raisebox{-.9pt} {0}}} }&\gate{\raisebox{.5pt}{\textcircled{\raisebox{-.9pt} {0}}} } &\gate{X_{0,2}} &\gate{\raisebox{.5pt}{\textcircled{\raisebox{-.9pt} {2}}} }&\qw\\
 &\lstick{}&\qw&\gate{\raisebox{.5pt}{\textcircled{\raisebox{-.9pt} {2}}} }\qwx[-1] &\gate{X_{0,2}} \qwx[-1]  &\gate{\raisebox{.5pt}{\textcircled{\raisebox{-.9pt} {0}}} } \qwx[-1]  &\gate{X_{0,2}} \qwx[-1] &\gate{\raisebox{.5pt}{\textcircled{\raisebox{-.9pt} {2}}} } \qwx[-1]&\gate{\raisebox{.5pt}{\textcircled{\raisebox{-.9pt} {2}}} }\qwx[-1]&\qw\\
 &\lstick{}&\qw&\gate{\raisebox{.5pt}{\textcircled{\raisebox{-.9pt} {1}}} } \qwx[-1] &\gate{\raisebox{.5pt}{\textcircled{\raisebox{-.9pt} {1}}} }\qwx[-1] & \gate{X_{0,1}} \qwx[-1] &\gate{\raisebox{.5pt}{\textcircled{\raisebox{-.9pt} {1}}} }\qwx[-1] &\gate{\raisebox{.5pt}{\textcircled{\raisebox{-.9pt} {1}}} } \qwx[-1]&\gate{X_{1,2} }\qwx[-1]&\qw\\}} 
 
 and 
 
 \textsc{Destop}:
 \\\centerline{\Qcircuit @C=1em @R=.7em {
 &\lstick{}&\qw&\gate{\raisebox{.5pt}{\textcircled{\raisebox{-.9pt} {2}}} } &\gate{\raisebox{.5pt}{\textcircled{\raisebox{-.9pt} {2}}} } &\qw\\
 &\lstick{}&\qw&\gate{\raisebox{.5pt}{\textcircled{\raisebox{-.9pt} {2}}} }\qwx[-1] &\gate{\raisebox{.5pt}{\textcircled{\raisebox{-.9pt} {2}}} }\qwx[-1]&\qw\\
 &\lstick{}&\qw&\gate{X_{0,2}} \qwx[-1] &\gate{X_{1,2}} \qwx[-1]&\qw\\}}\vspace{0.25cm}

Using the gates above, the circuit for the walk on $\mathrm{Cay(D_{25},\{a,b\})}$ is given by 

\vspace{0.25cm}

\centerline{\Qcircuit @C=1em @R=.7em {
 &\lstick{1}&\qw&\gate{C}& \gate{C^*}&\qw&\gate{\raisebox{.5pt}{\textcircled{\raisebox{-.9pt} {2}}} }&\qw &\gate{\raisebox{.5pt}{\textcircled{\raisebox{-.9pt} {0}}} }&\gate{\raisebox{.5pt}{\textcircled{\raisebox{-.9pt} {0}}} }&\gate{\raisebox{.5pt}{\textcircled{\raisebox{-.9pt} {0}}} }&\gate{\raisebox{.5pt}{\textcircled{\raisebox{-.9pt} {0}}} } &\qw\\ 
 &\lstick{2}&\qw&\qw&\gate{\raisebox{.5pt}{\textcircled{\raisebox{-.9pt} {2}}} }\qwx[-1]&\qw&\gate{X_{0,1}}\qwx[-1]&\qw&\gate{\raisebox{.5pt}{\textcircled{\raisebox{-.9pt} {0}}} }\qwx[-1]&\gate{\raisebox{.5pt}{\textcircled{\raisebox{-.9pt} {0}}} }\qwx[-1]&\gate{\raisebox{.5pt}{\textcircled{\raisebox{-.9pt} {1}}} }\qwx[-1]&\gate{\raisebox{.5pt}{\textcircled{\raisebox{-.9pt} {1}}} }\qwx[-1]&\qw\\
&\lstick{3}&\qw&\qw& \qw&\qw&\qw&\qw&\multigate{2}{\textsc{Increment}}\qwx[-1]&\multigate{2}{\textsc{Instop}}\qwx[-1]&\multigate{2}{\textsc{Decrement}}\qwx[-1]&\multigate{2}{\textsc{Destop}}\qwx[-1]&\qw\\
&\lstick{4}&\qw&\qw& \qw&\qw&\qw&\qw&\ghost{\textsc{Increment}}&\ghost{\textsc{Instop}}&\ghost{\textsc{Decrement}}&\ghost{\textsc{Destop}}&\qw\\
&\lstick{5}&\qw&\qw& \qw&\qw&\qw&\qw&\ghost{\textsc{Increment}}&\ghost{\textsc{Instop}}&\ghost{\textsc{Decrement}}&\ghost{\textsc{Destop}}&\qw\\}}

\vspace{0.25cm} 

Now, we shall prove that constructing quantum circuits with reflection states as qutrits does not affect the structure of the walk.  


\begin{theorem}
Let $U_\mathrm{Cay(D_N)}\in \C^{6N\times 6N}, 3^{n-1}\leq N\leq 3^{n}$ be the unitary matrix corresponding to the three state DTQW on $\mathrm{Cay(D_N,\{a,b\})}$ obtained theoretically as described in Section \ref{sec:cd}. Also, let $U_\mathrm{circ}\in\C^{3^{n+2}\times 3^{n+2}}$ be the unitary matrix corresponding to the qutrit circuit for the three state DTQW $\mathrm{Cay(D_N,\{a,b\})}$. Then there exists a permutation matrix $P$ such that  $$U_\mathrm{circ}=P\left[ 
\begin{array}{c|c} 
      {U}_{\mathrm{Cay(D_N)}} &  0 \\ 
      \hline 
       0 &  I_{(3^{n+2}-6N)} \\
    \end{array} 
    \right]P^{T}$$
\end{theorem}
\pf Let $N=3^n$. Then from the construction of $U_\mathrm{circ}$, we see that any state of the form $\ket{l}_3\ket{2}_3\ket{r_n}_3\hdots \ket{r_1}_3$ i.e. any state with $\ket{2}_3$ in the reflection state remains invariant to the circuit where the coin state $\ket{l}_3\in \{\ket{0}_3,\ket{1}_3,\ket{2}_3\}$, and $ 0\leq r\leq N-1$ such that $r=\sum_{j=1}^nr_j3^{j-1},r_j\in \{0,1,2\},j\in \{1,\hdots,n\}$. For the state $\ket{l}_3\ket{s}_3\ket{r_n}_3\hdots \ket{r_1}_3 ,s\in \{0,1\},r_j\in\{0,1,2\},j\in \{1,\hdots,n\}$, the circuit acts non-trivially.

Then, the state of the walker after time $t$ 
is given by ${\ket{{\phi(t)}}}=U_{\mathrm{circ}}^t\ket{\phi(0)}$ for initial state $\ket{\phi(0)}$. Further \begin{eqnarray*}
        \ket{{\phi}(t)}=\sum_{s=0}^2\sum_{{r=0}}^{N-1}\sum_{l\in \{0,1,2\}}\phi(l,s,r,t) \ket{l}\otimes\ket{s}_3\ket{r_n}_3\hdots \ket{r_1}_3= \sum_{{r=0}}^{N-1}\ket{{\phi}(r,t)}\otimes\ket{r_n}_3\hdots \ket{r_1}_3
    \end{eqnarray*} where $\sum_{s=0}^2\sum_{r=0}^{N-1}|\phi(l,s,r,t)|^2=1$ and $r=\sum_{j=1}^nr_j3^{j-1},r_j\in \{0,1,2\},j\in \{1,\hdots,n\}$. Further $\ket{{\phi}(r,t)}=\bmatrix{{\phi}(0,0,r,t)\\{\phi}(0,1,r,t)\\{\phi}(0,2,r,t)\\{\phi}(1,0,r,t)\\{\phi}(1,1,r,t)\\{\phi}(1,2,r,t)\\{\phi}(2,0,r,t)\\{\phi}(2,1,r,t)\\{\phi}(2,2,r,t)}$. If the coin operator is $C=[c_{ij}]\in\C^{3\times 3}$ then the probability that the walker will be at the vertex labelled $(s,r)$ at time $t$ is $\sum_{l=0,1,2}|{\phi}(l,s,r,t)|^2,$ where $s\in \{0,1\}$. Further, the state $\ket{l}_3\ket{2}_3\ket{r_n}_3\hdots \ket{r_1}_3$ is invariant in our circuit for all $0\leq r\leq N-1,0\leq l\leq 2$. Thus it is computationally easy to verify that \begin{eqnarray*}
        \ket{\phi(r,t+1)}=V_1\ket{\phi(r-1,t)}+V_2\ket{\phi(r+1,t)}+V_3\ket{\phi(r,t)}
    \end{eqnarray*} where \begin{eqnarray*}
        V_1&=&\bmatrix{c_{11} & 0 & 0 & c_{12} & 0 &0 &c_{13} & 0 & 0\\0 & 0 & 0 & 0 & 0 &0 &0 & 0 & 0\\0 & 0 & 0 & 0 & 0 &0 &0 & 0 & 0\\0 & 0 & 0 & 0 & 0 &0 &0 & 0 & 0\\0 & 0 & 0 & 0 & 0 &0 &0 & 0 & 0\\0 & 0 & 0 & 0 & 0 &0 &0 & 0 & 0\\0 & 0 & 0 & 0 & 0 &0 &0 & 0 & 0\\0 & 0 & 0 & 0 & 0 &0 &0 & 0 & 0\\0 & 0 & 0 & 0 & 0 &0 &0 & 0 & 0},\\ V_2&=&\bmatrix{0 & 0 & 0 & 0 & 0 &0 &0 & 0 & 0\\0& c_{11} & 0 & 0 & c_{12} & 0 &0 &c_{13} & 0\\0 & 0 & 0 & 0 & 0 &0 &0 & 0 & 0\\0 & 0 & 0 & 0 & 0 &0 &0 & 0 & 0\\0 & 0 & 0 & 0 & 0 &0 &0 & 0 & 0\\0 & 0 & 0 & 0 & 0 &0 &0 & 0 & 0\\0 & 0 & 0 & 0 & 0 &0 &0 & 0 & 0\\0 & 0 & 0 & 0 & 0 &0 &0 & 0 & 0\\0 & 0 & 0 & 0 & 0 &0 &0 & 0 & 0},\\ V_3&=&\bmatrix{0 & 0 & 0 & 0 & 0 &0 &0 & 0 & 0\\0& 0 & 0 & 0 & 0 & 0 &0 &0 & 0\\0 & 0 & 1 & 0 & 0 &0 &0 & 0 & 0\\c_{21} & 0 & 0 & c_{22} & 0 &0 &c_{23} & 0 & 0\\0 & c_{21} & 0 & 0 & c_{22} &0 &0 & c_{23} & 0\\0 & 0 & 0 & 0 & 0 &1 &0 & 0 & 0\\0 & c_{31} & 0 & 0 & c_{32} &0 &0 & c_{33} & 0\\c_{31} & 0 & 0 & c_{32} & 0 &0 &c_{32} & 0 & 0\\0 & 0 & 0 & 0 & 0 &0 &0 & 0 & 1}.
    \end{eqnarray*} 
    
    Further, using the Discrete Fourier (DFT) Transformation\cite{Nakahara2008} of $\ket{\phi(r,t)}$, we get  $$\ket{\Phi(k,t)}=\sum_{r=0}^{N-1}e^{-\iota 2\pi kr/N}\ket{\phi(r,t)}$$ where $0\leq k\leq N-1.$
Hence, we obtain \begin{eqnarray}
    \ket{(\Phi(k,t+1))} &=& \sum_{r=0}^{N-1}e^{-\iota 2\pi kr/N} V_1\ket{(\phi(r+1,t))}+\sum_{r=0}^{N-1}e^{-\iota 2\pi kr/N}V_2\ket{(\phi(r-1,t))} \nonumber \\ && +{\sum_{r=0}^{N-1}e^{-\iota 2\pi kr/N}V_3\ket{(\phi(r,t))}}. \nonumber
\end{eqnarray}
Consequently, we obtain the Fourier evolution matrix  $$\ket{(\Phi(k,t+1))}=U_{\mathrm{circ}}(k)\ket{\Phi(k,t)},$$

where 
    
   \begin{eqnarray*}
   U_{\mathrm{circ}}(k)= \bmatrix{c_{11}e^{-2\iota\pi k/N} & 0 & 0 & c_{12}e^{-2\iota\pi k/N}&0& 0&c_{13}e^{-2\iota\pi k/N}&0&0\\0&c_{11}e^{2\iota\pi k/N} & 0 &0 & c_{12}e^{2\iota\pi k/N}& 0&0&c_{13}e^{2\iota\pi k/N}&0\\0 & 0 & 1 & 0 & 0 &0 &0 & 0 & 0\\c_{21} & 0 & 0 & c_{22} & 0 &0 &c_{23} & 0 & 0\\0 & c_{21} & 0 & 0 & c_{22} &0 &0 & c_{23} & 0\\0 & 0 & 0 & 0 & 0 &1 &0 & 0 & 0\\0 & c_{31} & 0 & 0 & c_{32} &0 &0 & c_{33} & 0\\c_{31} & 0 & 0 & c_{32} & 0 &0 &c_{32} & 0 & 0\\0 & 0 & 0 & 0 & 0 &0 &0 & 0 & 1}     
   \end{eqnarray*}

From elementary computation, we observe that $$U_{circ}(k)=P\left[ 
\begin{array}{c|c} 
      {U}_{\mathrm{Cay(D_N)}}(k) &  0 \\ 
      \hline 
       0 &  I_3\\
    \end{array} 
    \right]P^{T}$$ holds true for all $k\in \{0,1,\hdots,N-1\}$ where $P$ is a $9\times 9$ permutation  matrix such that $P=P_{(34)}P_{(45)}P_{(57)}P_{(68)}$, and ${U}_{\mathrm{Cay(D_N)}}(k)\in\C^{6\times 6}$ is the Fourier evolution matrix of $U_{\mathrm{Cay(D_N)}}$ (see \cite{SarmaSarkar2023}). Note that $P_{(a,b)}$ denotes a transposition matrix i.e. an identity matrix whose $a$ and $b$-th rows are interchanged. 
    
    From Theorem 2.1 in \cite{SarmaSarkar2023}, we know that the spectra of $U_{\mathrm{circ}}(k)\mbox{ and }U_{\mathrm{Cay(D_N)}}(k),0\leq k\leq N-1$ are equal to that of $U_{\mathrm{circ}}$ and $U_{\mathrm{Cay(D_N)}}$ respectively. Hence, we obtain \begin{eqnarray*}
    U_{\mathrm{circ}} &=& (P\otimes I_{3^{n}})\left[ 
\begin{array}{c|c} 
      {{U}_{\mathrm{Cay(D_N)}}}_{6.3^{n}\times 6.3^{n}} &  0 \\ 
      \hline 
       0 &  I_{3^{n+2}-6.3^{n}}\\
    \end{array} 
    \right](P\otimes I_{3^{n}})^T \\
    &=& (P\otimes I_{3^{n}})\left[ 
\begin{array}{c|c} 
      {{U}_{\mathrm{Cay(D_N)}}}_{6.3^{n}\times 6.3^{n}} &  0 \\ 
      \hline 
       0 &  I_{3^{n+1}}\\
    \end{array} 
    \right](P\otimes I_{3^{n}})^T. 
    \end{eqnarray*} This concludes the proof for $N=3^n.$
    
    Next, consider $3^{n-1}\leq N< 3^n$. Then the additional states such as $\ket{l}_3\ket{s}_3\ket{r'_n}_3\hdots\ket{r'_1}_3$ are unaffected in the circuit where $s\in \{0,1\}$ and $r'_j\in \{0,1,2\}$ such that $\sum_{j=1}^nr'_j3^{j-1}>N$. 
     The rest of the proof is similar to the first case and after a bit of computations we obtain \begin{eqnarray*}        
   U_{\mathrm{circ}}=(P\otimes I_{3^{n}})\left[ 
\begin{array}{c|c|c} 
      {{U}_{\mathrm{Cay(D_N)}}}_{6N\times 6N} &  0 &0 \\ 
      \hline 
      0&I_{6.3^n-6N}&0\\
       \hline
       0 &0&I_{3^{n+2}-6N}\\
    \end{array} 
    \right](P\otimes I_{3^{n}})^T. \end{eqnarray*} This completes the proof. \hfill{$\square$}\\
    
Hence, we have successfully provided a qutrit quantum circuit for three state discrete time quantum walk on $\mathrm{Cay(D_N,\{a,b\})}$. In the next section, we shall provide somewhat similar circuit construction for three state lively quantum walks on cycle graphs.

\subsection{Qutrit circuit model for three-state lively DTQWs on $\mathrm{Cay(\mathbb{Z}_N,\{1,-1\})}$}

We now propose a qutrit quantum circuit model for the lively DTQW on cycle graphs \cite{Kajiwara2019}. We recall that the walker moves to the left if the coin state $\ket{0}_3$ and walker moves to the right if the coin state is $\ket{1}_3$ and jumps to vertex at distance $a(\mod N)$ if the coin state is $\ket{2}_3$ where $0< a\leq \lfloor \frac{N}{2}\rfloor$ is the liveliness parameter. For $a=0$, it is obvious that the walk becomes a standard three-state lazy quantum walk \cite{Kajiwara2019} on cycles. For further analysis of the walk, see \cite{Kajiwara2019,Sarkar2020}. 

Similar to the previous model, in our circuit model, we see that for the cycle of length $3^{n-1}\leq N\leq 3^n$, we require total $n+1$ qutrits. The first qutrit is required for the coin states and the rest $n$-qutrits are required for the vertices in order to represent the position states. 
 We first construct the simple model of three state lively quantum walks on cycles where the liveliness parameter is taken to be $0$ i.e. the standard three state lazy quantum walk. The circuit for three-state lazy DTQW on cycle of length $3^n$ is as follows.
 \vspace{0.25cm}
\\\centerline{\Qcircuit @C=1em @R=.7em {
 &\lstick{1}&\qw&\gate{C}& \qw&\qw&\qw&\qw &\gate{\raisebox{.5pt}{\textcircled{\raisebox{-.9pt} {0}}} }&\qw&\gate{\raisebox{.5pt}{\textcircled{\raisebox{-.9pt} {1}}} }&\qw\\ 
&\lstick{2}&\qw&\qw& \qw&\qw&\qw&\qw&\multigate{4}{\textsc{Decrement}}\qwx[-1]&\qw&\multigate{4}{\textsc{Increment}}\qwx[-1]&\qw\\
&\lstick{3}&\qw&\qw& \qw&\qw&\qw&\qw&\ghost{\textsc{Decrement}}&\qw&\ghost{\textsc{Increment}}&\qw\\
&\lstick{\vdots}&\qw&\qw& \qw&\qw&\qw&\qw&\ghost{\textsc{Decrement}}&\qw&\ghost{\textsc{Increment}}&\qw\\
&\lstick{n}&\qw&\qw& \qw&\qw&\qw&\qw&\ghost{\textsc{Decrement}}&\qw&\ghost{\textsc{Increment}}&\qw\\
&\lstick{n+1}&\qw&\qw& \qw&\qw&\qw&\qw&\ghost{\textsc{Decrement}}&\qw&\ghost{\textsc{Increment}}&\qw\\}}\vspace{0.25cm}\\ 
The observations obtained from the above circuit are as follows.
\begin{itemize}
    \item The walker has a non-zero probability of being found at all vertices of the cycle after $N$-time steps.
    \item The quantum circuit is again scalable.
\end{itemize}

Similarly, when $3^{n-1}\leq N<3^n$ we use multi-controlled qutrit $X$ gates in between. For example, let us look at a walk with $25$ vertices. Then we take the circuit \\\centerline{\Qcircuit @C=1em @R=.7em {
 &\lstick{1}&\qw&\gate{C}& \qw&\qw&\qw&\qw &\gate{\raisebox{.5pt}{\textcircled{\raisebox{-.9pt} {1}}} }&\gate{\raisebox{.5pt}{\textcircled{\raisebox{-.9pt} {1}}} }&\gate{\raisebox{.5pt}{\textcircled{\raisebox{-.9pt} {0}}} }&\gate{\raisebox{.5pt}{\textcircled{\raisebox{-.9pt} {0}}} }&\qw\\ 
&\lstick{2}&\qw&\qw& \qw&\qw&\qw&\qw&\multigate{2}{\textsc{Increment}}\qwx[-1]&\multigate{2}{\textsc{Instop}}\qwx[-1]&\multigate{2}{\textsc{Decrement}}\qwx[-1]&\multigate{2}{\textsc{Destop}}\qwx[-1]&\qw\\
&\lstick{3}&\qw&\qw& \qw&\qw&\qw&\qw&\ghost{\textsc{Increment}}&\ghost{\textsc{Instop}}&\ghost{\textsc{Decrement}}&\ghost{\textsc{Destop}}&\qw\\
&\lstick{4}&\qw&\qw& \qw&\qw&\qw&\qw&\ghost{\textsc{Increment}}&\ghost{\textsc{Instop}}&\ghost{\textsc{Decrement}}&\ghost{\textsc{Destop}}&\qw\\}} \vspace{0.25cm}

 For lively quantum walk on cycles with non-zero liveliness operator $a\leq \lfloor\frac{N}{2}\rfloor$ and $N=3^n$, the circuit needs to be modified slightly in the following manner. Define a new repeated circuit 
 $\textsc{RC}(a)$ as \\\centerline{$\underbrace{{{\Qcircuit @C=1em @R=.7em {
 &\lstick{1}& \qw&\multigate{2}{\textsc{Increment}}&\multigate{2}{\hdots}&\multigate{2}{\textsc{Increment}}&\qw\\&\lstick{\vdots}& \qw&\ghost{\textsc{Increment}}&\ghost{\hdots}&\ghost{\textsc{Increment}}&\qw\\&\lstick{n}& \qw&\ghost{\textsc{Increment}}&\ghost{\hdots}&\ghost{\textsc{Increment}}&\qw}}}}_{\mbox{a times }}$}\vspace{0.25cm} 

Hence, the circuit for three state lazy DTQWs on cycle of length $3^n$ is as follows.
\\\centerline{\Qcircuit @C=1em @R=.7em {
 &\lstick{1}&\qw&\gate{C}& \qw&\qw&\qw&\qw &\gate{\raisebox{.5pt}{\textcircled{\raisebox{-.9pt} {0}}} }&\qw&\gate{\raisebox{.5pt}{\textcircled{\raisebox{-.9pt} {1}}} }&\qw &\gate{\raisebox{.5pt}{\textcircled{\raisebox{-.9pt} {2}}} }&\qw\\ 
&\lstick{2}&\qw&\qw& \qw&\qw&\qw&\qw&\multigate{4}{\textsc{Decrement}}\qwx[-1]&\qw&\multigate{4}{\textsc{Increment}}\qwx[-1]&\qw&\multigate{4}{\textsc{RC}(a)}\qwx[-1]&\qw\\
&\lstick{3}&\qw&\qw& \qw&\qw&\qw&\qw&\ghost{\textsc{Decrement}}&\qw&\ghost{\textsc{Increment}}&\qw&\ghost{\textsc{RC}(a)}&\qw\\
&\lstick{\vdots}&\qw&\qw& \qw&\qw&\qw&\qw&\ghost{\textsc{Decrement}}&\qw&\ghost{\textsc{Increment}}&\qw&\ghost{\textsc{RC}(a)}&\qw\\
&\lstick{n}&\qw&\qw& \qw&\qw&\qw&\qw&\ghost{\textsc{Decrement}}&\qw&\ghost{\textsc{Increment}}&\qw&\ghost{\textsc{RC}(a)}&\qw\\
&\lstick{n+1}&\qw&\qw& \qw&\qw&\qw&\qw&\ghost{\textsc{Decrement}}&\qw&\ghost{\textsc{Increment}}&\qw&\ghost{\textsc{RC}(a)}&\qw\\}} \vspace{0.25cm}\\  
Similarly, when $3^{n-1}\leq N <3^n$, then one can add the gates \textsc{Instop} and \textsc{Destop} in the circuit to modify the periodic condition.

\begin{remark} Note from our circuit construction that the coin operators $C$ and $C^\star$ can be implemented though the elementary qutrit gates due to Theorem \ref{qutritZYZ}. However, the implementation of multi-controlled qutrit gates used to construct \textsc{Increment}, \textsc{Decrement}, \textsc{Instop} and \textsc{Destop} through elementary qutrit gates need further investigation. These circuits can also be viewed as $n$-qutrit Toffoli gates. In Section \ref{Sec:4}, we provide construction of qutri-circuits for block diagonal unitary matrices for $3\times 3$ diagonal blocks and any $n$-qutrit Toffoli gate is a special case. 
 \end{remark}

\section{Scalable qutrit circuit implementation of $3\times 3$ block diagonal unitary matrices}\label{Sec:4}

The quantum circuit models as proposed above utilize multi-controlled qutrit-$X$ gates of the following form.  \\
\begin{eqnarray}\label{circ1}
    \centerline{\Qcircuit @C=1em @R=.7em {
 &\lstick{1}&\qw&\qw& \gate{X} &\qw&\qw\\
 &\lstick{2}&\qw&\qw&\gate{\raisebox{.5pt}{\textcircled{\raisebox{-.9pt} {a}}} }\qwx[-1] &\qw&\qw\\
&\lstick{\vdots}&\qw& &\vdots & &\qw\\
&\lstick{n-1}&\qw&\qw& \gate{\raisebox{.5pt}{\textcircled{\raisebox{-.9pt} {a}}} }  &\qw &\qw\\
&\lstick{n}&\qw&\qw& \gate{\raisebox{.5pt}{\textcircled{\raisebox{-.9pt} {a}}} } \qwx[-1]&\qw&\qw\\}}
\end{eqnarray} where $a\in\{0,1,2\}.$ In this section, we develop a quantum circuit model of its implementation through single qutrit rotation gates and controlled qutrit-$X$ gates. 

First, we note that the matrix representation of the circuit 
\begin{eqnarray}\label{circ2}
\centerline{\Qcircuit @C=1em @R=.7em {
 &\lstick{1}&\qw&\qw& \gate{\raisebox{.5pt}{\textcircled{\raisebox{-.9pt} {a}}} } &\qw&\qw\\
 &\lstick{2}&\qw&\qw&\gate{\raisebox{.5pt}{\textcircled{\raisebox{-.9pt} {a}}} }\qwx[-1] &\qw&\qw\\
&\lstick{\vdots}&\qw& &\vdots & &\qw\\
&\lstick{n-1}&\qw&\qw& \gate{\raisebox{.5pt}{\textcircled{\raisebox{-.9pt} {a}}} } &\qw&\qw \\
&\lstick{n}&\qw&\qw& \gate{X} \qwx[-1]&\qw&\qw\\}}    
\end{eqnarray}
\vspace{0.25cm}\\
is a block diagonal special unitary matrix. In what follows, we construct an ancilla free quantum circuit for $n$-qutrit block-diagonal special unitary matrices having $3\times 3$ special unitary blocks using M-S gates and single-qutrit rotation gates. 

We emphasize that the obtained circuit is scalable i.e. we can construct circuit for $n+1$-qutrit block diagonal special unitary using circuits from $n$-qutrit block diagonal special unitary. The construction is also exact and the process follows similar to work done in \cite{SarmaSarkar2023}. It is also of note that since any unitary matrix $U$ can always be expressed as $e^{i\alpha}V$ for some $\alpha\in \R$ and $V$ is a special unitary matrix i.e. any unitary matrix is a special unitary matrix multiplied by a global phase, hence, we restrict our construction to special unitary matrices only. Using the construction of block diagonal unitary matrices and in turn, the circuit in equation (\ref{circ2}), we also provide a circuit construction for the qutrit gate in equation (\ref{circ1}).

\begin{definition} \label{def:mcgate}
For $n$-qutrit systems, a multi-controlled rotation gate is defined as 
$$\framebox[12cm][c]{\Qcircuit @C=1em @R=.7em {
 &\lstick{1}&\qw& \gate{\circ} &\gate{\circ} &\qw & \hdots &\gate{\circ} &\gate{\circ}&\qw\\
 &\lstick{2}&\qw&\gate{\circ}\qwx[-1] &\gate{\circ}\qwx[-1] &\qw & \hdots &\gate{\circ}\qwx[-1] &\gate{\circ}\qwx[-1] &\qw\\
&\lstick{\vdots}&\qw&\vdots&\vdots &\hdots & \hdots  &\vdots &\vdots &\qw\\
&\lstick{n-1}&\qw& \gate{\circ} &\gate{\circ} &\qw &\hdots &\gate{\circ} &\gate{\circ} &\qw\\
&\lstick{n}&\qw& \gate{R_a(\psi_1)} \qwx[-1]&\gate{R_a(\psi_2)} \qwx[-1] &\qw& \hdots &\gate{R_a(\psi_{3^{n-2}})} \qwx[-1]&\gate{R_a(\psi_{3^{n-1}})}\qwx[-1] &\qw\\}}$$
where \Qcircuit @C=1em @R=.7em {&\gate{\circ}&\qw\\} $\in\{$  \Qcircuit @C=1em @R=.7em{&\qw&\gate{\raisebox{.5pt}{\textcircled{\raisebox{-.9pt} {0}}} }&\qw\\},\Qcircuit @C=1em @R=.7em{&\qw&\gate{\raisebox{.5pt}{\textcircled{\raisebox{-.9pt} {1}}}}&\qw\\},\Qcircuit @C=1em @R=.7em{&\qw&\gate{\raisebox{.5pt}{\textcircled{\raisebox{-.9pt} {2}}}}&\qw\\}$\}$, $\psi_j\in \R, 1\leq j\leq 3^{n-1}$ and $$a\in\{X01,X02,X12,Y01,Y02,Y12,Z01,Z02,Z12\}.$$ Then the unitary matrix corresponding to the above circuit is given by $$F_n(R_a(\psi_1,\psi_2,\hdots,\psi_{3^{k-1}})) =\left[ \begin{array}{c|c|c|c} 
      R_a(\psi_1) &  0 &0 & 0\\ 
      \hline 
       0 &  0&\ddots &0\\
        \hline
       0 & 0& 0& R_a(\psi_{3^n-1})
    \end{array} 
    \right], $$
 which is a block diagonal matrices with $3\times 3$ rotation blocks. 
\end{definition}
In short, we denote the circuit in Definition \ref{def:mcgate} as


\begin{eqnarray}\label{multi1}
\framebox[8cm][c]{\Qcircuit @C=1em @R=.7em {
    &\lstick{1}&\qw& \gate{} &\qw\\
    &\lstick{2}&\qw&\gate{ }\qwx[-1]&\qw\\
    &\lstick{\vdots}&\qw&\gate{ }\qwx[-1]&\qw\\
    &\lstick{n-1}&\qw&\gate{ }\qwx[-1]&\qw\\
    &\lstick{n}&\qw&\gate{F_n(R_a(\Psi))}\qwx[-1]&\qw\\}}\end{eqnarray} such that $\Psi=(\psi_1,\hdots,\psi_{3^{n-1}})$.\\ 
    
    In order to demonstrate the scalability of such multi-controlled gates, win the following we construct $(n+1)$-qutrit multi-controlled gates that consists of  $n$-qutrit multi-controlled rotation gates and some additional two-qutrit control gates. It is to be noted that similar construction has been done in \cite{Di2013}, however, in this work, we provide a more general construction of generalized Toffoli gates mentioned in equation (\ref{circ1}) using the construction of multi-controlled qutrit X gates. 
    
    First, consider the following circuits. with $,\Phi=(\phi_1,\hdots,\phi_{3^{n-2}}),\Theta=(\theta_1,\hdots,\theta_{3^{n-2}}),\Gamma=(\gamma_1,\hdots,\gamma_{3^{n-2}}).$ 

\begin{eqnarray}\label{multiY1}
\framebox[17cm][c]{\Qcircuit @C=1em @R=.7em {
    &\lstick{1}&\qw&\qw&\gate{\raisebox{.5pt}{\textcircled{\raisebox{-.9pt} {2}}}}&\qw &\gate{\raisebox{.5pt}{\textcircled{\raisebox{-.9pt} {2}}}}&\qw&\qw&\qw&\gate{\raisebox{.5pt}{\textcircled{\raisebox{-.9pt} {1}}}}&\qw &\gate{\raisebox{.5pt}{\textcircled{\raisebox{-.9pt} {1}}}}&\qw&\qw&\qw\\
    &\lstick{2}&\qw&\qw&\qw&\gate{ }&\qw&\qw&\qw&\qw&\qw&\gate{ }&\qw&\qw&\gate{ }&\qw\\
    &\lstick{\vdots}&\qw&\qw&\qw&\gate{ }\qwx[-1]&\qw&\qw&\qw&\qw&\qw&\gate{ }\qwx[-1]&\qw&\qw&\gate{ }\qwx[-1]&\qw\\
    &\lstick{n-1}&\qw&\qw&\qw&\gate{ }\qwx[-1]&\qw&\qw&\qw&\qw&\qw&\gate{ }\qwx[-1]&\qw&\qw&\gate{ }\qwx[-1]&\qw\\
    &\lstick{n}&\qw&\qw&\gate{X_{0,1}}\qwx[-4]&\gate{F_{n-1}(R_a(\Gamma))}\qwx[-1]&\gate{X_{0,1}}\qwx[-4]&\qw&\qw&\qw&\gate{X_{0,1}}\qwx[-4]&\gate{F_{n-1}(R_a(\Phi))}\qwx[-1]&\gate{X_{0,1}}\qwx[-4]&\qw&\gate{F_{n-1}(R_a(\Theta))}\qwx[-1]&\qw\\}}\end{eqnarray}

  \begin{eqnarray}\label{multiY2}
\framebox[17cm][c]{\Qcircuit @C=1em @R=.7em {
    &\lstick{1}&\qw&\qw&\gate{\raisebox{.5pt}{\textcircled{\raisebox{-.9pt} {2}}}}&\qw &\gate{\raisebox{.5pt}{\textcircled{\raisebox{-.9pt} {2}}}}&\qw&\qw&\qw&\gate{\raisebox{.5pt}{\textcircled{\raisebox{-.9pt} {1}}}}&\qw &\gate{\raisebox{.5pt}{\textcircled{\raisebox{-.9pt} {1}}}}&\qw&\qw&\qw\\
    &\lstick{2}&\qw&\qw&\qw&\gate{ }&\qw&\qw&\qw&\qw&\qw&\gate{ }&\qw&\qw&\gate{ }&\qw\\
    &\lstick{\vdots}&\qw&\qw&\qw&\gate{ }\qwx[-1]&\qw&\qw&\qw&\qw&\qw&\gate{ }\qwx[-1]&\qw&\qw&\gate{ }\qwx[-1]&\qw\\
    &\lstick{n-1}&\qw&\qw&\qw&\gate{ }\qwx[-1]&\qw&\qw&\qw&\qw&\qw&\gate{ }\qwx[-1]&\qw&\qw&\gate{ }\qwx[-1]&\qw\\
    &\lstick{n}&\qw&\qw&\gate{X_{1,2}}\qwx[-4]&\gate{F_{n-1}(R_a(\Gamma))}\qwx[-1]&\gate{X_{1,2}}\qwx[-4]&\qw&\qw&\qw&\gate{X_{1,2}}\qwx[-4]&\gate{F_{n-1}(R_a(\Phi))}\qwx[-1]&\gate{X_{1,2}}\qwx[-4]&\qw&\gate{F_{n-1}(R_a(\Theta))}\qwx[-1]&\qw\\}}\end{eqnarray}

  \begin{eqnarray}\label{multiY3}
\framebox[17cm][c]{\Qcircuit @C=1em @R=.7em {
    &\lstick{1}&\qw&\qw&\gate{\raisebox{.5pt}{\textcircled{\raisebox{-.9pt} {2}}}}&\qw &\gate{\raisebox{.5pt}{\textcircled{\raisebox{-.9pt} {2}}}}&\qw&\qw&\qw&\gate{\raisebox{.5pt}{\textcircled{\raisebox{-.9pt} {1}}}}&\qw &\gate{\raisebox{.5pt}{\textcircled{\raisebox{-.9pt} {1}}}}&\qw&\qw&\qw\\
    &\lstick{2}&\qw&\qw&\qw&\gate{ }&\qw&\qw&\qw&\qw&\qw&\gate{ }&\qw&\qw&\gate{ }&\qw\\
    &\lstick{\vdots}&\qw&\qw&\qw&\gate{ }\qwx[-1]&\qw&\qw&\qw&\qw&\qw&\gate{ }\qwx[-1]&\qw&\qw&\gate{ }\qwx[-1]&\qw\\
    &\lstick{n-1}&\qw&\qw&\qw&\gate{ }\qwx[-1]&\qw&\qw&\qw&\qw&\qw&\gate{ }\qwx[-1]&\qw&\qw&\gate{ }\qwx[-1]&\qw\\
    &\lstick{n}&\qw&\qw&\gate{X_{0,2}}\qwx[-4]&\gate{F_{n-1}(R_a(\Gamma))}\qwx[-1]&\gate{X_{0,2}}\qwx[-4]&\qw&\qw&\qw&\gate{X_{0,2}}\qwx[-4]&\gate{F_{n-1}(R_a(\Phi))}\qwx[-1]&\gate{X_{0,2}}\qwx[-4]&\qw&\gate{F_{n-1}(R_a(\Theta))}\qwx[-1]&\qw\\}}\end{eqnarray} 
    
  where $$\psi_k=\begin{cases}
        \theta_j+\phi_j+\gamma_j  \mbox{ where } 1\leq j\leq 3^{n-2},k=j\\
        \theta_j-\phi_j+\gamma_j \mbox{ where } 1\leq j\leq 3^{n-2},k=j+3^{n-2}\\
        \theta_j+\phi_j-\gamma_j \mbox{ where } 1\leq j\leq 3^{n-2},k=j+2.3^{n-2}.
    \end{cases}$$

Then we have the following lemma.

\begin{lemma}\label{mqzyz}
 The following statements hold true:\begin{enumerate}
     \item The quantum circuits in equation (\ref{multi1}) and Equation (\ref{multiY1}) are equivalent when $a\in \{Y01,Z01\}$ 
     \item The quantum circuits in equation (\ref{multi1}) and Equation (\ref{multiY2}) are equivalent when $a\in \{Y12,Z12\}$ 
     \item The quantum circuits in equation (\ref{multi1}) and Equation (\ref{multiY3}) are equivalent when $a\in \{Y02,Z02\}$ 
 \end{enumerate}
   
\end{lemma}

\pf  We shall prove the first statement. The proofs for the rest of the statements are similar. The unitary matrix corresponding to the circuit in equation (\ref{multiY1}) is given by \begin{eqnarray*}
    \left(I_3\otimes \left[ 
\begin{array}{c|c|c} 
      R_{Y01}(\theta_1) &  0 &0 \\ 
      \hline 
       0 & \ddots &0\\
        \hline
        0& 0& R_{Y01}(\theta_{3^{n-2}})
    \end{array} 
    \right] \right)\left[ 
\begin{array}{c|c|c} 
      I_{3^{n-2}} &  O & O\\ 
      \hline 
     O& U & O\\
     \hline
     O& O & I_{3^{n-2}}
    \end{array} 
    \right] \left(I_3\otimes \left[ 
\begin{array}{c|c|c} 
      R_{Y01}(\phi_1) &  0 &0 \\ 
      \hline 
       0 & \ddots &0\\
        \hline
        0& 0& R_{Y01}(\phi_{3^{n-2}})
    \end{array} 
    \right] \right)\\\left[ 
\begin{array}{c|c|c} 
      I_{3^{n-2}} &  O & O\\ 
      \hline 
     O& U & O\\
     \hline
     O& O & I_{3^{n-2}}
    \end{array} 
    \right]\left[ 
\begin{array}{c|c|c} 
      I_{3^{n-2}} &  O & O\\ 
      \hline 
     O& I_{3^{n-2}} & O\\
     \hline
     O& O & U
    \end{array} 
    \right] \left(I_3\otimes \left[ 
\begin{array}{c|c|c} 
      R_{Y01}(\gamma_1) &  0 &0 \\ 
      \hline 
       0 & \ddots &0\\
        \hline
        0& 0& R_{Y01}(\gamma_{3^{n-2}})
    \end{array} 
    \right] \right)  \left[ 
\begin{array}{c|c|c} 
      I_{3^{n-2}} &  O & O\\ 
      \hline 
     O& I_{3^{n-2}} & O\\
     \hline
     O& O & U
    \end{array} 
    \right]
\end{eqnarray*}
     where $U_{3^{n-2}\times 3^{n-2}}=\left[ 
\begin{array}{c|c|c|c} 
      X_{0,1} &  0 &0 &0\\ 
      \hline 
       0 &  X_{0,1} &0&0\\
        \hline
       0 & 0& \ddots& 0\\
       \hline
       0 & 0& 0& X_{0,1}\\
    \end{array} 
    \right] $. This gives the matrix \begin{eqnarray*}
        \left[ 
\begin{array}{c|c|c} 
      R_{11}&0&0\\
      \hline
      0&R_{22}&0\\
      \hline
      0&0&R_{33}
    \end{array} 
    \right] 
    \end{eqnarray*} \\ where \begin{eqnarray*}
        R_{11}&=&\left[ 
\begin{array}{c|c|c} 
      R_{Y01}(\theta_1+\phi_1+\gamma_1) &  0 &0 \\ 
      \hline 
       0 &  \ddots &0\\
        \hline
        0& 0& R_{Y01}(\theta_{3^{n-2}}+\phi_{3^{n-2}}+\gamma_{3^{n-2}})
    \end{array} 
    \right]
    \end{eqnarray*}
    \begin{eqnarray*}
    R_{22} &=& \left[ 
\begin{array}{c|c|c} 
      X_{0,1} R_{Y01}(\phi_1) X_{0,1} R_{Y1}(\theta_1+\gamma_1) &  0 &0 \\ 
      \hline 
       0 &  \ddots &0\\
        \hline
        0& 0&  X_{0,1} R_{Y01}(\phi_{3^{n-2}}) X_{0,1} R_{Y01}(\theta_{3^{n-2}}+\gamma_{3^{n-2}})
    \end{array} 
    \right] \\
    &=& \left[ 
\begin{array}{c|c|c} 
       R_{Y01}(\theta_1-\phi_1+\gamma_1) &  0 &0 \\ 
      \hline 
       0 &  \ddots &0\\
        \hline
        0& 0&  R_{Y01}(\theta_{3^{n-2}}-\phi_{3^{n-2}}+\gamma_{3^{n-2}})
    \end{array} 
    \right].
    \end{eqnarray*}
     \begin{eqnarray*}
    R_{33} &=& \left[ 
\begin{array}{c|c|c} 
      X_{0,1} R_{Y01}(\gamma_1) X_{0,1} R_{Y01}(\theta_1+\phi_1) &  0 &0 \\ 
      \hline 
       0 &  \ddots &0\\
        \hline
        0& 0&  X_{0,1} R_{Y01}(\gamma_{3^{n-2}}) X_{0,1} R_{Y01}(\theta_{3^{n-2}}+\phi_{3^{n-2}})
    \end{array} 
    \right] \\
    &=& \left[ 
\begin{array}{c|c|c} 
       R_{Y01}(\theta_1+\phi_1-\gamma_1) &  0 &0 \\ 
      \hline 
       0 &  \ddots &0\\
        \hline
        0& 0&   R_{Y01}(\theta_{3^{n-2}}+\phi_{3^{n-2}}-\gamma_{3^{n-2}})
    \end{array} 
    \right].
    \end{eqnarray*} This completes the proof. \hfill{$\square$}

Next, let us consider the following circuits.
  \begin{eqnarray}\label{multiX1}
\framebox[17.5cm][c]{\Qcircuit @C=0.6em @R=.42em {
    &\lstick{1}&\gate{\raisebox{.5pt}{\textcircled{\raisebox{-.9pt} {2}}}}&\qw &\gate{\raisebox{.5pt}{\textcircled{\raisebox{-.9pt} {2}}}}&\gate{\raisebox{.5pt}{\textcircled{\raisebox{-.9pt} {1}}}}&\qw &\gate{\raisebox{.5pt}{\textcircled{\raisebox{-.9pt} {1}}}}&\qw&\qw\\
    &\lstick{2}&\qw&\gate{ }&\qw&\qw&\gate{ }&\qw&\gate{ }&\qw\\
    &\lstick{\vdots}&\qw&\gate{ }\qwx[-1]&\qw&\qw&\gate{ }\qwx[-1]&\qw&\gate{ }\qwx[-1]&\qw\\
    &\lstick{n-1}&\qw&\gate{ }\qwx[-1]&\qw&\qw&\gate{ }\qwx[-1]&\qw&\gate{ }\qwx[-1]&\qw\\
    &\lstick{n}&\gate{R_{Z{12}}(\pi)}\qwx[-4]&\gate{F_{n-1}(R_{X02}(\Gamma))}\qwx[-1]&\gate{R_{Z{12}}(\pi)}\qwx[-4]&\gate{R_{Z{12}}(\pi)}\qwx[-4]&\gate{F_{n-1}(R_{X02}(\Phi))}\qwx[-1]&\gate{R_{Z{12}}(\pi)}\qwx[-4]&\gate{F_{n-1}(R_{X02}(\Theta))}\qwx[-1]&\qw\\}}\end{eqnarray} 

   \begin{eqnarray}\label{multiX2}
\framebox[17.5cm][c]{\Qcircuit @C=0.6em @R=.42em {
    &\lstick{1}&\gate{\raisebox{.5pt}{\textcircled{\raisebox{-.9pt} {2}}}}&\qw &\gate{\raisebox{.5pt}{\textcircled{\raisebox{-.9pt} {2}}}}&\gate{\raisebox{.5pt}{\textcircled{\raisebox{-.9pt} {1}}}}&\qw &\gate{\raisebox{.5pt}{\textcircled{\raisebox{-.9pt} {1}}}}&\qw&\qw\\
    &\lstick{2}&\qw&\gate{ }&\qw&\qw&\gate{ }&\qw&\gate{ }&\qw\\
    &\lstick{\vdots}&\qw&\gate{ }\qwx[-1]&\qw&\qw&\gate{ }\qwx[-1]&\qw&\gate{ }\qwx[-1]&\qw\\
    &\lstick{n-1}&\qw&\gate{ }\qwx[-1]&\qw&\qw&\gate{ }\qwx[-1]&\qw&\gate{ }\qwx[-1]&\qw\\
    &\lstick{n}&\gate{R_{Z{12}}(\pi)}\qwx[-4]&\gate{F_{n-1}(R_{X01}(\Gamma))}\qwx[-1]&\gate{R_{Z{12}}(\pi)}\qwx[-4]&\gate{R_{Z{12}}(\pi)}\qwx[-4]&\gate{F_{n-1}(R_{X01}(\Phi))}\qwx[-1]&\gate{R_{Z{12}}(\pi)}\qwx[-4]&\gate{F_{n-1}(R_{X01}(\Theta))}\qwx[-1]&\qw\\}}\end{eqnarray}

  \begin{eqnarray}\label{multiX3}
\framebox[17.5cm][c]{\Qcircuit @C=0.6em @R=.42em {
    &\lstick{1}&\gate{\raisebox{.5pt}{\textcircled{\raisebox{-.9pt} {2}}}}&\qw &\gate{\raisebox{.5pt}{\textcircled{\raisebox{-.9pt} {2}}}}&\gate{\raisebox{.5pt}{\textcircled{\raisebox{-.9pt} {1}}}}&\qw &\gate{\raisebox{.5pt}{\textcircled{\raisebox{-.9pt} {1}}}}&\qw&\qw\\
    &\lstick{2}&\qw&\gate{ }&\qw&\qw&\gate{ }&\qw&\gate{ }&\qw\\
    &\lstick{\vdots}&\qw&\gate{ }\qwx[-1]&\qw&\qw&\gate{ }\qwx[-1]&\qw&\gate{ }\qwx[-1]&\qw\\
    &\lstick{n-1}&\qw&\gate{ }\qwx[-1]&\qw&\qw&\gate{ }\qwx[-1]&\qw&\gate{ }\qwx[-1]&\qw\\
    &\lstick{n}&\gate{R_{Z{02}}(\pi)}\qwx[-4]&\gate{F_{n-1}(R_{X12}(\Gamma))}\qwx[-1]&\gate{R_{Z{02}}(\pi)}\qwx[-4]&\gate{R_{Z{02}}(\pi)}\qwx[-4]&\gate{F_{n-1}(R_{X12}(\Phi))}\qwx[-1]&\gate{R_{Z{02}}(\pi)}\qwx[-4]&\gate{F_{n-1}(R_{X12}(\Theta))}\qwx[-1]&\qw\\}}\end{eqnarray} 
    
  where $$\psi_k=\begin{cases}
        \theta_j+\phi_j+\gamma_j, \, 1\leq j\leq 3^{n-2},k=j\\
        \theta_j-\phi_j+\gamma_j,1\leq j\leq 3^{n-2},k=j+3^{n-2}\\
        \theta_j+\phi_j-\gamma_j, 1\leq j\leq 3^{n-2},k=j+2.3^{n-2}.
    \end{cases}$$

    Then we have the following corollary. 

\begin{corollary}\label{mqx}
 The following statements hold true:\begin{enumerate}
     \item The quantum circuits in equation (\ref{multi1}) and Equation (\ref{multiX1}) are equivalent when $a=X02$ 
     \item The quantum circuits in equation (\ref{multi1}) and Equation (\ref{multiX2}) are equivalent when $a=X01$ 
     \item The quantum circuits in equation (\ref{multi1}) and Equation (\ref{multiX3}) are equivalent when $a= X12$ 
 \end{enumerate}
   
\end{corollary}
\pf Follows similar to the proof of Lemma\ref{mqzyz}. $\hfill\square$

Now using Lemma \ref{mqzyz}, we show how to construct scalable qutrit circuits for block diagonal special unitaries. Let $U\in SU(3^n)$ be a block diagonal special unitary matrix with $3\times 3$ special unitary diagonal blocks. In other words, 
\begin{eqnarray}\label{blkdg1}
U&=&\left[ 
\begin{array}{cccc} 
      U_1(\Lambda_1) &   & & \\ 
       &  U_2(\Lambda_2) & & \\
       &  & \ddots &   \\
        & & &U_{3^{n-1}}(\Lambda_{3^{n-1}})
    \end{array} 
    \right] \end{eqnarray}   where $\Lambda_j:=(\theta_1^{(j)},\phi_1^{(j)},\psi_1^{(j)},\theta_2^{(j)},\phi_2^{(j)},\psi_2^{(j)},\theta_3^{(j)},\phi_3^{(j)},\psi_3^{(j)})$ and each $U_j(\Lambda_j)\in SU(3)$. Then clearly from {theorem \ref{qutritZYZ}}, we have\begin{eqnarray}\label{blkdg}
U&=&\underbrace{\left[ 
\begin{array}{cccc} 
      M_1(-\theta_1^{(1)},-\phi_1^{(1)},\psi_1^{(1)}) &   & & \\ 
       &  M_1(-\theta_1^{(2)},-\phi_1^{(2)},\psi_1^{(2)}) & & \\
       &  & \ddots &   \\
        & & &M_1(-\theta_1^{(3^{n-1})},-\phi_1^{(3^{n-1})},\psi_1^{(3^{n-1})})
    \end{array} 
    \right]}_{U_{1}}\\&&\underbrace{\left[ 
\begin{array}{cccc} 
      M_2(-\theta_2^{(1)},-\phi_2^{(1)},\psi_2^{(1)}) &   & & \\ 
       &  M_2(-\theta_2^{(2)},-\phi_2^{(2)},\psi_2^{(2)}) & & \\
       &  & \ddots &   \\
        & & &M_2(-\theta_2^{(3^{n-1})},-\phi_2^{(3^{n-1})},\psi_2^{(3^{n-1})})
    \end{array} 
    \right]}_{U_{2}}\\&&\underbrace{\left[ 
\begin{array}{cccc} 
      M_3(-\theta_3^{(1)},-\phi_3^{(1)},\psi_3^{(1)}) &   & & \\ 
       &  M_2(-\theta_3^{(2)},-\phi_3^{(2)},\psi_3^{(2)}) & & \\
       &  & \ddots &   \\
        & & &M_2(-\theta_3^{(3^{n-1})},-\phi_3^{(3^{n-1})},\psi_3^{(3^{n-1})})
    \end{array} 
    \right]}_{U_{3}}\end{eqnarray} such that      \begin{eqnarray}\label{small1}
    U=U_1U_2U_3
\end{eqnarray} where $U_1,U_2,U_3$ are block diagonal special unitary matrices with each block belonging to $SU(3)$. In order to construct qutrit circuit of $U$, we shall look at the circuits for each of its components on the RHS in equation \ref{small1}.\\

\begin{theorem}\label{componentzyz}
The qutrit circuit for $U_1$ in equation (\ref{small1}) is given by  \begin{eqnarray}\label{multiblkdiag1}
\framebox[16cm][c]{\Qcircuit @C=1em @R=.7em {
    &\lstick{1}&\qw& \gate{} &\qw&\qw& \gate{} &\qw&\qw& \gate{} &\qw&\\
    &\lstick{2}&\qw&\gate{ }\qwx[-1]&\qw&\qw&\gate{ }\qwx[-1]&\qw&\qw&\gate{ }\qwx[-1]&\qw\\
    &\lstick{\vdots}&\qw&\gate{ }\qwx[-1]&\qw&\qw&\gate{ }\qwx[-1]&\qw&\qw&\gate{ }\qwx[-1]&\qw\\
    &\lstick{n-1}&\qw&\gate{ }\qwx[-1]&\qw&\qw&\gate{ }\qwx[-1]&\qw&\qw&\gate{ }\qwx[-1]&\qw\\
    &\lstick{n}&\qw&\gate{F_n(R_{Z02}(\overline{\Lambda_2^{(1)}}))}\qwx[-1]&\qw&\qw&\gate{F_n(R_{Y02}(\overline{\Theta}^{(1)}))}\qwx[-1]&\qw&\qw&\gate{F_n(R_{Z02}(\overline{\Lambda_1^{(1)}}))}\qwx[-1]&\qw\\}}\end{eqnarray} such that \begin{eqnarray*}
        \overline{\Theta^{(1)}}&=&(-\theta_1^{(1)},\hdots,-\theta_1^{({3^{n-1}})}),\\\overline{\Lambda_1^{(1)}}&=&(\frac{-\phi_1^{(1)}+\psi_1^{(1)}}{2},\hdots,\frac{-\phi_{1}^{({({3^{n-1}})})}+\psi_{1}^{({({3^{n-1}})})}}{2}),\\\overline{\Lambda_2^{(1)}}&=&(\frac{-\phi_1^{(1)}-\psi_1^{(1)}}{2},\hdots,\frac{-\phi_{1}^{({3^{n-1}})}-\psi_{1}^{({({3^{n-1}})})}}{2})
    \end{eqnarray*}
    \end{theorem}

\pf Follows from equation (\ref{M02}) and construction of qutrit block diagonal rotations. \hfill{$\square$}

In a similar way, we obtain the following.

\begin{theorem}\label{componentzyz2}
The qutrit circuit for $U_2$ in equation \ref{small1} is  \begin{eqnarray}\label{multiblkdiag2}
\framebox[16cm][c]{\Qcircuit @C=1em @R=.7em {
    &\lstick{1}&\qw& \gate{} &\qw&\qw& \gate{} &\qw&\qw& \gate{} &\qw&\\
    &\lstick{2}&\qw&\gate{ }\qwx[-1]&\qw&\qw&\gate{ }\qwx[-1]&\qw&\qw&\gate{ }\qwx[-1]&\qw\\
    &\lstick{\vdots}&\qw&\gate{ }\qwx[-1]&\qw&\qw&\gate{ }\qwx[-1]&\qw&\qw&\gate{ }\qwx[-1]&\qw\\
    &\lstick{n-1}&\qw&\gate{ }\qwx[-1]&\qw&\qw&\gate{ }\qwx[-1]&\qw&\qw&\gate{ }\qwx[-1]&\qw\\
    &\lstick{n}&\qw&\gate{F_n(R_{Z01}(\overline{\Lambda_2^{(2)}}))}\qwx[-1]&\qw&\qw&\gate{F_n(R_{Y01}(\overline{\Theta}^{(2)}))}\qwx[-1]&\qw&\qw&\gate{F_n(R_{Z01}(\overline{\Lambda_1^{(2)}}))}\qwx[-1]&\qw\\}}\end{eqnarray} such that \begin{eqnarray*}
        \overline{\Theta^{(2)}}&=&(-\theta_2^{(1)},\hdots,-\theta_2^{({3^{n-1}})}),\\\overline{\Lambda_1^{(2)}}&=&(\frac{\psi_2^{(1)}}{2},\hdots,\frac{\psi_{2}^{({({3^{n-1}})})}}{2}),\\\overline{\Lambda_2^{(2)}}&=&(\frac{-\psi_2^{(1)}}{2},\hdots,\frac{-\psi_{2}^{({({3^{n-1}})})}}{2})
    \end{eqnarray*}
    \end{theorem}
\pf Follows from equation (\ref{M01}) and construction of qutrit block diagonal rotations. \hfill{$\square$}

\begin{theorem}\label{componentzyz3}
The qutrit circuit for $U_3$ in equation \ref{small1} is  \begin{eqnarray}\label{multiblkdiag3}
\framebox[16cm][c]{\Qcircuit @C=1em @R=.7em {
    &\lstick{1}&\qw& \gate{} &\qw&\qw& \gate{} &\qw&\qw& \gate{} &\qw&\\
    &\lstick{2}&\qw&\gate{ }\qwx[-1]&\qw&\qw&\gate{ }\qwx[-1]&\qw&\qw&\gate{ }\qwx[-1]&\qw\\
    &\lstick{\vdots}&\qw&\gate{ }\qwx[-1]&\qw&\qw&\gate{ }\qwx[-1]&\qw&\qw&\gate{ }\qwx[-1]&\qw\\
    &\lstick{n-1}&\qw&\gate{ }\qwx[-1]&\qw&\qw&\gate{ }\qwx[-1]&\qw&\qw&\gate{ }\qwx[-1]&\qw\\
    &\lstick{n}&\qw&\gate{F_n(R_{Z12}(\overline{\Lambda_2^{(3)}}))}\qwx[-1]&\qw&\qw&\gate{F_n(R_{Y12}(\overline{\Theta}^{(3)}))}\qwx[-1]&\qw&\qw&\gate{F_n(R_{Z12}(\overline{\Lambda_1^{(3)}}))}\qwx[-1]&\qw\\}}\end{eqnarray} such that \begin{eqnarray*}
        \overline{\Theta^{(3)}}&=&(-\theta_3^{(1)},\hdots,-\theta_3^{({3^{n-1}})}),\\\overline{\Lambda_1^{(3)}}&=&(\frac{-\phi_3^{(1)}+\psi_3^{(1)}}{2},\hdots,\frac{-\phi_{3}^{({({3^{n-1}})})}+\psi_{3}^{({({3^{n-1}})})}}{2}),\\\overline{\Lambda_2^{(3)}}&=&(\frac{-\phi_3^{(1)}-\psi_3^{(1)}}{2},\hdots,\frac{-\phi_{3}^{({3^{n-1}})}-\psi_{3}^{({({3^{n-1}})})}}{2}).
    \end{eqnarray*}
    \end{theorem}

\pf The proof follows from equation (\ref{M12}) and construction of the qutrit block diagonal rotations. \hfill{$\square$}


From the discussions so far, we are now ready to provide a scalable circuit construction of multi-controlled qutrit $X$ gates using single-qutrit gates and controlled qutrit-$X$ gates. Let us take the following multi-controlled $n$-qutrit X gate used in construction of quantum qutrit circuits of three-state DTQWs of Cayley graphs discussed before.

Note that the gate
   \begin{eqnarray}\label{multiToff}
\framebox[8cm][c]{\Qcircuit @C=1em @R=.7em {
    &\lstick{1}&\qw&\qw& \gate{\raisebox{.5pt}{\textcircled{\raisebox{-.9pt} {2}}}} &\qw&\qw\\
    &\lstick{2}&\qw&\qw&\gate{\raisebox{.5pt}{\textcircled{\raisebox{-.9pt} {2}}}}\qwx[-1]&\qw&\qw\\
    &\lstick{\vdots}&\qw& &\vdots & &\qw\\
    &\lstick{n-1}&\qw&\qw&\gate{\raisebox{.5pt}{\textcircled{\raisebox{-.9pt} {2}}}}&\qw&\qw\\
    &\lstick{n}&\qw&\qw&\gate{X_{+1}}\qwx[-1]&\qw&\qw\\}}\end{eqnarray}
 is equivalent to the following circuit    \begin{eqnarray}\label{multiToff1}
\framebox[15cm][c]{\Qcircuit @C=1em @R=.7em {
    &\lstick{1}&\qw& \gate{} &\gate{}&\qw&\qw&\qw\\
    &\lstick{2}&\qw&\gate{ }\qwx[-1]&\gate{ }\qwx[-1]&\qw&\qw&\qw\\
    &\lstick{\vdots}&\qw&\gate{ }\qwx[-1]&\gate{ }\qwx[-1]&\qw&\qw&\qw\\
    &\lstick{n-1}&\qw&\gate{ }\qwx[-1]&\gate{ }\qwx[-1]&\qw&\qw&\qw\\
    &\lstick{n}&\qw&\gate{F_n(R_{Y12}(0,\hdots,0,-\frac{\pi}{2}))}\qwx[-1]&\gate{F_n(R_{Y01}(0,\hdots,0,-\frac{\pi}{2}))}\qwx[-1]&\qw&\qw&\qw\\}}\end{eqnarray}

Similarly, the $n$-qutrit gate
\begin{eqnarray}\label{multiToff2}
\framebox[8cm][c]{\Qcircuit @C=1em @R=.7em {
    &\lstick{1}&\qw&\qw& \gate{\raisebox{.5pt}{\textcircled{\raisebox{-.9pt} {0}}}} &\qw&\qw\\
    &\lstick{2}&\qw&\qw&\gate{\raisebox{.5pt}{\textcircled{\raisebox{-.9pt} {0}}}}\qwx[-1]&\qw&\qw\\
    &\lstick{\vdots}&\qw& &\vdots & &\qw\\
    &\lstick{n-1}&\qw&\qw&\gate{\raisebox{.5pt}{\textcircled{\raisebox{-.9pt} {0}}}}&\qw&\qw\\
    &\lstick{n}&\qw&\qw&\gate{X_{+1}}\qwx[-1]&\qw&\qw\\}}\end{eqnarray} is equivalent to     \begin{eqnarray}\label{multiToff21}
\framebox[15cm][c]{\Qcircuit @C=1em @R=.7em {
    &\lstick{1}&\qw& \gate{} &\gate{}&\qw&\qw&\qw\\
    &\lstick{2}&\qw&\gate{ }\qwx[-1]&\gate{ }\qwx[-1]&\qw&\qw&\qw\\
    &\lstick{\vdots}&\qw&\gate{ }\qwx[-1]&\gate{ }\qwx[-1]&\qw&\qw&\qw\\
    &\lstick{n-1}&\qw&\gate{ }\qwx[-1]&\gate{ }\qwx[-1]&\qw&\qw&\qw\\
    &\lstick{n}&\qw&\gate{F_n(R_{Y12}(-\frac{\pi}{2},0,\hdots,0))}\qwx[-1]&\gate{F_n(R_{Y01}(-\frac{\pi}{2},0,\hdots,0))}\qwx[-1]&\qw&\qw&\qw\\}}\end{eqnarray}

Further, the gate \begin{eqnarray}\label{multiToff3}
\framebox[8cm][c]{\Qcircuit @C=1em @R=.7em {
    &\lstick{1}&\qw&\qw& \gate{\raisebox{.5pt}{\textcircled{\raisebox{-.9pt} {2}}}} &\qw&\qw\\
    &\lstick{2}&\qw&\qw&\gate{\raisebox{.5pt}{\textcircled{\raisebox{-.9pt} {2}}}}\qwx[-1]&\qw&\qw\\
    &\lstick{\vdots}&\qw& &\vdots & &\qw\\
    &\lstick{n-1}&\qw&\qw&\gate{\raisebox{.5pt}{\textcircled{\raisebox{-.9pt} {2}}}}&\qw&\qw\\
    &\lstick{n}&\qw&\qw&\gate{X_{+2}}\qwx[-1]&\qw&\qw\\}}\end{eqnarray} is equivalent to

    \begin{eqnarray}\label{multiToff31}
    {\Qcircuit @C=0.5em @R=.3em {
    &\lstick{1}&\qw& \gate{} &\gate{}&\gate{}&\gate{}&\qw\\
    &\lstick{2}&\qw&\gate{ }\qwx[-1]&\gate{ }\qwx[-1]&\gate{ }\qwx[-1]&\gate{ }\qwx[-1]&\qw\\
    &\lstick{\vdots}&\qw&\gate{ }\qwx[-1]&\gate{ }\qwx[-1]&\gate{ }\qwx[-1]&\gate{ }\qwx[-1]&\qw\\
    &\lstick{n-1}&\qw&\gate{ }\qwx[-1]&\gate{ }\qwx[-1]&\gate{ }\qwx[-1]&\gate{ }\qwx[-1]&\qw\\
    &\lstick{n}&\qw&\gate{F_n(R_{Z12}(0,\hdots,0,-\frac{\pi}{2}))}\qwx[-1]&\gate{F_n(R_{Y12}(0,\hdots,0,-\frac{\pi}{2}))}\qwx[-1]&\gate{F_n(R_{Z12}(0,\hdots,0,\frac{\pi}{2}))}\qwx[-1]&\gate{F_n(R_{Y02}(0,\hdots,0,-\frac{\pi}{2}))}\qwx[-1]&\qw\\}}\end{eqnarray}

In a similar way, finally, the gate \begin{eqnarray}\label{multiToff41}
\framebox[8cm][c]{\Qcircuit @C=1em @R=.7em {
    &\lstick{1}&\qw&\qw& \gate{\raisebox{.5pt}{\textcircled{\raisebox{-.9pt} {0}}}} &\qw&\qw\\
    &\lstick{2}&\qw&\qw&\gate{\raisebox{.5pt}{\textcircled{\raisebox{-.9pt} {0}}}}\qwx[-1]&\qw&\qw\\
    &\lstick{\vdots}&\qw& &\vdots & &\qw\\
    &\lstick{n-1}&\qw&\qw&\gate{\raisebox{.5pt}{\textcircled{\raisebox{-.9pt} {0}}}}&\qw&\qw\\
    &\lstick{n}&\qw&\qw&\gate{X_{+2}}\qwx[-1]&\qw&\qw\\}}\end{eqnarray} is equivalent to

    \begin{eqnarray}\label{multiToff42}
    {\Qcircuit @C=0.5em @R=.3em {
    &\lstick{1}&\qw& \gate{} &\gate{}&\gate{}&\gate{}&\qw\\
    &\lstick{2}&\qw&\gate{ }\qwx[-1]&\gate{ }\qwx[-1]&\gate{ }\qwx[-1]&\gate{ }\qwx[-1]&\qw\\
    &\lstick{\vdots}&\qw&\gate{ }\qwx[-1]&\gate{ }\qwx[-1]&\gate{ }\qwx[-1]&\gate{ }\qwx[-1]&\qw\\
    &\lstick{n-1}&\qw&\gate{ }\qwx[-1]&\gate{ }\qwx[-1]&\gate{ }\qwx[-1]&\gate{ }\qwx[-1]&\qw\\
    &\lstick{n}&\qw&\gate{F_n(R_{Z12}(-\frac{\pi}{2},0,\hdots,0))}\qwx[-1]&\gate{F_n(R_{Y12}(-\frac{\pi}{2},0,\hdots,0))}\qwx[-1]&\gate{F_n(R_{Z12}(\frac{\pi}{2},0,\hdots,0))}\qwx[-1]&\gate{F_n(R_{Y02}(-\frac{\pi}{2},0,\hdots,0))}\qwx[-1]&\qw\\}}\end{eqnarray}

So far we have constructed ancilla free circuits of the $n$-qutrit block diagonal special unitaries with special unitary blocks. However, the quantum circuit for DTQWs as derived above require the quantum gates given by
\begin{eqnarray}\label{InvmultiToff}
\framebox[8cm][c]{\Qcircuit @C=1em @R=.7em {
 &\lstick{1}&\qw&\qw& \gate{X} &\qw&\qw\\
 &\lstick{2}&\qw&\qw&\gate{\raisebox{.5pt}{\textcircled{\raisebox{-.9pt} {a}}} }\qwx[-1] &\qw&\qw\\
&\lstick{\vdots}&\qw& &\vdots & &\qw\\
&\lstick{n-1}&\qw&\qw& \gate{\raisebox{.5pt}{\textcircled{\raisebox{-.9pt} {a}}} }  &\qw &\qw\\
&\lstick{n}&\qw&\qw& \gate{\raisebox{.5pt}{\textcircled{\raisebox{-.9pt} {a}}} } \qwx[-1]&\qw&\qw\\}}\end{eqnarray}

where $X\in\{X_{+1}, X_{+2}\}, a\in\{0,2\}.$ Hence, it is imperative that we express the qutrit gates in equation (\ref{InvmultiToff}), 
in terms of multi-controlled $n$-qutrit-$X$ gates. The main problem boils down to find unitary matrices $P_1,P_2$  such that 
\begin{eqnarray}\label{cnot}
\framebox[8cm][c]{\Qcircuit @C=0.25cm @R=.25cm {
\lstick{}&\multigate{1}{P_2}&\gate{\raisebox{.5pt}{\textcircled{\raisebox{-.9pt} {a}}} }&\multigate{1}{P_1}&\qw\\
\lstick{}&\ghost{P_2}&\gate{X}\qwx[-1]&\ghost{P_1}&\qw\\}}\end{eqnarray} implements \begin{eqnarray}\label{Invcnot}
\framebox[8cm][c]{\Qcircuit @C=0.25cm @R=.25cm {
\lstick{}&\qw&\gate{X}&\qw\\
\lstick{}&\qw&\gate{\raisebox{.5pt}{\textcircled{\raisebox{-.9pt} {a}}} }\qwx[-1]&\qw\\}}\end{eqnarray} for $a\in \{0,1,2\}$. 

To resolve this we have the following theorem. 

\begin{theorem}\label{swapeq}
    Let $P\in U(9)$ be a permuation matrix given by $P=P_{(3,8)}P_{(6,7)}$ where $P_{(i,j)}$ represents a two cycle permutation with $i$-th and $j$-th rows interchanged. Then setting $P_1=P_2=P,$ the circuits in equations (\ref{cnot}) and (\ref{Invcnot}) are equivalent.  
   Then the circuits   \begin{eqnarray*} \framebox[8cm][c]{\Qcircuit @C=0.25cm @R=.25cm { \lstick{}&\multigate{1}{P}&\gate{\raisebox{.5pt}{\textcircled{\raisebox{-.9pt} {2}}} }&\multigate{1}{P}&\qw\\ \lstick{}&\ghost{P}&\gate{X_{+1}}\qwx[-1]&\ghost{P}&\qw\\}}      &\mbox{ and }&  \framebox[8cm][c]{\Qcircuit @C=0.25cm @R=.25cm { \lstick{}&\multigate{1}{P}&\gate{\raisebox{.5pt}{\textcircled{\raisebox{-.9pt} {2}}} }&\multigate{1}{P}&\qw\\ \lstick{}&\ghost{P}&\gate{X_{+2}}\qwx[-1]&\ghost{P}&\qw\\}}         \end{eqnarray*} are equivalent to the gates \begin{eqnarray*} \framebox[8cm][c]{\Qcircuit @C=0.25cm @R=.25cm {\lstick{}&\qw&\gate{X_{+2}}&\qw\\\lstick{}&\qw&\gate{\raisebox{.5pt}{\textcircled{\raisebox{-.9pt} {2}}} }\qwx[-1]&\qw\\}} &\mbox{and}& \framebox[8cm][c]{\Qcircuit @C=0.25cm @R=.25cm {\lstick{}&\qw&\gate{X_{+1}}&\qw\\\lstick{}&\qw&\gate{\raisebox{.5pt}{\textcircled{\raisebox{-.9pt} {2}}} }\qwx[-1]&\qw\\}}\end{eqnarray*} respectively.

Further the circuit of $P$ is given by   \begin{eqnarray*}
\framebox[8cm][c]{\Qcircuit @C=0.25cm @R=.25cm {
\lstick{}&\qw&\gate{\raisebox{.5pt}{\textcircled{\raisebox{-.9pt} {2}}} }&\gate{X_{01}}&\multigate{1}{\textsc{SWAP}_{1,2}}&\multigate{1}{\textsc{SWAP}_{0,2}}&\qw\\
\lstick{}&\qw&\gate{X_{01}}\qwx[-1]&\gate{\raisebox{.5pt} {\textcircled{\raisebox{-.9pt} {2}}} }\qwx[-1]&\ghost{\textsc{SWAP}_{1,2}}&\ghost{\textsc{SWAP}_{0,2}}&\qw\\}} \end{eqnarray*} 
\end{theorem}
\pf The proof follows by mapping the input and output states for the circuits. \hfill{$\square$}

Thus observe that the $n$-qutrit circuit, where $n$ is odd, given by \begin{eqnarray}\label{InvToff2}
\framebox[8cm][c]{\Qcircuit @C=1em @R=.7em {
 &\lstick{1}&\qw&\qw& \gate{X_{+1}} &\qw&\qw\\
 &\lstick{2}&\qw&\qw&\gate{\raisebox{.5pt}{\textcircled{\raisebox{-.9pt} {2}}} }\qwx[-1] &\qw&\qw\\
&\lstick{\vdots}&\qw& &\vdots & &\qw\\
&\lstick{n-1}&\qw&\qw& \gate{\raisebox{.5pt}{\textcircled{\raisebox{-.9pt} {2}}} }  &\qw &\qw\\
&\lstick{n}&\qw&\qw& \gate{\raisebox{.5pt}{\textcircled{\raisebox{-.9pt} {2}}} } \qwx[-1]&\qw&\qw\\}}\end{eqnarray} 
can be written as \begin{eqnarray}
    \framebox[12.35cm][c]{\Qcircuit @C=0.25cm @R=.25cm {
\lstick{1}&\multigate{1}{P}&\qw&\qw&\qw&\qw&\gate{\raisebox{.5pt}{\textcircled{\raisebox{-.9pt} {2}}} }&\qw&\qw&\qw&\qw&\multigate{1}{P}&\qw\\
\lstick{2}&\ghost{P}&\multigate{1}{P}&\qw&\qw&\qw&\gate{\raisebox{.5pt}{\textcircled{\raisebox{-.9pt} {2}}} }\qwx[-1]&\qw&\qw&\qw&\multigate{1}{P}&\ghost{P}&\qw\\
\lstick{3}&\qw&\ghost{P}&\ddots&\qw&\qw&\gate{\raisebox{.5pt}{\textcircled{\raisebox{-.9pt} {2}}} }\qwx[-1]&\qw&\qw&{\iddots}&\ghost{P}&\qw&\qw\\
\lstick{\vdots}&\qw&\ddots& &\ddots&\ddots&\vdots&\iddots& &\iddots&\iddots& &\qw\\
\lstick{n-1}&\qw&\qw&\qw&\ddots&\multigate{1}{P}&\gate{\raisebox{.5pt}{\textcircled{\raisebox{-.9pt} {2}}} }&\multigate{1}{P}&\iddots&\qw&\qw&\qw&\qw\\
\lstick{n}&\qw&\qw&\qw&\qw&\ghost{P}&\gate{X_{+2}}\qwx[-1]&\ghost{P}&\qw&\qw&\qw&\qw&\qw\\}}
\end{eqnarray}
Finally, recall that the circuit

{\centerline{\framebox[8cm][c]{\Qcircuit @C=1em @R=.7em {
 &\lstick{}&\qw&\gate{\raisebox{.5pt}{\textcircled{\raisebox{-.9pt} {0}}} }& \qw\\
 &\lstick{}&\qw& \gate{X} \qwx[-1]& \qw\\}}}
 is equivalent to 
  
  \centerline{\framebox[8cm][c]{\Qcircuit @C=1em @R=.7em {
 &\lstick{}&\qw&\gate{X_{+2}} &\gate{\raisebox{.5pt}{\textcircled{\raisebox{-.9pt} {2}}} }&\gate{X_{+1}}  &\qw\\
 &\lstick{}&\qw&\qw &\gate{X} \qwx[-1]& \qw&\qw\\}}}
 \vspace{0.25cm}
 
 Thus, from our discussion so far,  the qutrit gate \begin{eqnarray}\label{InvToff3}
\framebox[8cm][c]{\Qcircuit @C=1em @R=.7em {
 &\lstick{1}&\qw&\qw& \gate{X_{+1}} &\qw&\qw\\
 &\lstick{2}&\qw&\qw&\gate{\raisebox{.5pt}{\textcircled{\raisebox{-.9pt} {0}}} }\qwx[-1] &\qw&\qw\\
&\lstick{\vdots}&\qw& &\vdots & &\qw\\
&\lstick{n-1}&\qw&\qw& \gate{\raisebox{.5pt}{\textcircled{\raisebox{-.9pt} {0}}} }  &\qw &\qw\\
&\lstick{n}&\qw&\qw& \gate{\raisebox{.5pt}{\textcircled{\raisebox{-.9pt} {0}}} } \qwx[-1]&\qw&\qw\\}}\end{eqnarray} 
is equivalent to  
 \begin{eqnarray}
    \framebox[13.25cm][c]{\Qcircuit @C=0.25cm @R=.25cm {
\lstick{1}&\gate{X_{+2}}&\multigate{1}{P}&\qw&\qw&\qw&\qw&\gate{\raisebox{.5pt}{\textcircled{\raisebox{-.9pt} {2}}} }&\qw&\qw&\qw&\qw&\multigate{1}{P}&\gate{X_{+1}}&\qw\\
\lstick{2}&\gate{X_{+2}}&\ghost{P}&\multigate{1}{P}&\qw&\qw&\qw&\gate{\raisebox{.5pt}{\textcircled{\raisebox{-.9pt} {2}}} }\qwx[-1]&\qw&\qw&\qw&\multigate{1}{P}&\ghost{P}&\gate{X_{+1}}&\qw\\
\lstick{3}&\gate{X_{+2}}&\qw&\ghost{P}&\ddots&\qw&\qw&\gate{\raisebox{.5pt}{\textcircled{\raisebox{-.9pt} {2}}} }\qwx[-1]&\qw&\qw&\iddots&\ghost{P}&\qw&\gate{X_{+1}}&\qw\\
\lstick{\vdots}&\vdots&\hdots&\ddots& &\ddots&\ddots&\vdots&\iddots &\iddots& &\iddots&\hdots&\vdots&\qw\\
\lstick{n-1}&\gate{X_{+2}}&\qw&\qw&\qw&\ddots&\multigate{1}{P}&\gate{\raisebox{.5pt}{\textcircled{\raisebox{-.9pt} {2}}} }&\multigate{1}{P}&\iddots&\qw&\qw&\qw&\gate{X_{+1}}&\qw\\
\lstick{n}&\qw&\qw&\qw&\qw&\qw&\ghost{P}&\gate{X_{+2}}\qwx[-1]&\ghost{P}&\qw&\qw&\qw&\qw&\qw&\qw\\}}
\end{eqnarray}

When $n$ is even, however, the circuit \ref{InvToff2} is equivalent to \begin{eqnarray}
    \framebox[12.35cm][c]{\Qcircuit @C=0.25cm @R=.25cm {
\lstick{1}&\multigate{1}{P}&\qw&\qw&\qw&\qw&\gate{\raisebox{.5pt}{\textcircled{\raisebox{-.9pt} {2}}} }&\qw&\qw&\qw&\qw&\multigate{1}{P}&\qw\\
\lstick{2}&\ghost{P}&\multigate{1}{P}&\qw&\qw&\qw&\gate{\raisebox{.5pt}{\textcircled{\raisebox{-.9pt} {2}}} }\qwx[-1]&\qw&\qw&\qw&\multigate{1}{P}&\ghost{P}&\qw\\
\lstick{3}&\qw&\ghost{P}&\ddots&\qw&\qw&\gate{\raisebox{.5pt}{\textcircled{\raisebox{-.9pt} {2}}} }\qwx[-1]&\qw&\qw&{\iddots}&\ghost{P}&\qw&\qw\\
\lstick{\vdots}&\qw&\ddots& &\ddots&\ddots&\vdots&\iddots& &\iddots&\iddots& &\qw\\
\lstick{n-1}&\qw&\qw&\qw&\ddots&\multigate{1}{P}&\gate{\raisebox{.5pt}{\textcircled{\raisebox{-.9pt} {2}}} }&\multigate{1}{P}&\iddots&\qw&\qw&\qw&\qw\\
\lstick{n}&\qw&\qw&\qw&\qw&\ghost{P}&\gate{X_{+1}}\qwx[-1]&\ghost{P}&\qw&\qw&\qw&\qw&\qw\\}}
\end{eqnarray}

Hence, the gate \ref{InvToff3} is equivalent to  \begin{eqnarray}
        \framebox[13.25cm][c]{\Qcircuit @C=0.25cm @R=.25cm {
\lstick{1}&\gate{X_{+2}}&\multigate{1}{P}&\qw&\qw&\qw&\qw&\gate{\raisebox{.5pt}{\textcircled{\raisebox{-.9pt} {2}}} }&\qw&\qw&\qw&\qw&\multigate{1}{P}&\gate{X_{+1}}&\qw\\
\lstick{2}&\gate{X_{+2}}&\ghost{P}&\multigate{1}{P}&\qw&\qw&\qw&\gate{\raisebox{.5pt}{\textcircled{\raisebox{-.9pt} {2}}} }\qwx[-1]&\qw&\qw&\qw&\multigate{1}{P}&\ghost{P}&\gate{X_{+1}}&\qw\\
\lstick{3}&\gate{X_{+2}}&\qw&\ghost{P}&\ddots&\qw&\qw&\gate{\raisebox{.5pt}{\textcircled{\raisebox{-.9pt} {2}}} }\qwx[-1]&\qw&\qw&\iddots&\ghost{P}&\qw&\gate{X_{+1}}&\qw\\
\lstick{\vdots}&\vdots&\hdots&\ddots& &\ddots&\ddots&\vdots&\iddots &\iddots& &\iddots&\hdots&\vdots&\qw\\
\lstick{n-1}&\gate{X_{+2}}&\qw&\qw&\qw&\ddots&\multigate{1}{P}&\gate{\raisebox{.5pt}{\textcircled{\raisebox{-.9pt} {2}}} }&\multigate{1}{P}&\iddots&\qw&\qw&\qw&\gate{X_{+1}}&\qw\\
\lstick{n}&\qw&\qw&\qw&\qw&\qw&\ghost{P}&\gate{X_{+1}}\qwx[-1]&\ghost{P}&\qw&\qw&\qw&\qw&\qw&\qw\\}}
\end{eqnarray}
This concludes the construction of quantum circuit model of the DTQWs on Cayley graphs considered in this paper  using single qutrit rotation gates and controlled qutrit gates. 

\begin{remark}
We observe that the qutrit circuit models developed above can also be derived for other DTQW models on other graphs when the walk can be studied by converting the time-space into Fourier space.     
\end{remark}

\subsection{Circuit complexity of the models}

In this section we discuss the circuit complexity of the quantum circuit models for the DTQWs developed above. This includes the number of elementary gates that are needed to implement the circuits for the cycle graphs i.e. $\mathrm{Cay(\mathbb{Z}_N,\{1,-1\})}$ and the Cayley graphs $\mathrm{Cay(D_N,\{a,b\})}$ where $3^{n-1}\leq N\leq 3^n.$ Note that a time-step of a DTQW is same as one layer in the circuit model for the implementation of $t$-step quantum walks, $t\geq 1.$

Moreover, the primary gates that are used in the design of the circuits are multi-controlled $X$  gates (generalized Toffoli gates) which correspond to the construction of the circuits for block diagonal unitary matrices, and the gates \textsc{Increment}, \textsc{Decrement}, \textsc{Destop}, \textsc{Instop}, and \textsc{RC}. 

First we have the following result for multi-controlled $X$ gates.


\begin{theorem}\label{count1}
   The multi-controlled $X$ gates described in equations (\ref{circ1}) and (\ref{circ2}), where $X\in \{X_{+1},X_{+2}\}$ need $O(4.3^{n-1})$ two-qutrit controlled $X$ gates and $O(2.3^{n-1})$ one-qutrit rotation gates. 
\end{theorem}
\pf Follows from using simple induction on the number of gates and the proof directly follows from equation (\ref{multiToff21}), Theorem \ref{mqzyz}, construction of qutrit SWAP gates and Theorem \ref{swapeq}.  $\hfill{\square}$

From the construction of multi-controlled qutrit gates, it is obvious that that more number of two-qutrit controlled $X$ gates or M-S gates are always used if we want more number of standard basis elements to remain invariant under multi-controlled qutrit gate transformation. Hence, it is obvious that a circuit like \begin{eqnarray}\label{multiToff4}
\framebox[8cm][c]{\Qcircuit @C=1em @R=.7em {
    &\lstick{1}&\qw& \gate{\raisebox{.5pt}{\textcircled{\raisebox{-.9pt} {a}}}} &\qw&\qw\\
    &\lstick{2}&\qw&\gate{\raisebox{.5pt}{\textcircled{\raisebox{-.9pt} {a}}}}\qwx[-1]&\qw&\qw\\
    &\lstick{3}&\qw&\gate{\raisebox{.5pt}{\textcircled{\raisebox{-.9pt} {a}}}}\qwx[-1]&\qw&\qw\\
    &\lstick{\vdots}& \qw&\vdots& &\qw\\
    &\lstick{n-1}&\qw&\gate{\raisebox{.5pt}{\textcircled{\raisebox{-.9pt} {a}}}}&\qw&\qw\\
    &\lstick{n}&\qw&\gate{X_{+2}}\qwx[-1]&\qw&\qw\\}}\end{eqnarray}
 comprises of more number of two-qutrit controlled $X$ gates or M-S gates than that a circuit like \begin{eqnarray}\label{multiToff5}
\framebox[8cm][c]{\Qcircuit @C=1em @R=.7em {
    &\lstick{1}&\qw& \gate{\raisebox{.5pt}{\textcircled{\raisebox{-.9pt} {a}}}} &\qw&\qw\\
    &\lstick{2}&\qw&\qw\qwx[-1]&\qw&\qw\\
    &\lstick{3}&\qw&\gate{\raisebox{.5pt}{\textcircled{\raisebox{-.9pt} {a}}}}\qwx[-1]&\qw&\qw\\
    &\lstick{\vdots}& \qw&\vdots& &\qw\\
    &\lstick{n-1}&\qw&\gate{\raisebox{.5pt}{\textcircled{\raisebox{-.9pt} {a}}}}&\qw&\qw\\
    &\lstick{n}&\qw&\gate{X_{+2}}\qwx[-1]&\qw&\qw\\}}\end{eqnarray} where $a\in \{0,1,2\}$.\\

We also note that, the circuit 
 \begin{eqnarray}\label{Inc2}
\framebox[8cm][c]{\Qcircuit @C=1em @R=.7em {
    &\lstick{1}&\qw& \gate{\raisebox{.5pt}{\textcircled{\raisebox{-.9pt} {a}}}} &\qw\\
    &\lstick{2}&\qw&\multigate{3}{\textsc{Increment}}\qwx[-1]&\qw\\
    &\lstick{\vdots}&\qw&\ghost{\textsc{Increment}}&\qw\\
    &\lstick{n}&\qw&\ghost{\textsc{Increment}}&\qw\\
    &\lstick{n+1}&\qw&\ghost{\textsc{Increment}}&\qw\\}}\end{eqnarray} is equivalent to 
     \begin{eqnarray}\label{Inc3}
\framebox[12cm][c]{\Qcircuit @C=1em @R=.7em {
  &\lstick{1}&\qw&\gate{\raisebox{.5pt}{\textcircled{\raisebox{-.9pt} {a}}} } &\gate{\raisebox{.5pt}{\textcircled{\raisebox{-.9pt} {a}}} } & \gate{\raisebox{.5pt}{\textcircled{\raisebox{-.9pt} {a}}} } &\gate{\raisebox{.5pt}{\textcircled{\raisebox{-.9pt} {a}}} } &\gate{\raisebox{.5pt}{\textcircled{\raisebox{-.9pt} {a}}} }&\qw\\
 &\lstick{2}&\qw&\gate{X_{+1}}\qwx[-1] &\qw & \qw &\qw &\qw&\qw\\
 &\lstick{3}&\qw&\gate{\raisebox{.5pt}{\textcircled{\raisebox{-.9pt} {2}}} }\qwx[-1] &\gate{X_{+1}}\qwx[-2]  & \ddots & \ddots &\ddots &\qw\\
 &\lstick{\vdots}&\qw&\vdots &\vdots &\gate{\ddots X_{+1}}\qwx[-3] & \vdots\ddots &\hdots &\qw\\
 &\lstick{n}&\qw&\gate{\raisebox{.5pt}{\textcircled{\raisebox{-.9pt} {2}}} } &\gate{\raisebox{.5pt}{\textcircled{\raisebox{-.9pt} {2}}} }  &\gate{\raisebox{.5pt}{\textcircled{\raisebox{-.9pt} {2}}} } \qwx[-1] &\gate{X_{+1}}\qwx[-4]  &\qw&\qw\\
 &\lstick{n+1}&\qw&\gate{\raisebox{.5pt}{\textcircled{\raisebox{-.9pt} {2}}} } \qwx[-1]&\gate{\raisebox{.5pt}{\textcircled{\raisebox{-.9pt} {2}}} } \qwx[-1]& \gate{\raisebox{.5pt}{\textcircled{\raisebox{-.9pt} {2}}} } \qwx[-1] &\gate{\raisebox{.5pt}{\textcircled{\raisebox{-.9pt} {2}}} }\qwx[-1] &\gate{X_{+1}}\qwx[-5]&\qw\\}}\end{eqnarray}

Besides, we know that \begin{eqnarray*}
    \framebox[6cm][c]{\Qcircuit @C=1em @R=.7em {
  &\lstick{}&\qw&\gate{\raisebox{.5pt}{\textcircled{\raisebox{-.9pt} {1}}} }&\qw \\}}
\end{eqnarray*} is equivalent to \begin{eqnarray*}
    \framebox[6cm][c]{\Qcircuit @C=1em @R=.7em {
  &\lstick{}&\qw&\gate{X_{+1}}&\gate{\raisebox{.5pt}{\textcircled{\raisebox{-.9pt} {2}}} }&\gate{X_{+2}}&\qw \\}}
\end{eqnarray*} Similar argument is also valid when we replace 
\textsc{Increment} by \textsc{Decrement} as well. Further, we see that constructing a general one-qutrit gate (special unitary matrix) requires $9$ one-qutrit rotation gates. Thus we have the following theorem.

\begin{theorem}\label{Dicayley}
    A single layer (one time-step) of DTQW on $\mathrm{Cay}(D_N,\{a,b\})$ where $N=3^n$ requires $O(8n3^{n+1}+2)$  two-qutrit controlled $X$ gates and $O(4.3^{n+1})$ one-qutrit rotation gates.
\end{theorem}

\pf Follows from the argument above and Theorem \ref{count1}. $\hfill{\square}$

Now since constructing gates like \textsc{Destop} and \textsc{Instop} require a sequence of multi-controlled qutrit $X$ gates, it is clear from Theorem \ref{count1} that the number of two-qutrit controlled $X$ gates required for constructing such gates is $O(4p(n)3^{n-1})$ for some polynomial $p$. Hence, we have the following corollary.

\begin{corollary}\label{Dicayleyless}
    A single layer (time-step) of DTQW on $\mathrm{Cay}(D_N,\{a,b\}),$ $3^{n-1}\leq N<3^n$ requires $O(8(n+p(n))3^{n+1})$  two-qutrit controlled $X$ gates and $O(4(n+q(n))3^{n+1})$  one-qutrit rotation gates, where $p, q$ are some polynomials.
\end{corollary}

\pf Follows from the arguments above, Theorem \ref{count1}, and Theorem \ref{Dicayley}.  $\hfill\square$

Similar results can be obtained for three-state lively DTQWs on cycles i.e. $\mathrm{Cay}(\Z_N,\{1,-1\})$, however, it is to be observed that for non-zero liveliness parameter $a$, by the same argument used so far, the gate \textsc{RC}$(a)$ requires $O(4na3^{n-1})$ two-qutrit controlled $X$ gates. Further, unlike DTQW on Cayley graphs of Dihedral groups which is a $n+2$-qutrit circuit, this circuit is a unitary matrix of order $3^{n+1}\times 3^{n+1}$. Hence, we obtain the following results.
  
\begin{corollary}\label{cyclecount1}
    A single layer (one time-step) of three-state lively DTQW on $\mathrm{Cay}(\Z_N,\{1,-1\}),$ $N=3^n,a\leq \lfloor\frac{N}{2}\rfloor$, $O((8n+4na)3^{n})$  two-qutrit controlled $X$ gates and $O(4na3^{n})$  one-qutrit rotation gates.
\end{corollary}

\pf The proof follows from the argument above, Theorem 
 \ref{count1}, and Theorem\ref{Dicayley}. $\hfill{\square}$

\begin{corollary}\label{cyclecount2}
    A single layer (time-step) of three-state lively DTQW on $\mathrm{Cay}(\Z_N,\{1,-1\}),$ $3^{n-1}\leq N<3^n,a\leq \lfloor\frac{N}{2}\rfloor$ requires $O(4(na+f(n))3^{n})$  two-qutrit controlled $X$ gates and $O(4(na+g(n))3^{n})$  one-qutrit rotation gates, where $f,g$ are some polynomials.
\end{corollary}

\pf Follows from the argument above, Theorem \ref{count1}, Theorem\ref{Dicayley}. $\hfill\square$

Finally, we have the following theorem.

\begin{theorem}\label{blkdiagcount}
Let $U\in SU(3^n)$ be a block diagonal special unitary matrix with $3\times 3$ special unitary diagonal blocks. Then the quantum circuit for $U$ requires $O(2.3^{n+1})$ two-qutrit controlled $X$ gates and $O(3^{n+1})$ one-qutrit rotation gates.  
\end{theorem}
\pf The proof follows from theorem \ref{mqzyz} and equation (\ref{blkdg}). $\hfill{\square}$

In the following section we  numerically simulate these quantum circuits by inducing generic noise models in order to replicate its output which would be obtained in a noisy quantum computer.

\section{Noisy simulation of the circuit model}\label{sec:5}

In this section, we report numerical simulation results based on the circuit models as obtained above. The simulations are performed incorporating noise models in order to mimic a noisy quantum computer such as noisy intermediate scale quantum (NISC) computers for implementation of the qutrit model of the DTQWs on $\mathrm{Cay(D_N,\{a,b\})}$ and $\mathrm{Cay(\mathbb{Z}_N,\{1,-1\})}$. These noise include gate error and idle error, which are standard practice for numerical simulation of quantum circuits \cite{Gokhale2019}. Further, in order to investigate and verify the simulations results and the analytical results obtained in the literature, we consider a family of one-parameter coins for the DTQWs, known as generalized Grover coins. See \cite{Sarkar2020} for a more on generalized Grover coins.

The generalized Grover coin matrices of order $3$  are orthogonal matrices that can be expressed as linear sum of permutation matrices. These matrices are divided into four classes as described below.



\begin{eqnarray}
 \mathcal{X}_\theta &=& \left\{\bmatrix{\frac{2\cos{\theta}+1}{3} & \frac{1-\cos{\theta}}{3}+\frac{\sin{\theta}}{\sqrt{3}} & \frac{1-\cos{\theta}}{3}-\frac{\sin{\theta}}{\sqrt{3}}\\\frac{1-\cos{\theta}}{3}-\frac{\sin{\theta}}{\sqrt{3}} & \frac{2\cos{\theta}+1}{3} & \frac{1-\cos{\theta}}{3}+\frac{\sin{\theta}}{\sqrt{3}}\\\frac{1-\cos{\theta}}{3}+\frac{\sin{\theta}}{\sqrt{3}}&\frac{1-\cos{\theta}}{3}-\frac{\sin{\theta}}{\sqrt{3}}&\frac{2\cos{\theta}+1}{3}} : \, -\pi<\theta\leq\pi \right\}, \label{Xthetaclass} \\
 \mathcal{Y}_\theta &=& \left\{\bmatrix{\frac{2\cos{\theta}-1}{3} & \frac{-1-\cos{\theta}}{3}+\frac{\sin{\theta}}{\sqrt{3}} & \frac{-1-\cos{\theta}}{3}-\frac{\sin{\theta}}{\sqrt{3}}\\\frac{-1-\cos{\theta}}{3}-\frac{\sin{\theta}}{\sqrt{3}} & \frac{2\cos{\theta}-1}{3} & \frac{-1-\cos{\theta}}{3}+\frac{\sin{\theta}}{\sqrt{3}}\\\frac{-1-\cos{\theta}}{3}+\frac{\sin{\theta}}{\sqrt{3}}&\frac{-1-\cos{\theta}}{3}-\frac{\sin{\theta}}{\sqrt{3}}&\frac{2\cos{\theta}-1}{3}} : \, -\pi<\theta\leq\pi \right\}, \label{Ythetaclass} \\
 \mathcal{Z}_\theta &=& \left\{\bmatrix{\frac{2\cos{\theta}+1}{3} & \frac{1-\cos{\theta}}{3}+\frac{\sin{\theta}}{\sqrt{3}} & \frac{1-\cos{\theta}}{3}-\frac{\sin{\theta}}{\sqrt{3}}\\\frac{1-\cos{\theta}}{3}+\frac{\sin{\theta}}{\sqrt{3}} & \frac{1-\cos{\theta}}{3}-\frac{\sin{\theta}}{\sqrt{3}} & \frac{2\cos{\theta}+1}{3}\\\frac{1-\cos{\theta}}{3}-\frac{\sin{\theta}}{\sqrt{3}} & \frac{2\cos{\theta}+1}{3}& \frac{1-\cos{\theta}}{3}+\frac{\sin{\theta}}{\sqrt{3}}}: \, -\pi<\theta\leq\pi \right\},  \label{Zthetaclass} \\
 \mathcal{W}_\theta &=& \left\{\bmatrix{\frac{2\cos{\theta}-1}{3} & \frac{-1-\cos{\theta}}{3}+\frac{\sin{\theta}}{\sqrt{3}} & \frac{-1-\cos{\theta}}{3}-\frac{\sin{\theta}}{\sqrt{3}}\\\frac{-1-\cos{\theta}}{3}+\frac{\sin{\theta}}{\sqrt{3}} & \frac{-1-\cos{\theta}}{3}-\frac{\sin{\theta}}{\sqrt{3}} & \frac{2\cos{\theta}-1}{3}\\\frac{-1-\cos{\theta}}{3}-\frac{\sin{\theta}}{\sqrt{3}} & \frac{2\cos{\theta}-1}{3}& \frac{-1-\cos{\theta}}{3}+\frac{\sin{\theta}}{\sqrt{3}}}: \, -\pi<\theta\leq\pi \right\},  \label{Wthetaclass}
\end{eqnarray}

The Grover matrix of order $3$ given by $$\mathsf{G}=\bmatrix{\frac{-1}{3} & \frac{2}{3}  & \frac{2}{3} \\ \frac{2}{3}  & \frac{-1}{3}  & \frac{2}{3} \\\frac{2}{3} &\frac{2}{3} &\frac{-1}{3} } \in \mathcal{X}_{\theta}$$ by setting $\theta=\pi.$

We also corroborate our results on localization pertaining to DTQWs on Cayley graphs obtained in \cite{SarmaSarkar2023}. All simulations have been done in MATLAB2019a  on a system with 16GB RAM, Intel(R) Core(TM) i5 - 035G1 CPU @1.00GHz1.19GHz processor.


\subsection{Noise models}

The noise in the gate and idle errors are incorporated through the use of Kraus Operators. Kraus operators are a set of positive semi-definite matrices $\{E_j|j\subset \mathbb{N} \}$, such that the time evolution of a quantum system with initial state $\sigma = \ket{\psi} \bra{\psi}$ is expressed as a
function $E(\sigma)=\sum_jE_j\sigma E_j^{\star}$, where $\star$ denotes the conjugate transpose. We follow the operators discussed in \cite{Gokhale2019,Ramzan2012}.

\subsubsection{Gate Noise}
For qutrits , taking $X_{+1}=\bmatrix{0&0&1\\1&0&0\\0&1&0}$ and $Z_3=\bmatrix{1&0&0\\0&\exp{(i2\pi/3)}&0\\0&0&\exp{(i4\pi/3)}}$, the depolarizer gate error for single qutrit gate is defined as\begin{eqnarray}\label{Error1}
    E_{G1}(\sigma)=(I-\sum_{j,k\in \{0,1,2\}}p_{jk}\sigma)+\sum_{j,k\in \{0,1,2\}}p_{jk}(X_{+1}^jZ_3^{k})\sigma(X_{+1}^jZ_3^{k})^\star.
\end{eqnarray} 

We assume the probabilities of all the error terms to be equal to $p_1$. Then the equation (\ref{Error1}) can be rewritten as \begin{eqnarray}\label{Error2}
    E_{G1}(\sigma)=(I-9p_{1}\sigma)+\sum_{j,k\in \{0,1,2\}}p_{1}(X_{+1}^jZ_3^{k})\sigma(X_{+1}^jZ_3^{k})^*
\end{eqnarray}

For $n$-qutrit gates, the gate error is defined as 
\begin{eqnarray}\label{Errorn}
    E_{Gn}(\sigma)=(I-3^{2n}p_{1}\sigma)+\sum_{a_1,a_2,\hdots,a_n,b_1,b_2,\hdots b_n\in \{0,1,2\}} p_{1}E_{a_1b_1a_2b_2\hdots a_nb_n}\sigma E_{a_1b_1a_2b_2\hdots a_nb_n}^\star 
\end{eqnarray} where $E_{a_1b_1a_2b_2\hdots a_nb_n}=\otimes_{j=1}^nX_{+1}^{a_j}Z_3^{b_j}.$

\subsubsection{Idle Error}

Idle errors occur from decoherence of a quantum system that arises with interaction with the environment. Idle errors such as amplitude and phase damping errors are incorporated for excited qutrits. Some variants of idle errors are presented here.

\begin{enumerate}
    \item {\bf Amplitude damping:} In a mathematical sense, amplitude damping idle errors are represented using the expression \begin{eqnarray}\label{Errordamp}
    K_A(\sigma)=\sum_{j\in\{0,1,2\}}K_j\sigma K_j^{*}
\end{eqnarray}  set of Kraus operators $\{K_j\}$ such that $\sum_jK_jK_j^*=I$ and \begin{eqnarray}
&& K_0(t) = \bmatrix{1&0&0\\0&\sqrt{\exp{(-r_1 t)}}&0\\0&0&\sqrt{\exp{(-r_2 t)}}},K_1(t)=\bmatrix{0&\sqrt{1-\exp{(-r_1 t)}}&0\\0&0&0\\0&0&0}, \\ && K_2(t)=\bmatrix{0&0&\sqrt{1-\exp{(-r_2 t)}}\\0&0&0\\0&0&0},
\end{eqnarray} where $r_1,r_2, >0.$

\item {\bf Phase damping:}
Similar to amplitude damping, the phase damping idle errors are represented using the expression \begin{eqnarray}\label{Errorphasedamp}
    E_A(\sigma)=\sum_{j\in\{0,1\}}K_j\sigma K_j^{*}
\end{eqnarray}  set of Kraus operators $\{K_j\}$ such that $\sum_jK_jK_j^*=I$ and $K_0(t)=\sqrt{\exp{(-r_1 t)}}I,K_1(t)=\sqrt{1-\exp{(-r_1 t)}}Z_3,$ where $r_1>0.$ 
\end{enumerate}

Now, we shall incorporate these errors in to our circuit and numerically simulate it in order to see the effect of various errors on localization of the walk. 

\subsection{Numerical simulation of three-state DTQW on  $\mathrm{Cay(D_{27},\{a,b\})}$}
We corroborate our results from \cite{SarmaSarkar2023} on time-averaged probability of three state quantum walks and periodicity of $\mathrm{Cay(D_{N},\{a,b\})}$ where $N=27$. We recall that the unitary matrix corresponding to increment and decrement gates are \begin{eqnarray*}
    \textsc{Increment} =\left[ 
\begin{array}{c|c} 
      0_{1\times 26} &  1 \\ 
      \hline 
       I_{26} &  0_{26\times 1}\\
    \end{array} 
    \right],\,\,
    \textsc{Decrement}=\left[ 
\begin{array}{c|c} 
      0_{26\times 1} &  I_{26} \\ 
      \hline 
        1 &  0_{1\times 26}\\
    \end{array} 
    \right]
\end{eqnarray*}

The unitary matrix for the circuit  \\\centerline{\Qcircuit @C=1em @R=.7em {
 &\lstick{1}& \qw&\gate{\raisebox{.5pt}{\textcircled{\raisebox{-.9pt} {0}}} }&\qw\\
  &\lstick{2}& \qw&\gate{\raisebox{.5pt}{\textcircled{\raisebox{-.9pt} {0}}} }\qwx[-1]&\qw\\
 &\lstick{3}& \qw&\multigate{2}{\textsc{Increment}}\qwx[-1]&\qw\\&\lstick{4}& \qw&\ghost{\textsc{Increment}}&\qw\\&\lstick{5}& \qw&\ghost{\textsc{Increment}}&\qw}} is given by \begin{eqnarray*}
    U_{In}=\left[ 
\begin{array}{c|c} 
      \textsc{Increment} &  0_{27\times 216} \\ 
      \hline 
       0_{216\times 27} &  I_{216}\\
    \end{array} 
    \right]
\end{eqnarray*}

Similarly, the unitary matrix for the circuit  \\\centerline{\Qcircuit @C=1em @R=.7em {
 &\lstick{1}& \qw&\gate{\raisebox{.5pt}{\textcircled{\raisebox{-.9pt} {0}}} }&\qw\\
  &\lstick{2}& \qw&\gate{\raisebox{.5pt}{\textcircled{\raisebox{-.9pt} {1}}} }\qwx[-1]&\qw\\
 &\lstick{3}& \qw&\multigate{2}{\textsc{Decrement}}\qwx[-1]&\qw\\&\lstick{4}& \qw&\ghost{\textsc{Decrement}}&\qw\\&\lstick{5}& \qw&\ghost{\textsc{Increment}}&\qw}} is given by \begin{eqnarray*}
    U_{De}=\left[ 
\begin{array}{c|c|c} 
      I_{27} & O_{27\times 27} & O_{27\times 189}\\ 
      \hline 
       O_{27\times 27} &  \textsc{Decrement} & O_{27\times 189}\\
       \hline 
       O_{189\times 27} &  O_{27\times 27} & I_{189}\\
    \end{array} 
    \right].
\end{eqnarray*}

In Figure \ref{fig:Timeavamp}, we plot the time-averaged probability of finding a particle at all vertices of $\mathrm{Cay(D_{27},\{a,b\})}$ for the initial coin state $\ket{0}_3$ and starting vertex $(1,0)$. In the first figure the coin $C$ is taken from the class $\mathcal{X}_\theta$ where $\theta=\pi$ i.e. the coin is the Grover coin $\mathsf{G}$. In the subsequent figures, the coin $C$ is taken from the class $\mathcal{Y}_\theta$ where $\theta=\pi/2$, $C\in \mathcal{Z}_\theta,\theta=\pi/3$ and $C\in \mathcal{W}_\theta, \theta=-\pi/4$ respectively. We incorporate the gate error and amplitude damping idle error in order to replicate the output which would be typically found in a quantum computer. We incorporate varying degrees of error and see how much the result deviates from the result obtained via noiseless simulation. Firstly we run a noiseless simulation, then we incorporate high values $(O(10^{-2}))$ of gate error and idle error with the error parameters chosen at random from uniform distribution. We also simulate the circuit by incorporating only gate error and only channel error to see the effect of individual errors on the simulation. Finally we simulate the circuit with very small values ($O(10^{-6})$) of gate and idle errors. We notice that for high value of errors, the maximum probability of finding the walker on the graph is not necessarily found at the starting vertex or its reflection vertex as observed in noiseless simulation. It is also to be noted that the error parameters $r_1,r_2,p_1$ are chosen randomly via uniform distribution between $0$ and $10^{-\epsilon}$ to generate an error of order $O(10^{-\epsilon-1})$.

\begin{figure}[H]
    \centering
     \subfigure[$C\in \mathcal{X}_\theta,\theta=\pi$]{\includegraphics[height=3.5 cm,width=8 cm]{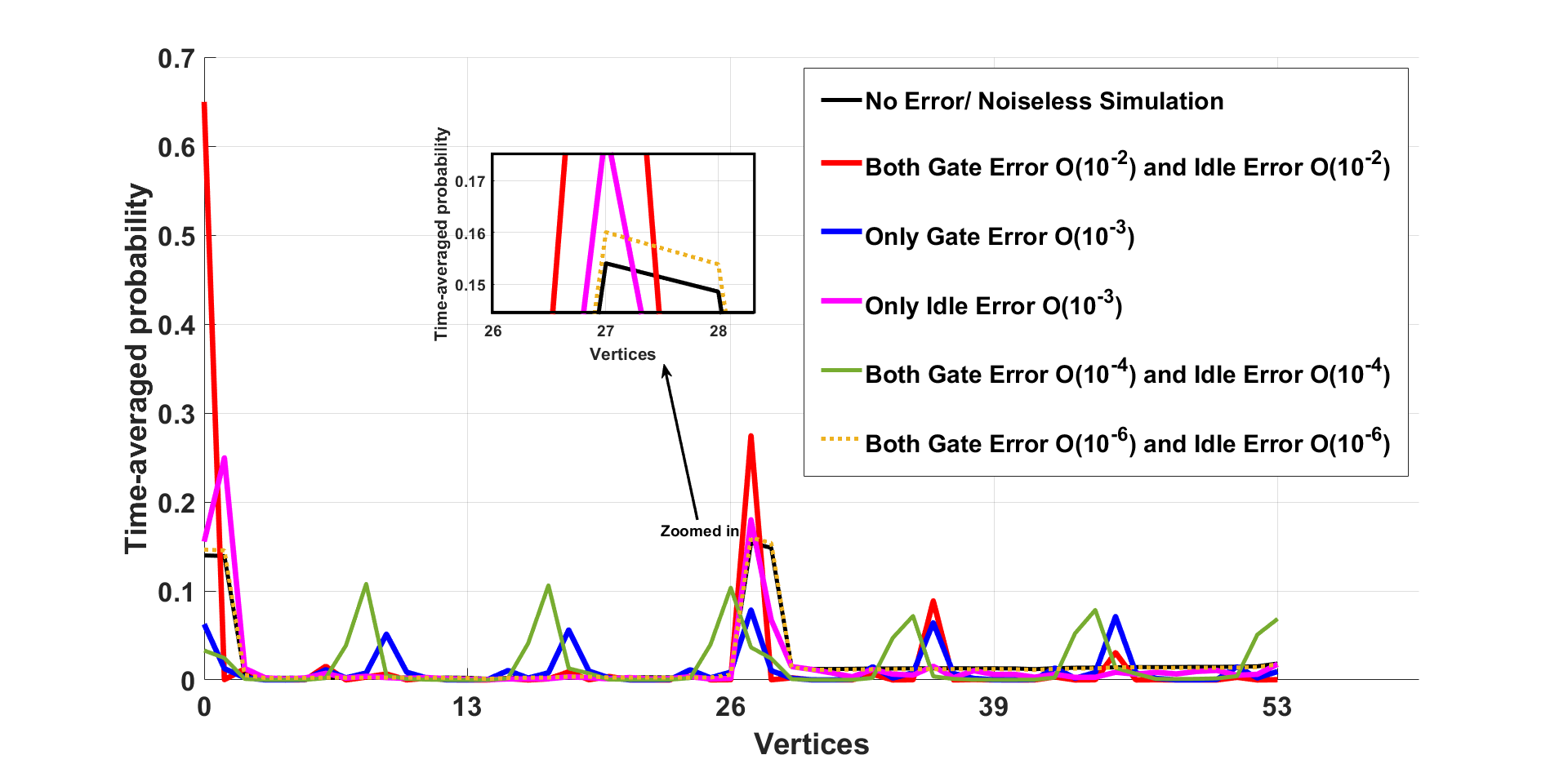}}
   \subfigure[$C\in \mathcal{Y}_\theta,\theta=\pi/2$]{\includegraphics[height=3.5 cm,width=8 cm]{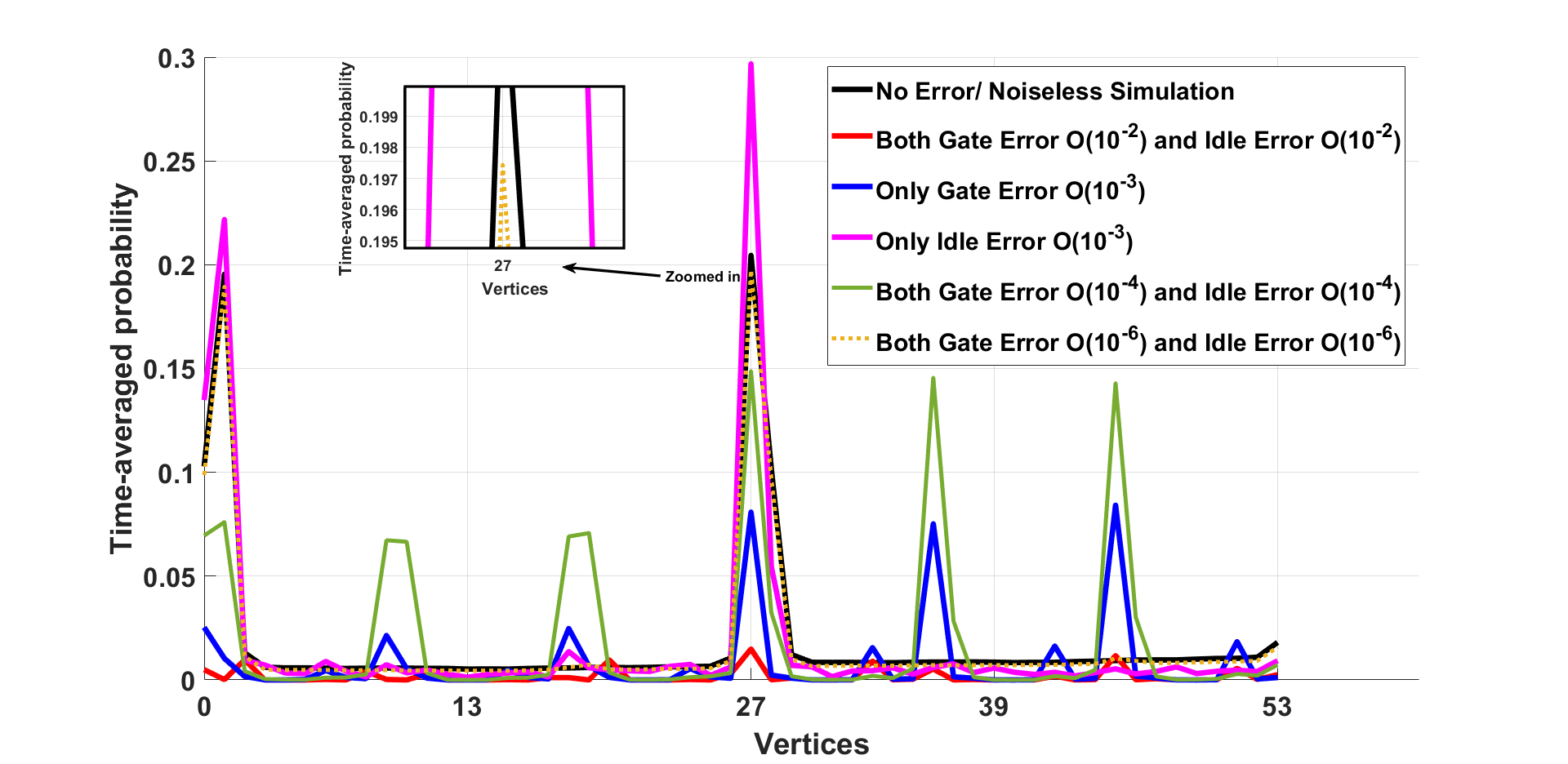}}
   \subfigure[$C\in \mathcal{Z}_\theta,\theta=\pi/3$]{\includegraphics[height=3.5 cm,width=8 cm]{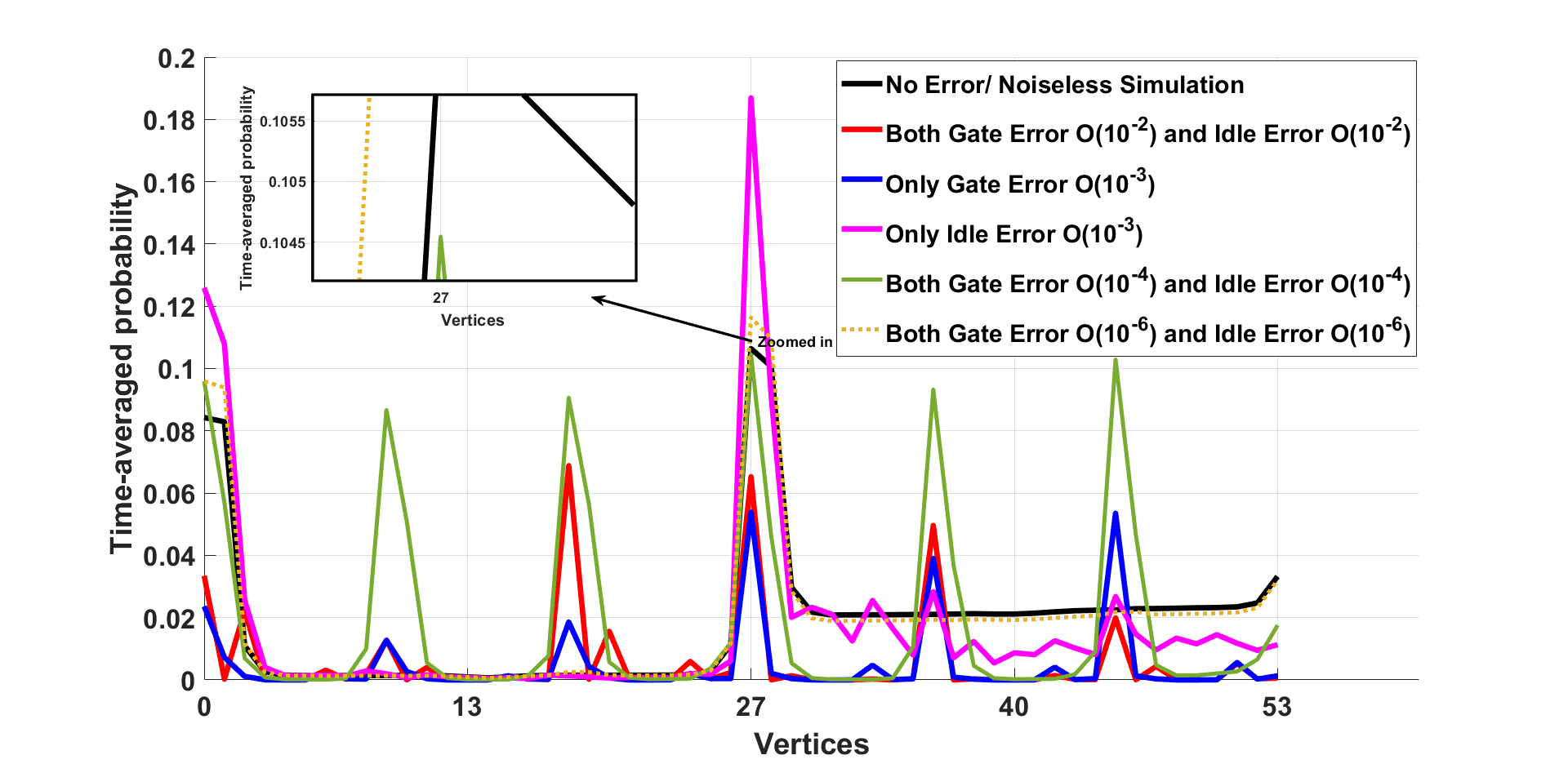}}
   \subfigure[$C\in \mathcal{W}_\theta,\theta=-\pi/4$]{\includegraphics[height=3.5 cm,width=8 cm]{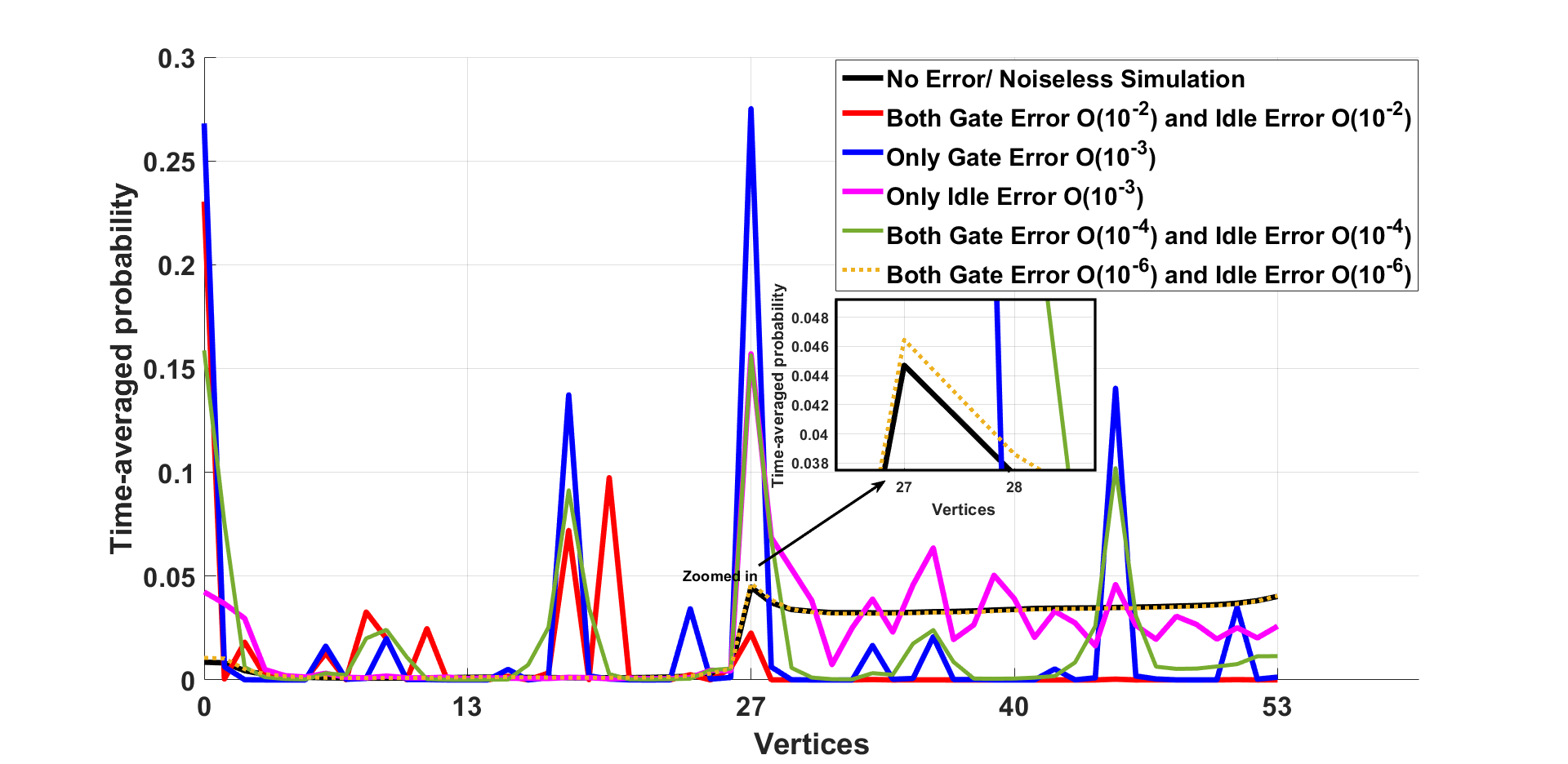}}
    \caption{Time averaged probability for $\mathrm{Cay(D_{27},\{a,b\})}$ taking coins from the classes $,\mathcal{X}_\theta,\mathcal{Y}_\theta,\mathcal{Z}_\theta,\mathcal{W}_\theta$ with initial position $(1,0)$ and initial coin state $\ket{0}_3$. The time step is taken up to $300$. The generic depolarizer gate noise and amplitude damping idle noise is incorporated in the circuit.}\label{fig:Timeavamp}
\end{figure}

Now, in Figure \ref{fig:Timeavph} we plot the time-averaged probability at various positions of the walk on $\mathrm{Cay(D_{27},\{a,b\})}$ with coin state $\ket{0}_3$ and starting vertex $(1,0)$ and coin is taken from the classes $\mathcal{X}_\theta,\mathcal{Y}_\theta,\mathcal{Z}_\theta,\mathcal{W}_\theta$ for several values of $\theta$ as described in Figure $\ref{fig:Timeavamp}$. The only difference is that we incorporate phase damping instead of the amplitude damping as the idle error and the error parameters are chosen from uniform distribution.

\begin{figure}[H]
    \centering
     \subfigure[$C\in \mathcal{X}_\theta,\theta=\pi$]{\includegraphics[height=3.5 cm,width=8 cm]{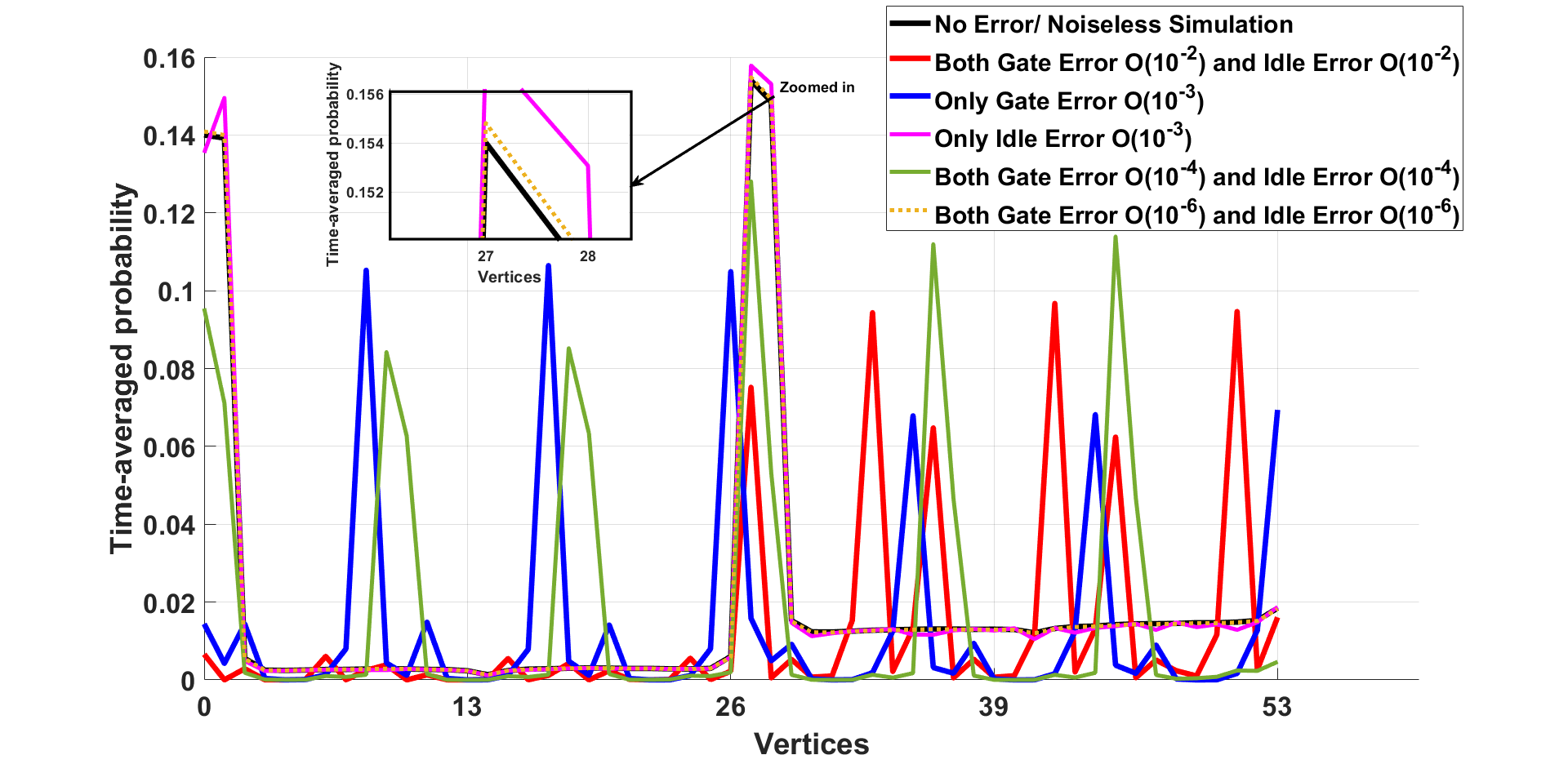}}
   \subfigure[$C\in \mathcal{Y}_\theta,\theta=\pi/2$]{\includegraphics[height=3.5 cm,width=8 cm]{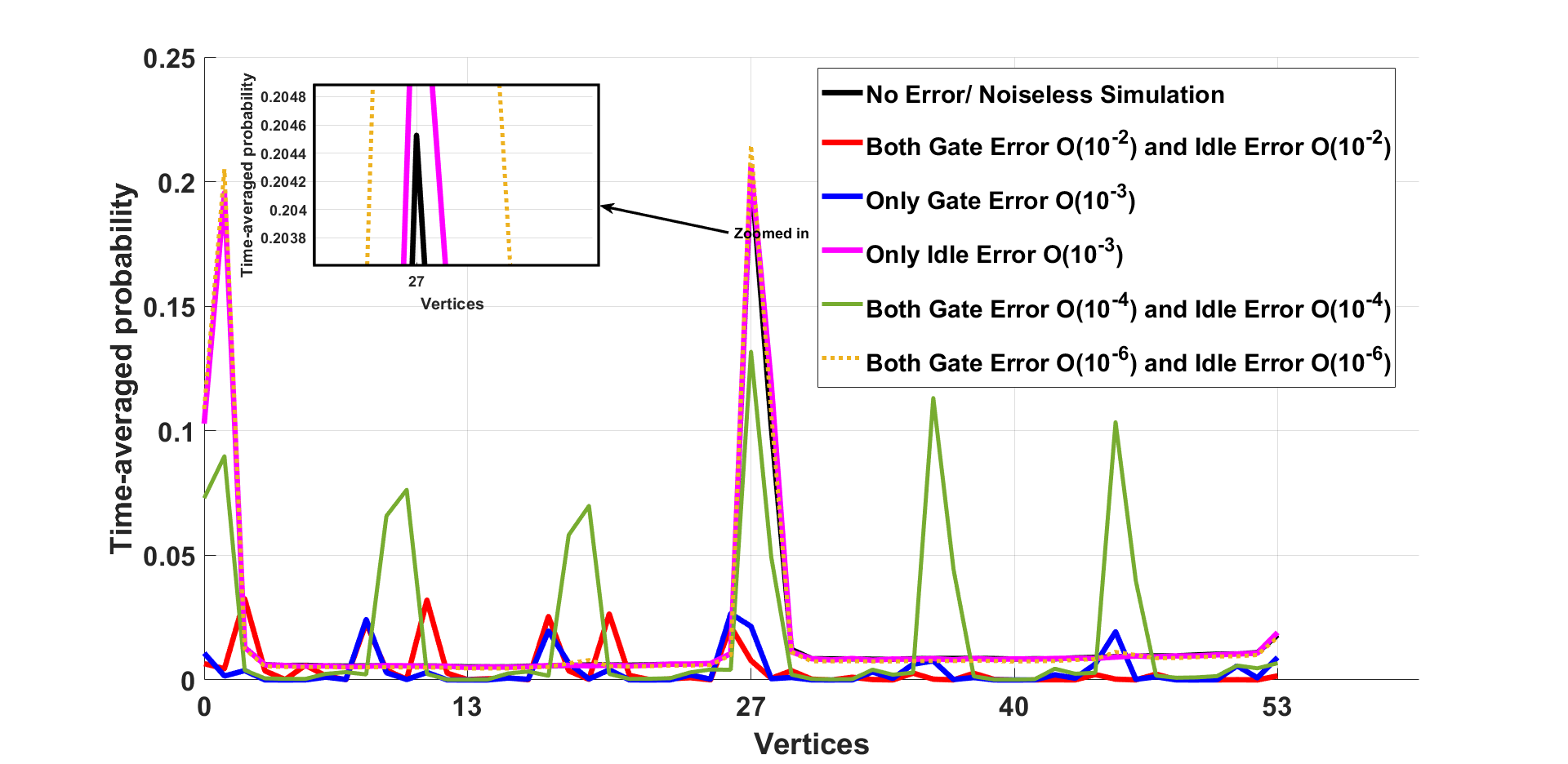}}
   \subfigure[$C\in \mathcal{Z}_\theta,\theta=\pi/3$]{\includegraphics[height=3.5 cm,width=8 cm]{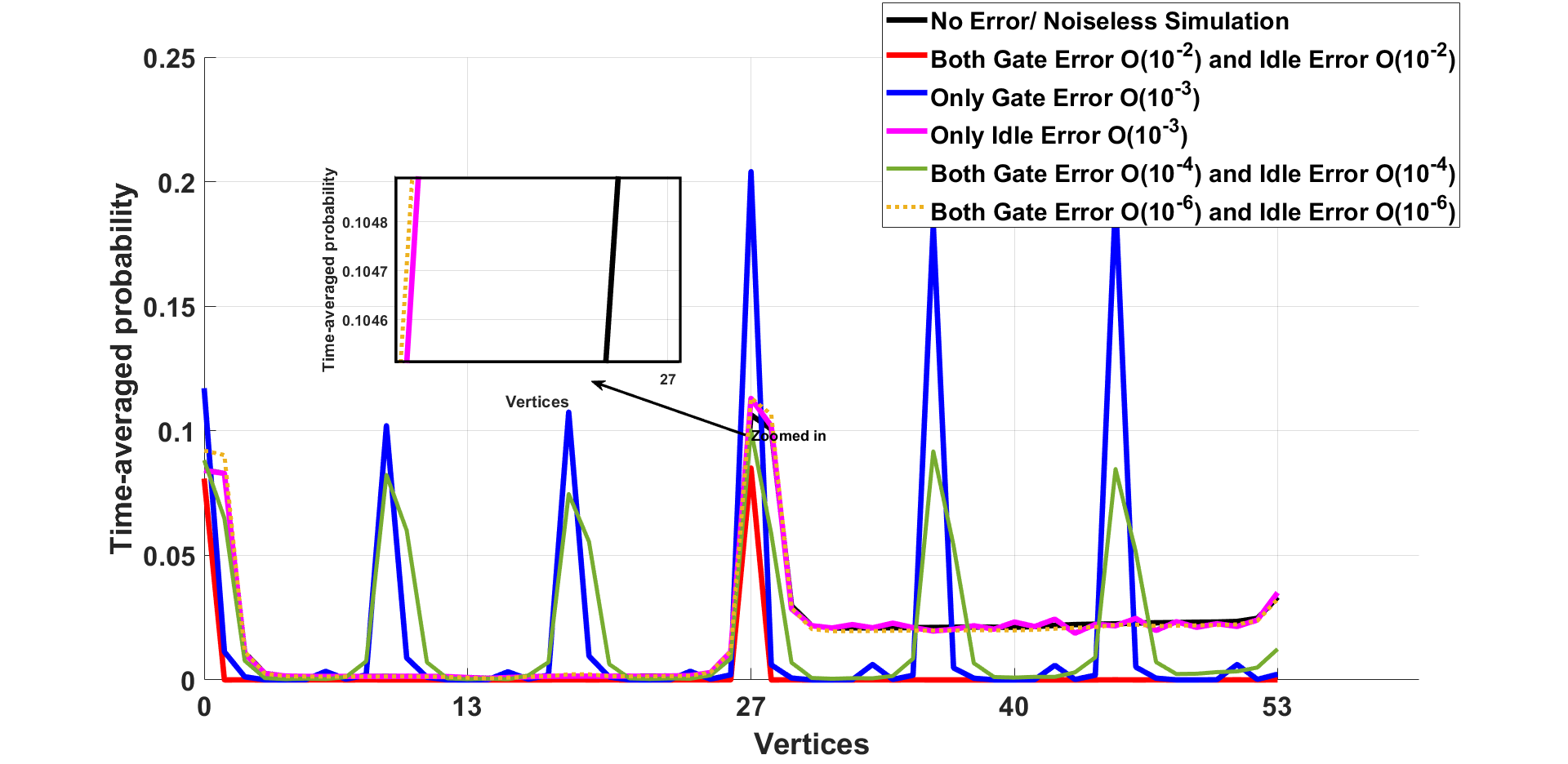}}
   \subfigure[$C\in \mathcal{W}_\theta,\theta=-\pi/4$]{\includegraphics[height=3.5 cm,width=8 cm]{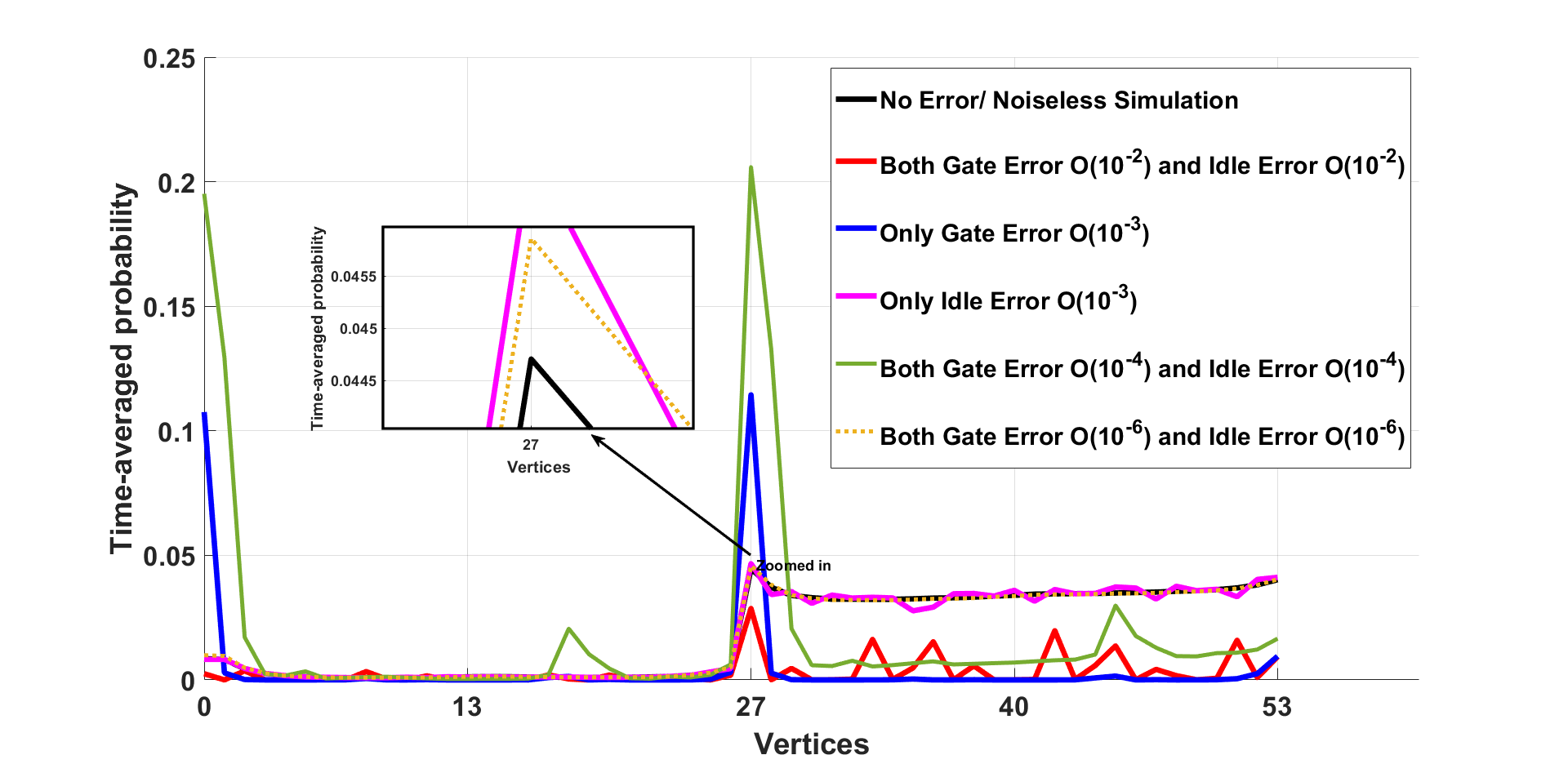}}
    \caption{Time averaged probability for $\mathrm{Cay(D_{27},\{a,b\})}$ taking coins from the classes $,\mathcal{X}_\theta,\mathcal{Y}_\theta,\mathcal{Z}_\theta,\mathcal{W}_\theta$ with initial position $(1,0)$ and initial coin state $\ket{0}_3$. The time step is taken up to $300$. The generic depolarizer gate noise and phase damping idle noise is incorporated in the circuit. The error parameters are chosen from uniform distribution.}\label{fig:Timeavph}
\end{figure}


\subsection{Numerical simulation of three-state lazy DTQW on $\mathrm{Cay(\Z_{27},\{1,-1\})}$}

We perform numerical simulations pertaining to three-state lazy DTQWs on cycle graphs with $27$ vertices. 
In Figure \ref{fig:Timeavampcy}  we plot the time-averaged probability of finding the quantum walker at vertices of $\mathrm{Cay(\mathbb{Z}_{27},\{1,-1\})}$ for the initial coin state $\frac{1}{\sqrt{3}}(\ket{0}_3+\ket{1}_3+\ket{2}_3)$ and starting vertex $0$. In the first figure the coin $C$ is taken from the class $\mathcal{X}_\theta$ where $\theta=\pi$ i.e. the coin is the Grover coin $\mathsf{G}$. In the subsequent figures, the coin $C$ is taken from the class $\mathcal{Y}_\theta$ where $\theta=\pi/2$, $C\in \mathcal{Z}_\theta,\theta=\pi/3$ and $C\in \mathcal{W}_\theta, \theta=-\pi/4$ respectively. We again introduce various gate and idle errors into our circuits similar to that used in Figure\ref{fig:Timeavamp} and Figure \ref{fig:Timeavph}. A similar phenomenon is observed where for high value of errors, the walker may not localize with high probability around the starting vertex.

\begin{figure}[H]
    \centering
     \subfigure[$C\in \mathcal{X}_\theta,\theta=\pi$]{\includegraphics[height=3.5 cm,width=8 cm]{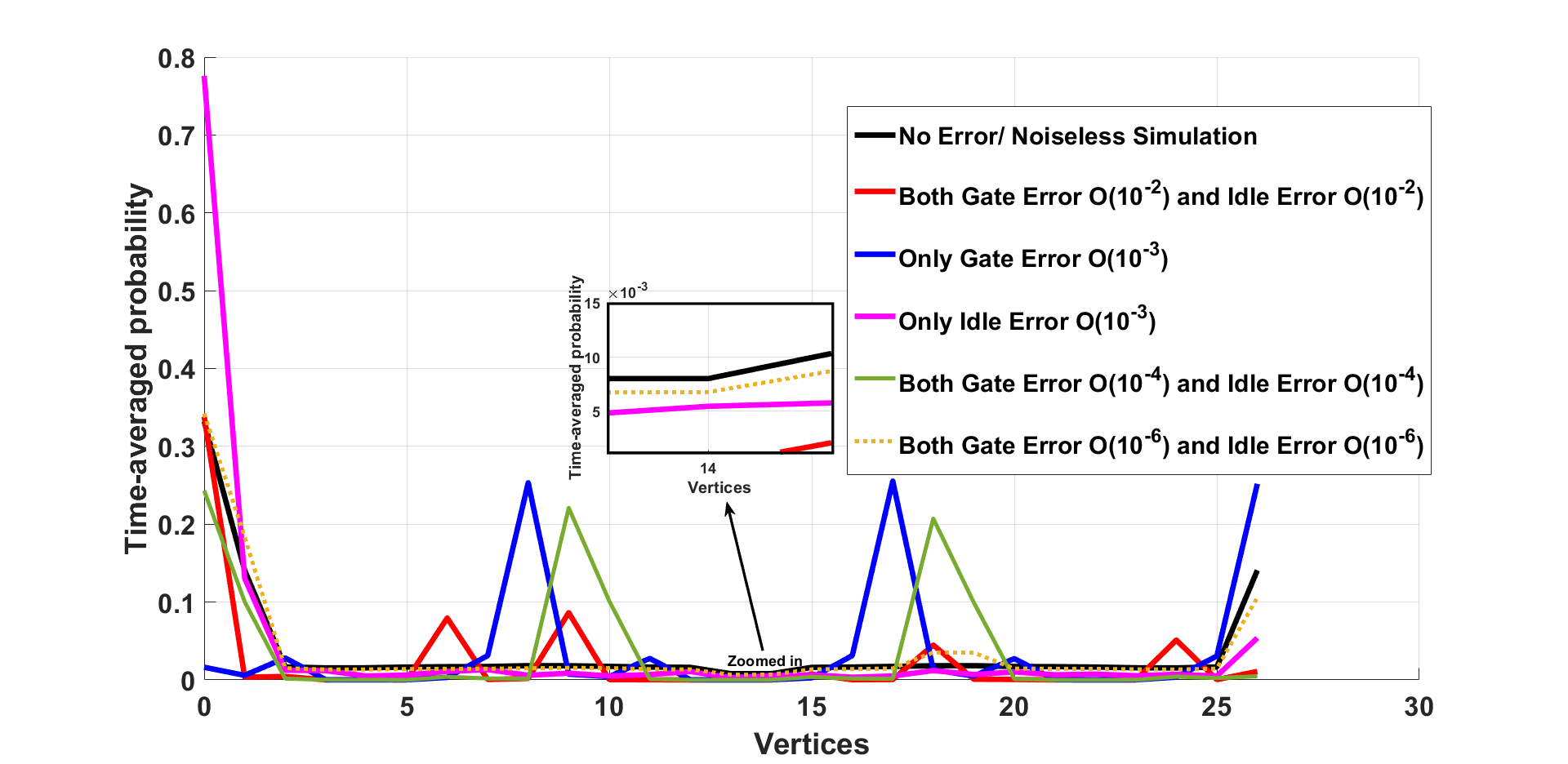}}
   \subfigure[$C\in \mathcal{Y}_\theta,\theta=\pi/2$]{\includegraphics[height=3.5 cm,width=8 cm]{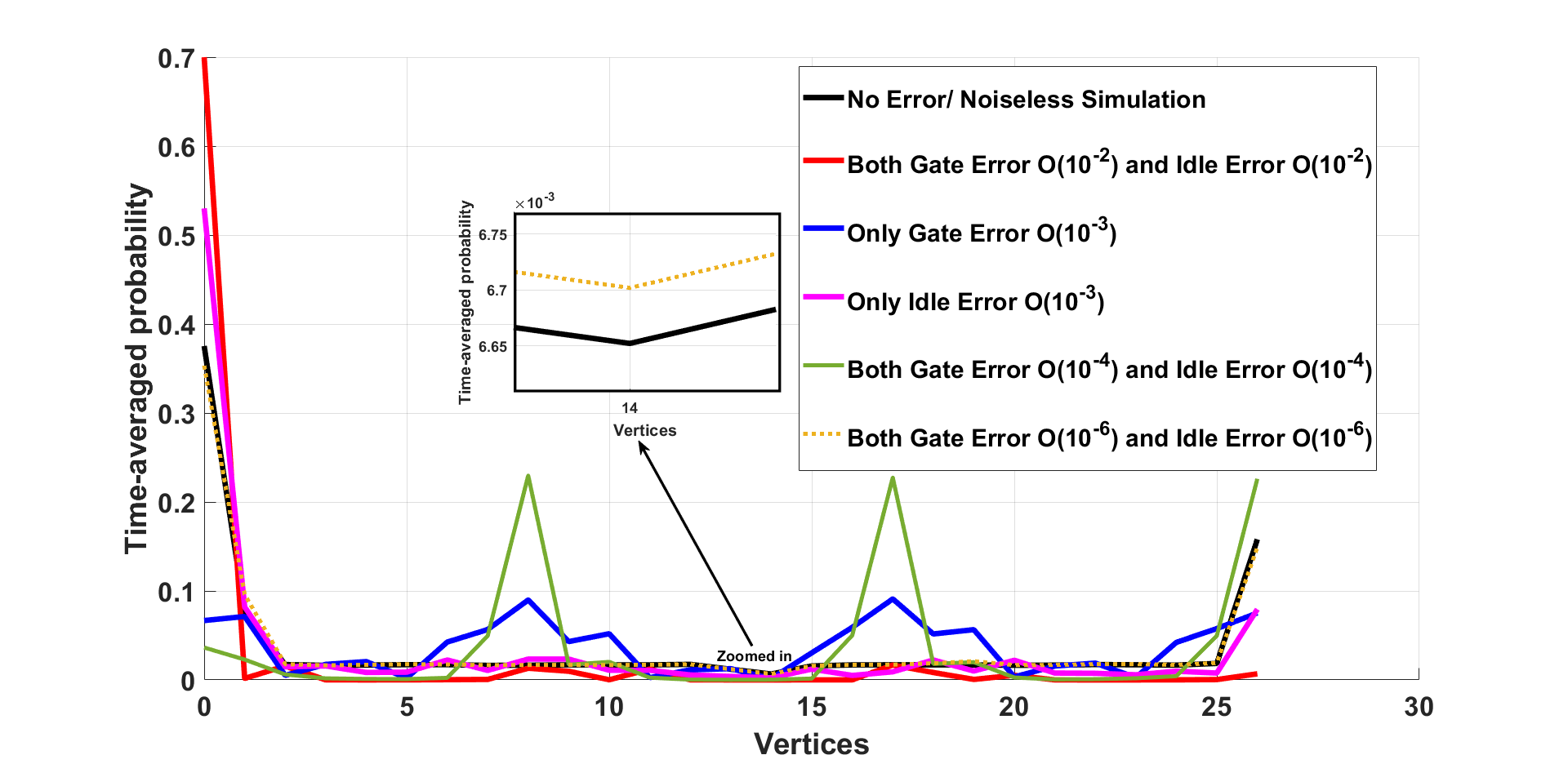}}
   \subfigure[$C\in \mathcal{Z}_\theta,\theta=\pi/3$]{\includegraphics[height=3.5 cm,width=8 cm]{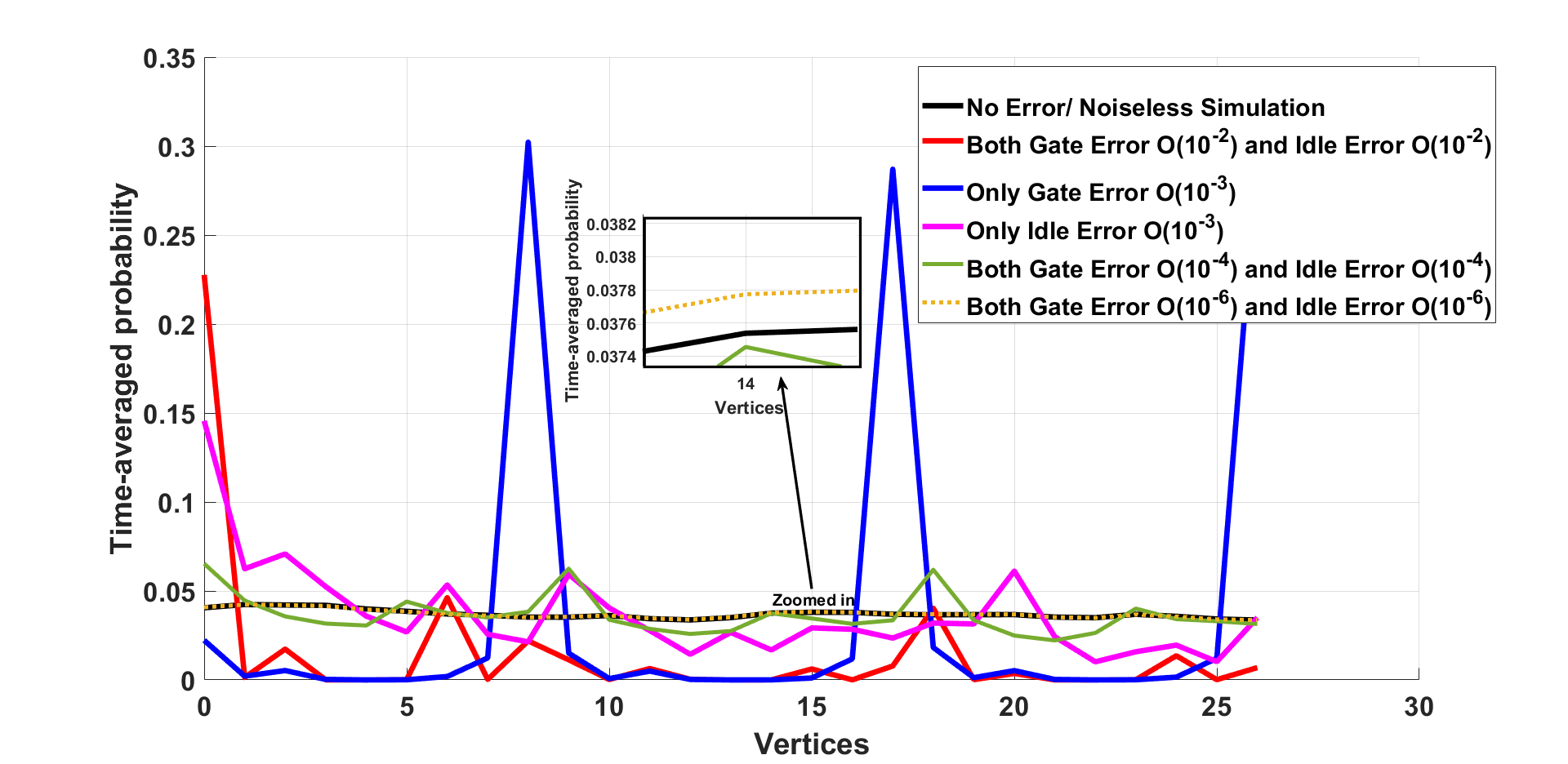}}
   \subfigure[$C\in \mathcal{W}_\theta,\theta=-\pi/4$]{\includegraphics[height=3.5 cm,width=8 cm]{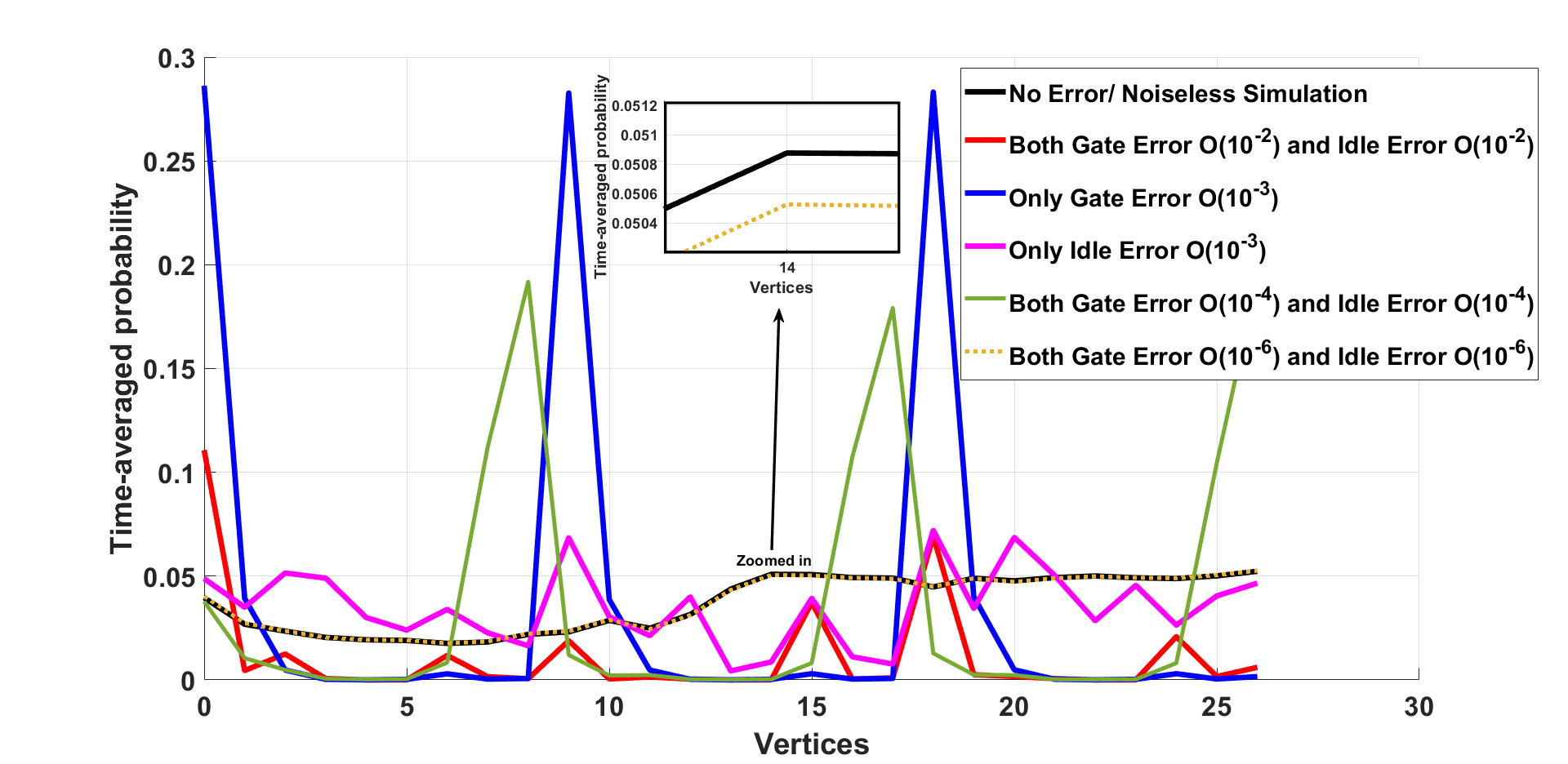}}
    \caption{Time averaged probability for $\mathrm{Cay(\mathbb{Z}_{27},\{1,-1\})}$ taking coins from the classes $,\mathcal{X}_\theta,\mathcal{Y}_\theta,\mathcal{Z}_\theta,\mathcal{W}_\theta$ with initial position $0$ and initial coin state $\frac{1}{\sqrt{3}}(\ket{0}_3+\ket{1}_3+\ket{2}_3)$. The time step is taken up to $300$. The generic depolarizer gate noise and amplitude damping idle noise is incorporated in the circuit. The error parameters are chosen from uniform distribution.}\label{fig:Timeavampcy}
\end{figure}

 In Figure \ref{fig:Timeavphcy},  we plot same the time-averaged probability of finding a particle at all vertices of $\mathrm{Cay(\mathbb{Z}_{27},\{1,-1\})}$ for the initial coin state $\frac{1}{\sqrt{3}}(\ket{0}_3+\ket{1}_3+\ket{2}_3)$ and starting vertex $0$ and coin similar to \ref{fig:Timeavampcy}. The idle error is taken to be phase damping instead of amplitude damping.

\begin{figure}[H]
    \centering
     \subfigure[$C\in \mathcal{X}_\theta,\theta=\pi$]{\includegraphics[height=3.5 cm,width=8 cm]{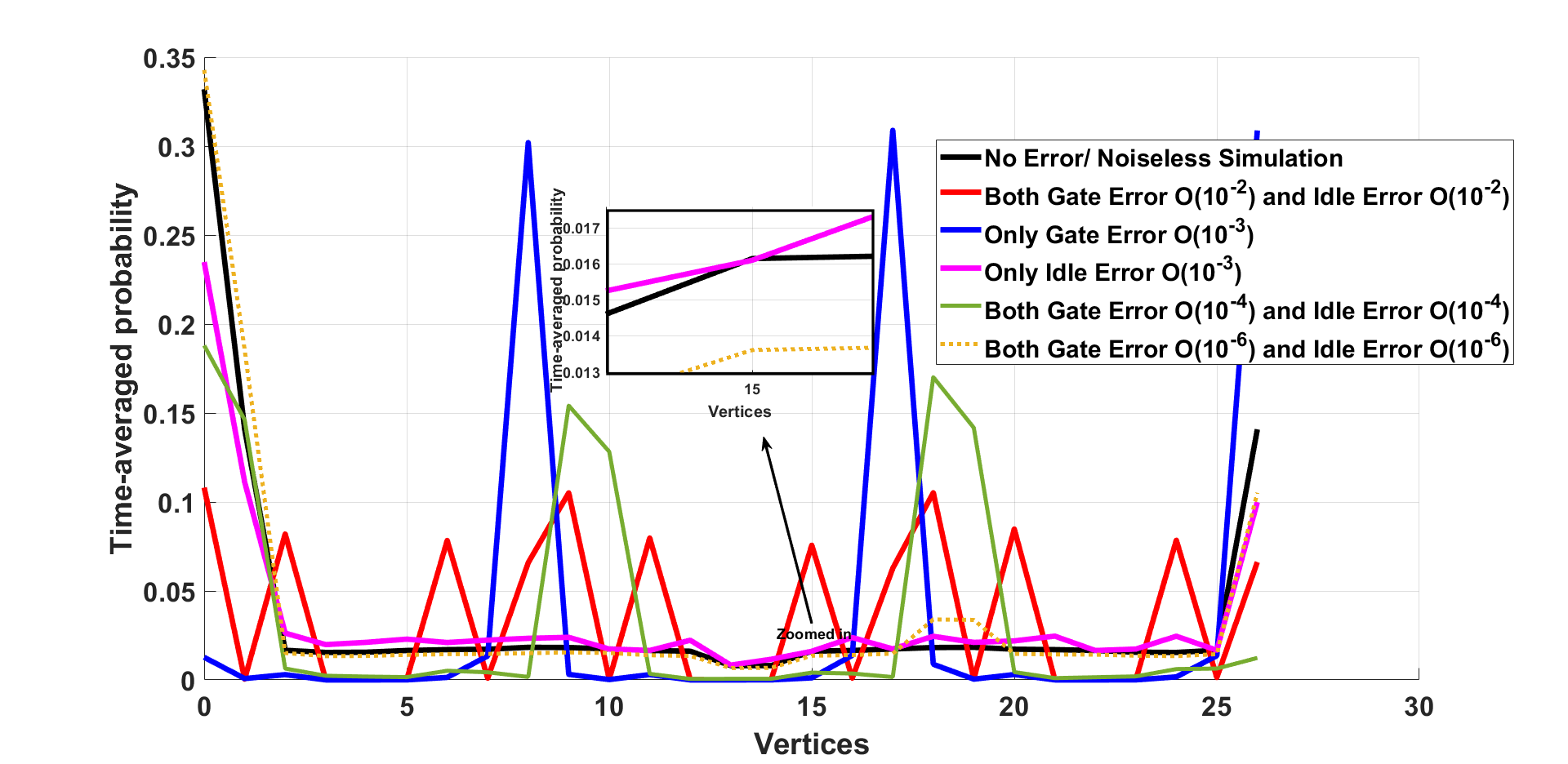}}
   \subfigure[$C\in \mathcal{Y}_\theta,\theta=\pi/2$]{\includegraphics[height=3.5 cm,width=8 cm]{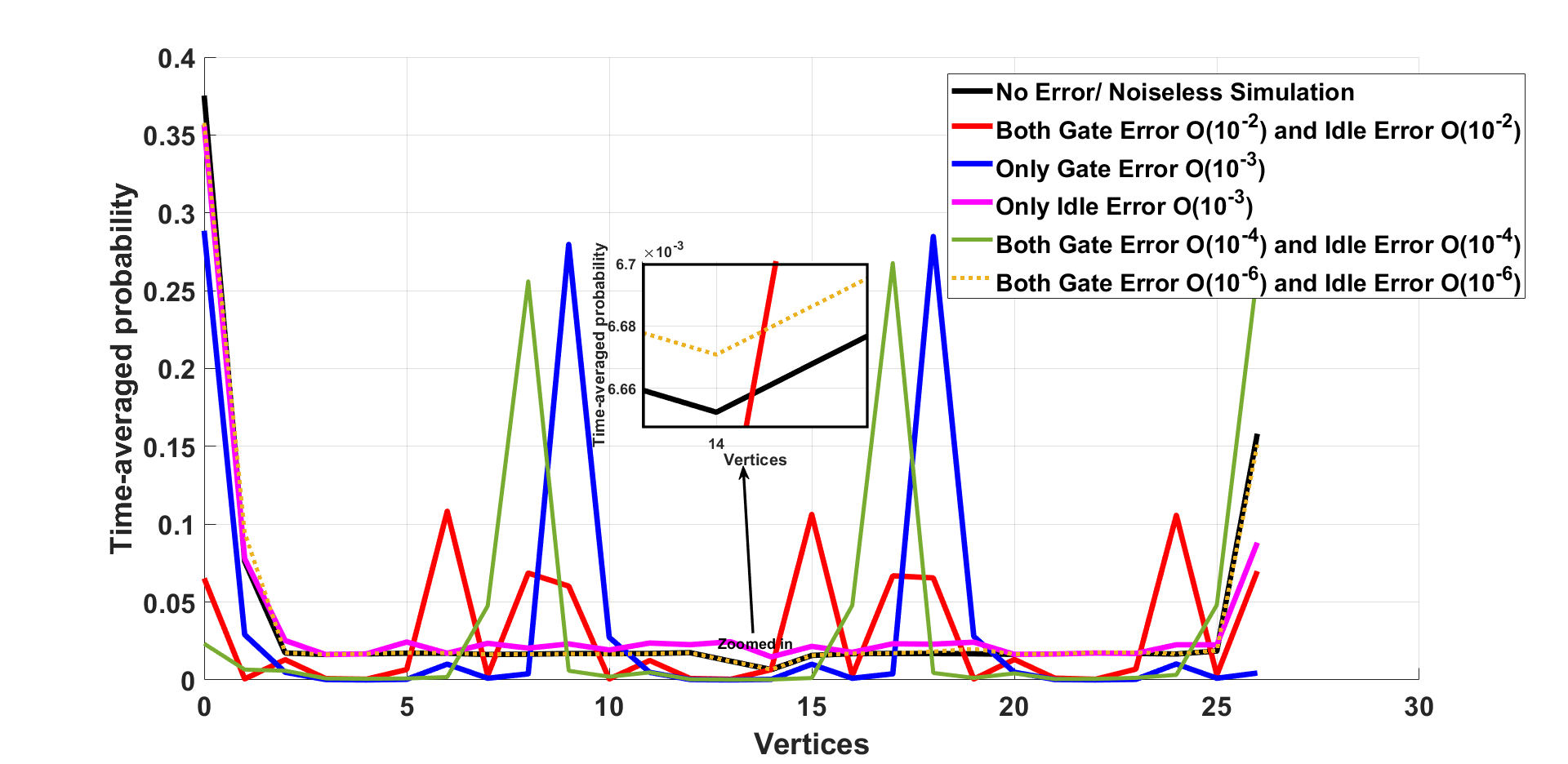}}
   \subfigure[$C\in \mathcal{Z}_\theta,\theta=\pi/3$]{\includegraphics[height=3.5 cm,width=8 cm]{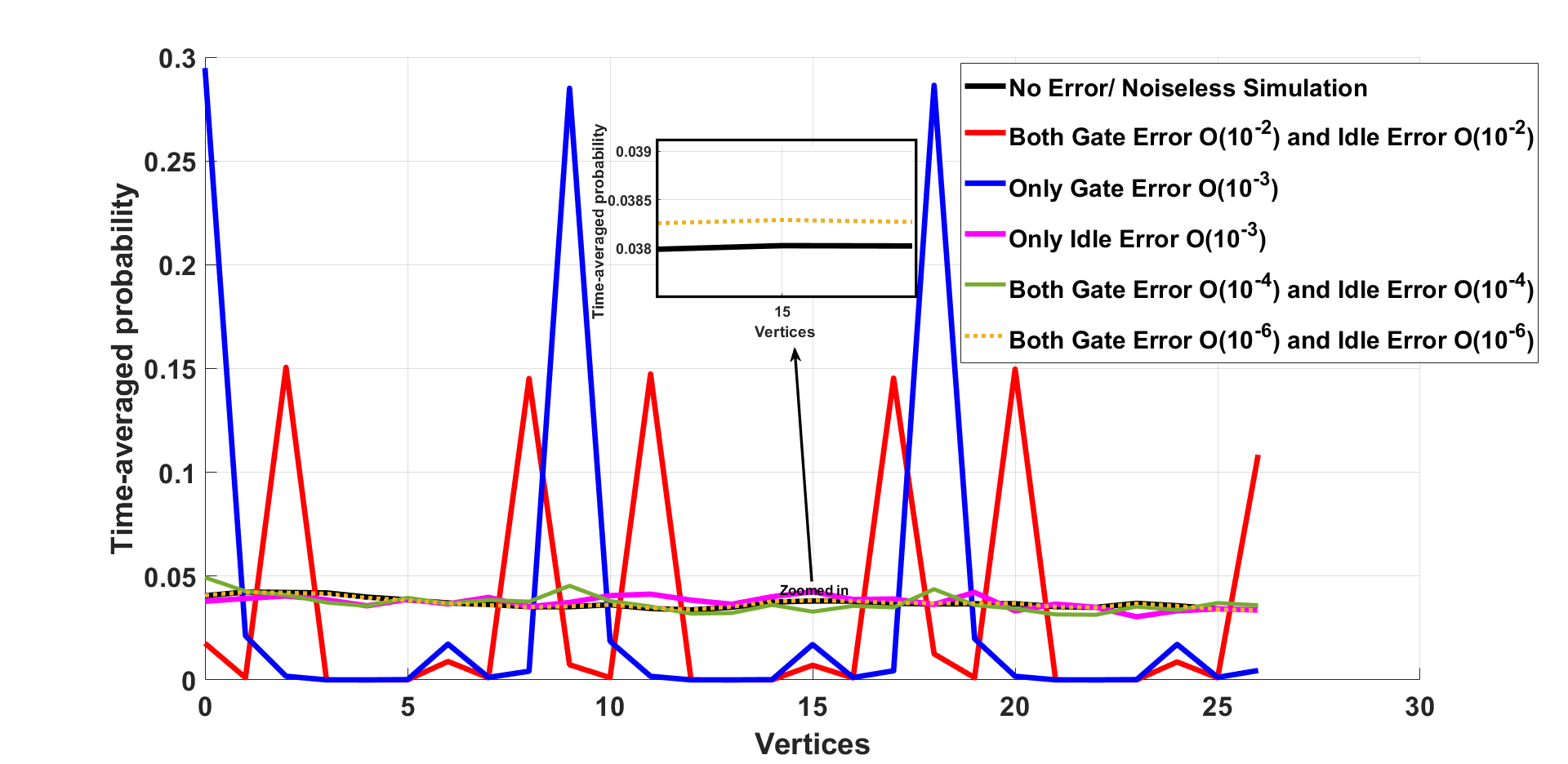}}
   \subfigure[$C\in \mathcal{W}_\theta,\theta=-\pi/4$]{\includegraphics[height=3.5 cm,width=8 cm]{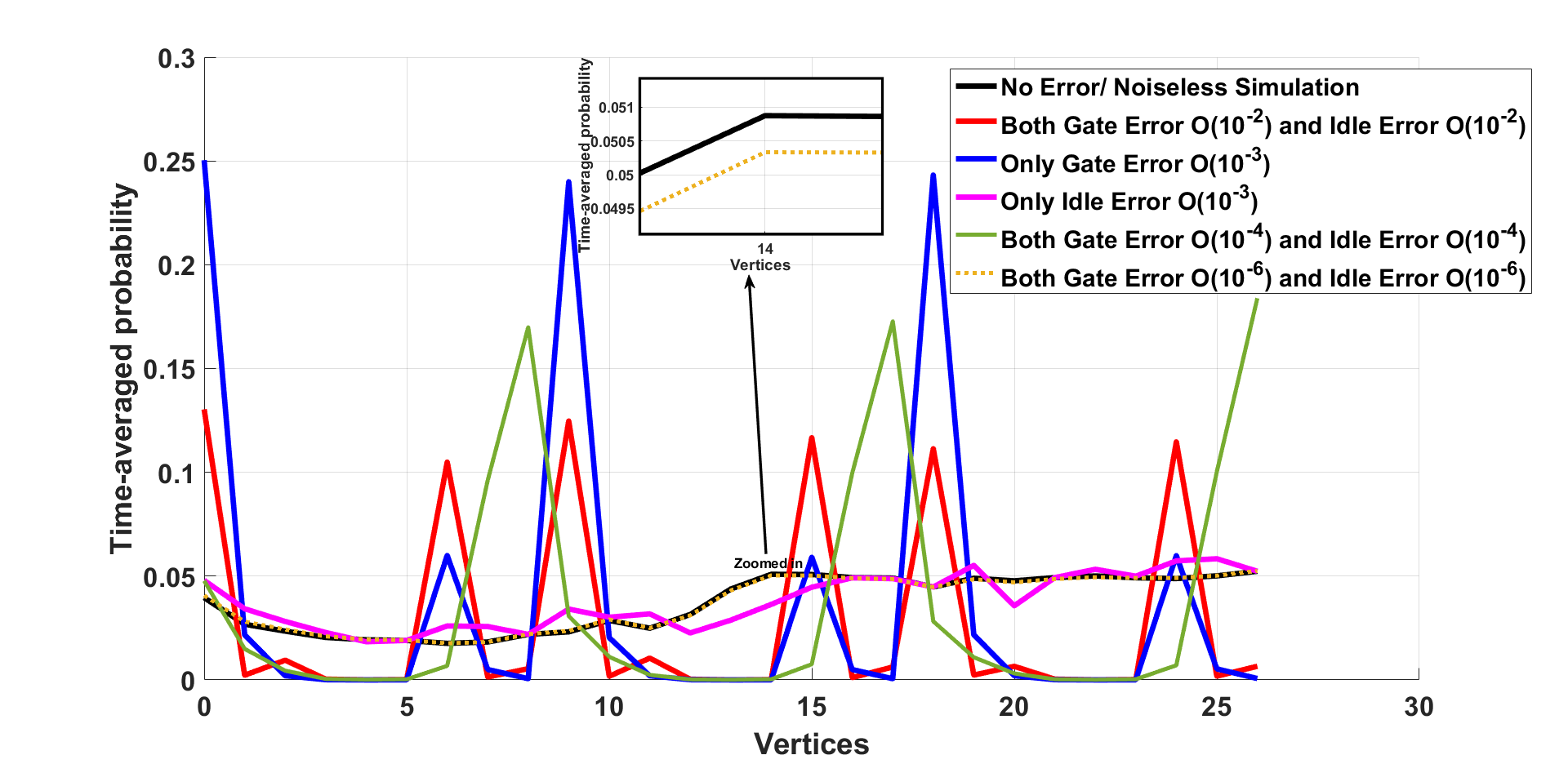}}
    \caption{Time averaged probability for $\mathrm{Cay(\mathbb{Z}_{27},\{1,-1\})}$ taking coins from the classes $,\mathcal{X}_\theta,\mathcal{Y}_\theta,\mathcal{Z}_\theta,\mathcal{W}_\theta$ with initial position $0$ and initial coin state $\frac{1}{\sqrt{3}}(\ket{0}_3+\ket{1}_3+\ket{2}_3)$. The time step is taken up to $300$. The generic depolarizer gate noise and phase damping idle noise is incorporated in the circuit. The error parameters are chosen from the uniform distribution.}\label{fig:Timeavphcy}
\end{figure}

\begin{remark}
We also have performed numerical simulation based on the noise models for the cycle graph and Cayley graph corresponding to Dihedral group with different number of nodes and different initial coin states. Those results valiadate the theoretical results obtained in \cite{SarmaSarkar2023}.     
\end{remark}


Now in order to gain further insights from the above simulation results for the time-averaged probability distributions, we compare it for the noise models and noiseless simulation results. We consider the \textit{Kullback–Leibler divergence} (KL divergence) and the \textit{total variation distance} (TVD) for finding the distance between the probability distributions (in the unit bits). First we recall these measures as follows.


Let $P, Q$ be two probability distributions for a discrete random variable $X.$ Let $\mathcal{X}$ denote the range set of $X.$ Then:

\begin{itemize}
    \item KL-divergence \cite{KullbackLeiber1951, Mackay2004}: $$
    D_{KL}(P\|Q) = \sum_{x\in \mathcal{X}}P(x)\log_2\left(\frac{P(x)}{Q(x)}\right)
    = -\sum_{x\in \mathcal{X}}P(x)\log_2\left(\frac{Q(x)}{P(x)}\right).$$
    
    \item Total variation distance \cite{Levin2017}: $$TVD_{(P,Q)}=\frac{1}{2} \sum_{x\in \mathcal{X}} |P(x)-Q(x)|$$
\end{itemize}




We shall denote the time-averaged probability distribution of the walker obtained via noiseless circuit as $P_{\mbox{Ideal}},$ and  $P_{\mbox{Noisy}}$ for the time-averaged probability distribution obtained in a noisy quantum circuit.  

In Figure \ref{fig:Timeavampcomp}, we plot the KL divergence $D_{KL}(P_{\mbox{Ideal}}\|P_{\mbox{Noisy}})$ and the total variation distance $TVD_{(P_{\mbox{Ideal}},P_{\mbox{Noisy}})}$ between the noiseless and noisy time-average probability distributions for finding the walker over the vertices of $\mathrm{Cay(D_{27},\{a,b\})}$ for the initial coin state $\ket{0}_3$ and starting vertex $(1,0)$ as shown in Figure \ref{fig:Timeavamp} and Figure \ref{fig:Timeavph}. In the first figure the coin $C$ is taken from the class $\mathcal{X}_\theta$ where $\theta=\pi$ i.e. the coin is the Grover coin $\mathsf{G}$. In the subsequent figures, the coin $C$ is taken from the class $\mathcal{Y}_\theta$ where $\theta=\pi/2$, $C\in \mathcal{Z}_\theta,\theta=\pi/3$ and $C\in \mathcal{W}_\theta, \theta=-\pi/4$ respectively. Further, amplitude damping idle error and the generic depolarizer gate noise are chosen as noise models. It is observed from the figure that the gate error plays a significant role in deviating from noiseless distribution. It is also interesting to notice that the idle error sometimes mitigates the effects of gate errors as we see that for several figures the second bars/columns are longer than the first despite the second columns containing no idle errors. The effect of gate noise is further emphasized from the third and fourth bars/columns in the figure where we see that a small gate error coupled with a small idle error provides a 'more noisy' distribution as compared to a distribution obtained via a larger idle error with no gate error. Further, comparing second and third columns in the Figure \ref{fig:Timeavampcomp} is another justification that gate errors are more significant than idle errors.  

\begin{figure}[H]
    \centering
     \subfigure[$C\in \mathcal{X}_\theta,\theta=\pi$]{\includegraphics[height=3.5 cm,width=8 cm]{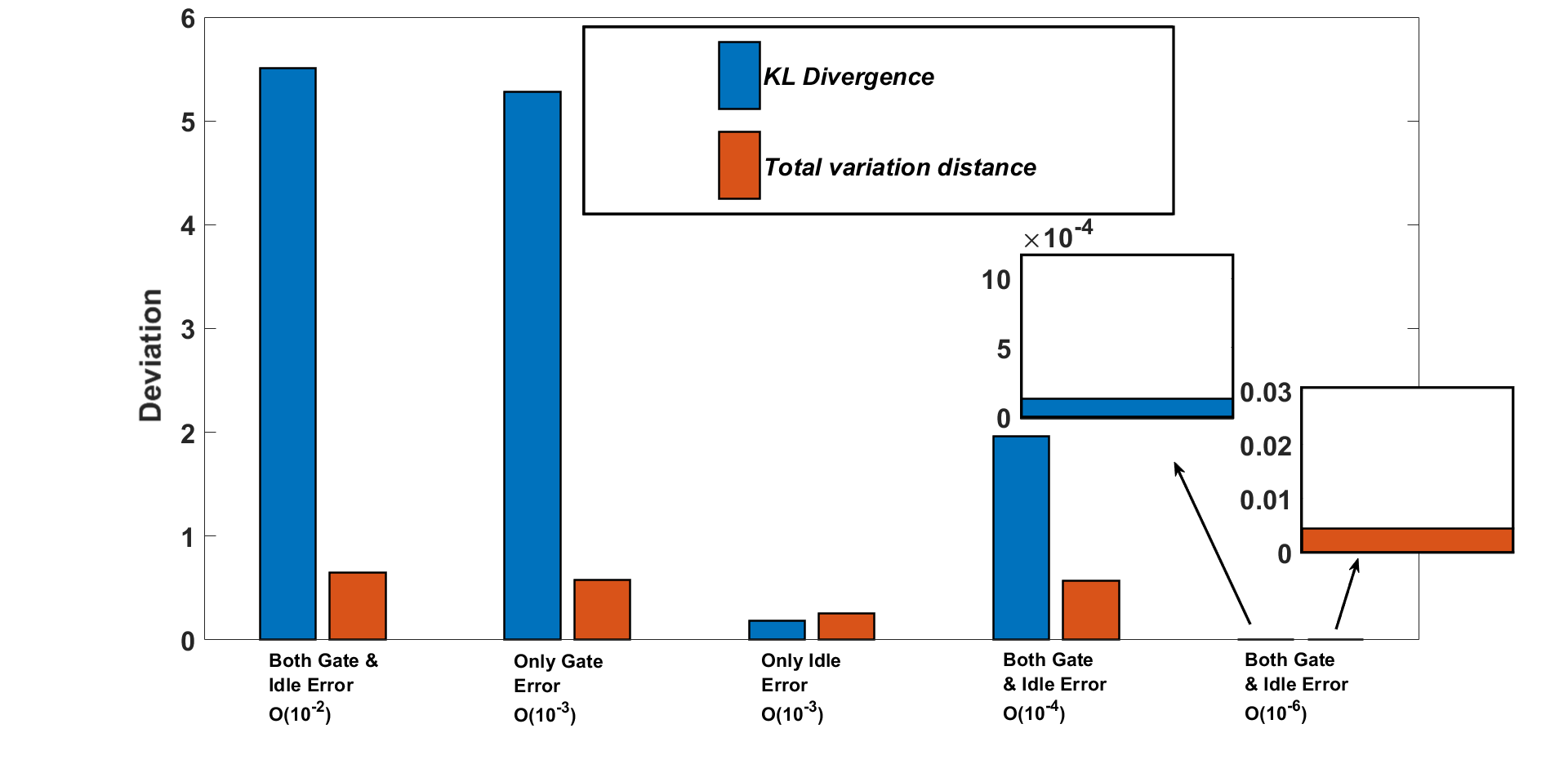}}
   \subfigure[$C\in \mathcal{Y}_\theta,\theta=\pi/2$]{\includegraphics[height=3.5 cm,width=8 cm]{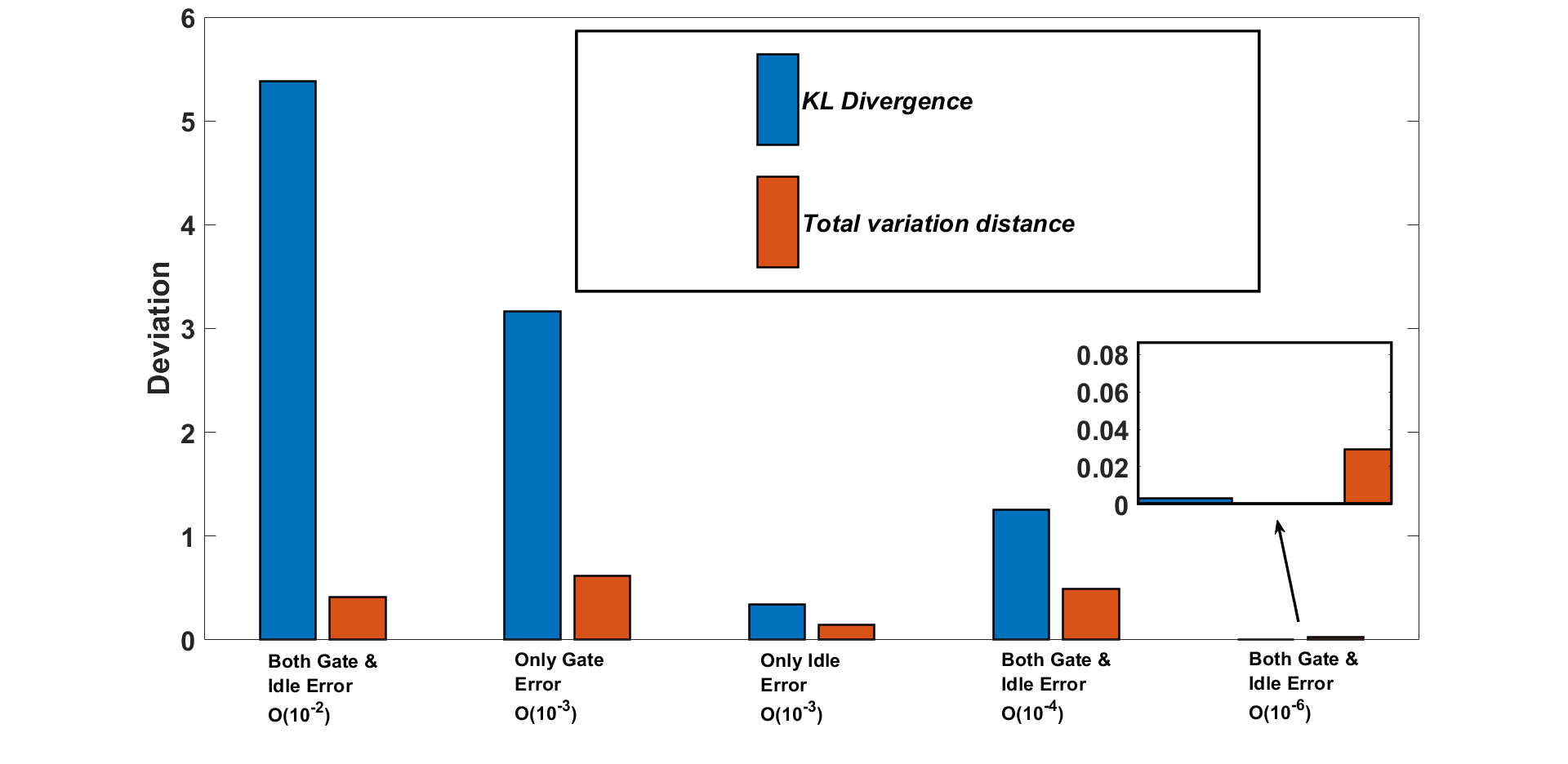}}
   \subfigure[$C\in \mathcal{Z}_\theta,\theta=\pi/3$]{\includegraphics[height=3.5 cm,width=8 cm]{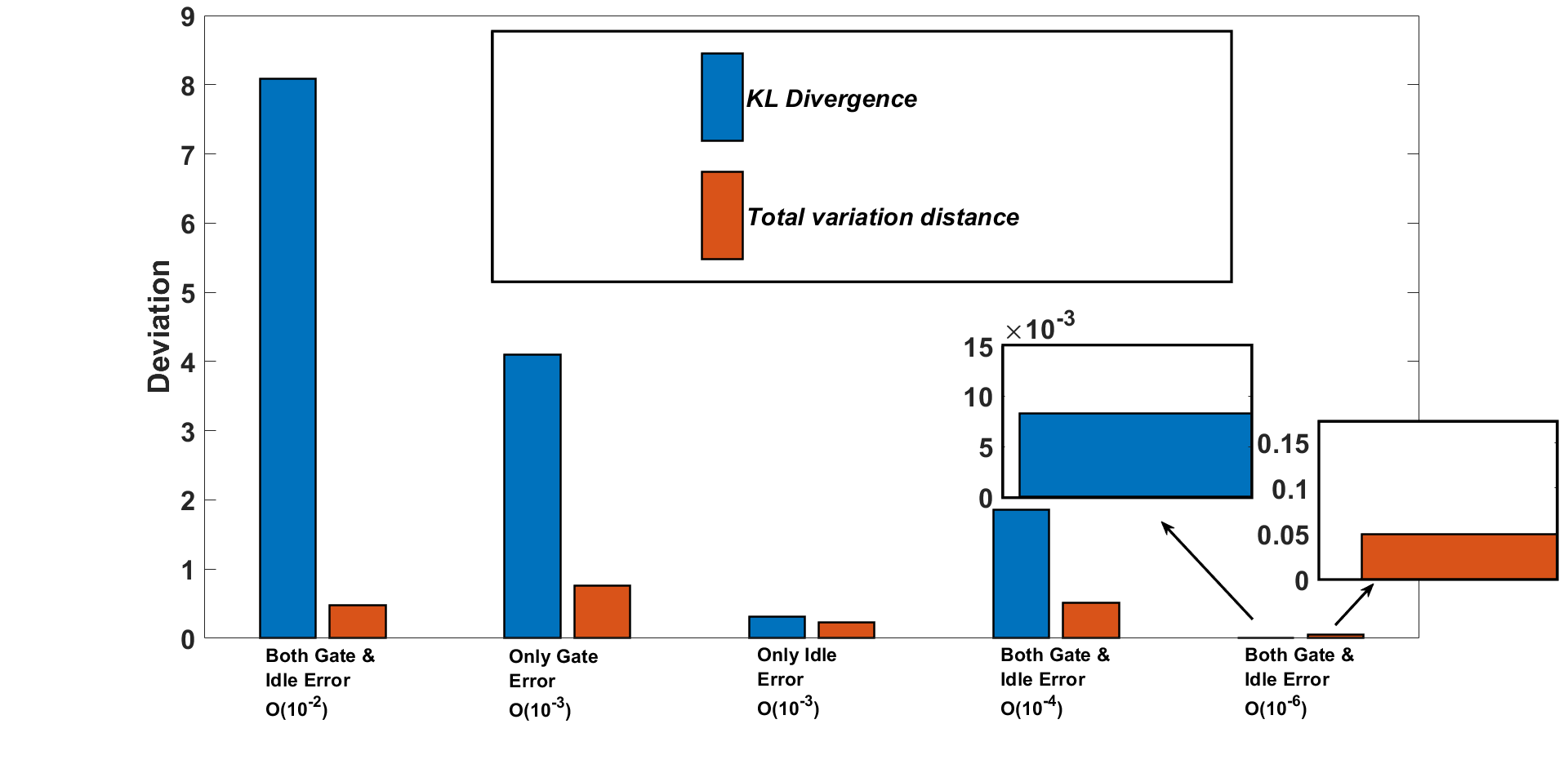}}
   \subfigure[$C\in \mathcal{W}_\theta,\theta=-\pi/4$]{\includegraphics[height=3.5 cm,width=8 cm]{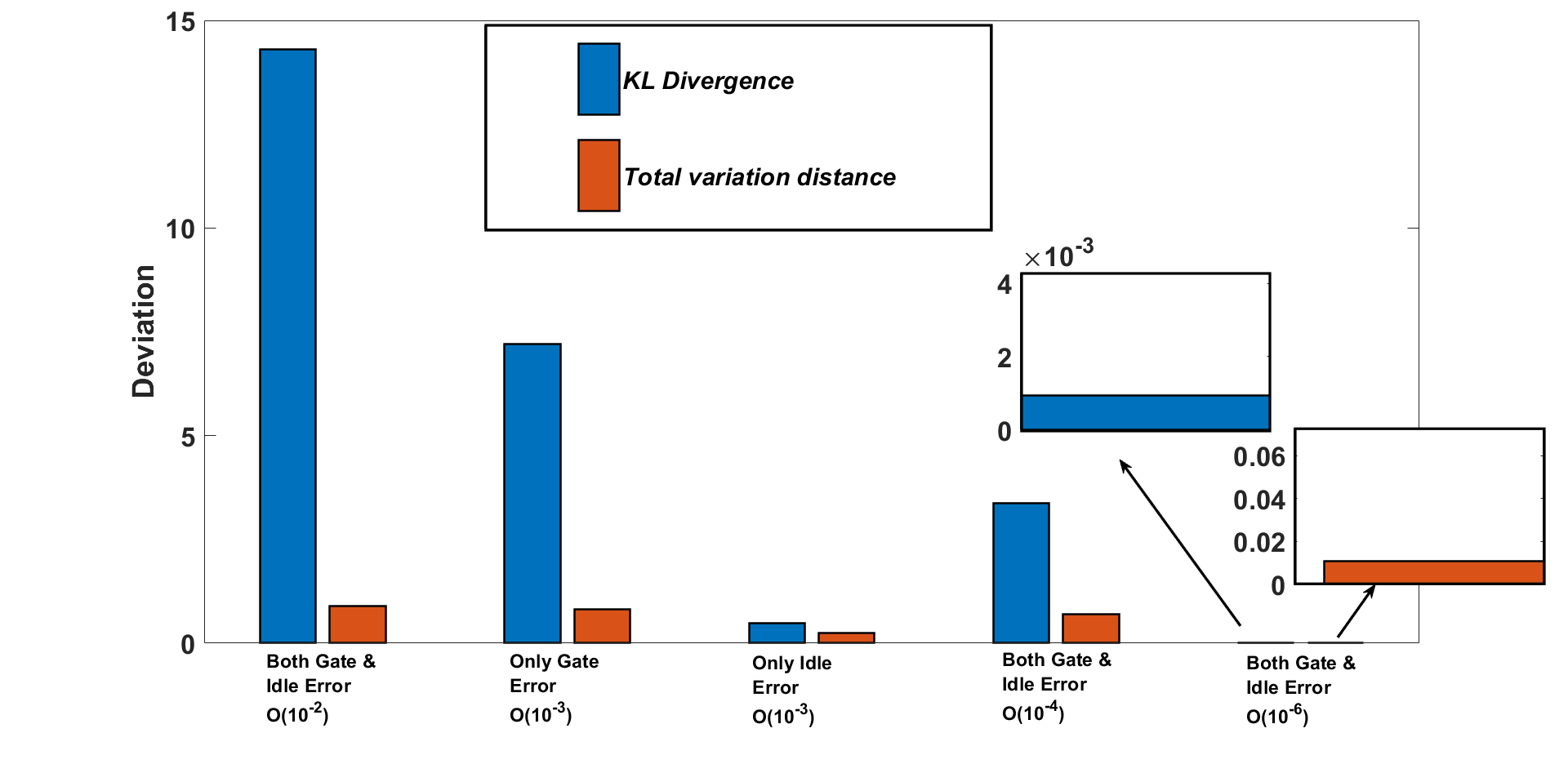}}
    \caption{Comparison $D_{KL}(P_{\mbox{Ideal}}\|P_{\mbox{Noisy}})$ and $TVD_{(P_{\mbox{Ideal}},P_{\mbox{Noisy}})}$ between the time-averaged probability distributions obtained through noiseless and noisy circuits for DTQWs in $\mathrm{Cay(D_{27},\{a,b\})}$ taking coins from the classes $,\mathcal{X}_\theta,\mathcal{Y}_\theta,\mathcal{Z}_\theta,\mathcal{W}_\theta$ with initial position $(1,0)$ and initial coin state $\ket{0}_3$. The time step is taken up to $300$. The generic depolarizer gate noise and amplitude damping idle noise is incorporated in the circuit. The error parameters are chosen from uniform distribution.}\label{fig:Timeavampcomp}
\end{figure}

Now, in Figure \ref{fig:Timeavphcomp} we plot same comparison between noisy and noiseless distributions via KL divergence and TVD. However, the idle error comprises of phase damping instead of amplitude damping. Observations are similar to Figure \ref{fig:Timeavampcomp} as discussed above.

\begin{figure}[H]
    \centering
     \subfigure[$C\in \mathcal{X}_\theta,\theta=\pi$]{\includegraphics[height=3.5 cm,width=8 cm]{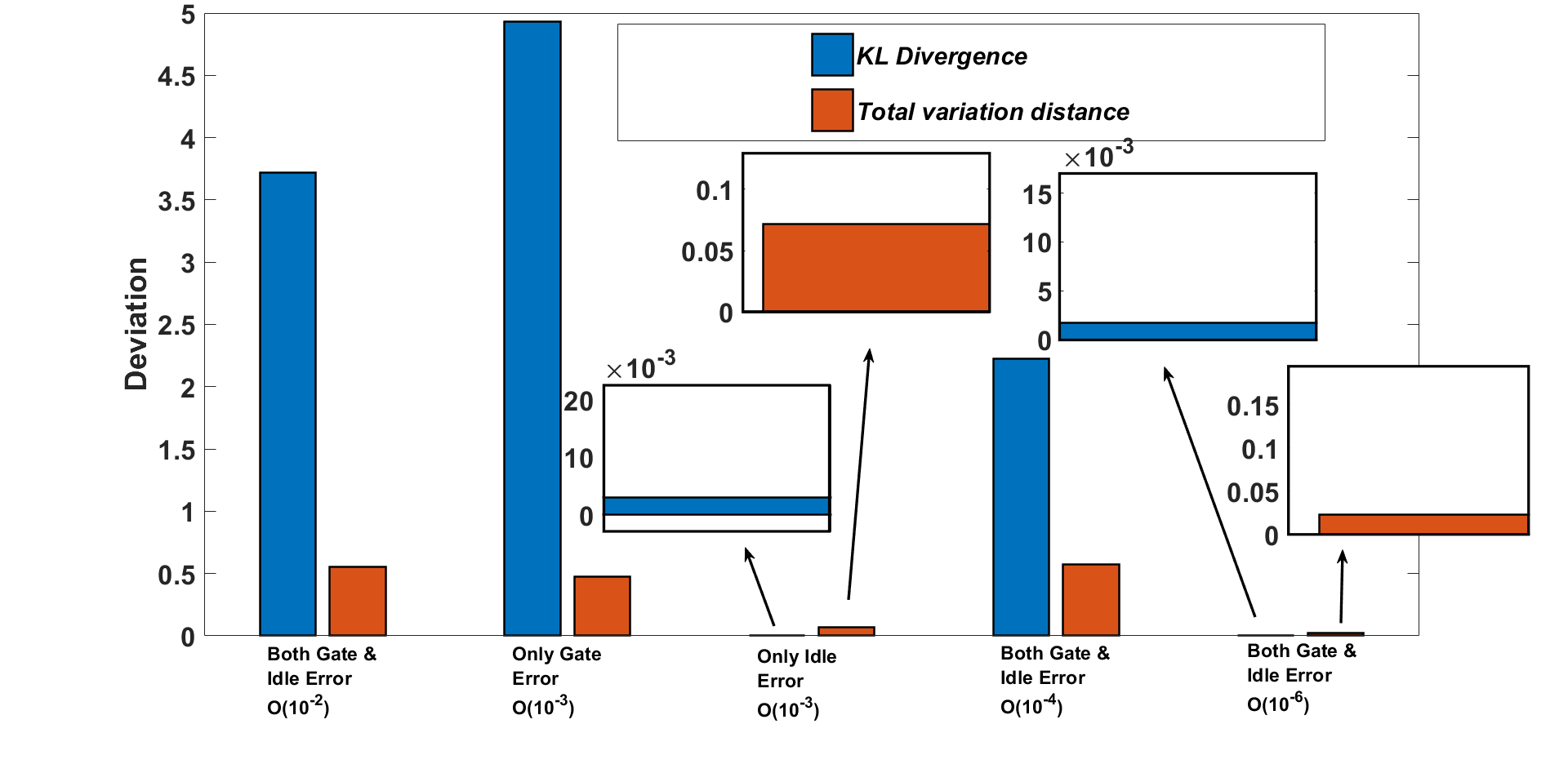}}
   \subfigure[$C\in \mathcal{Y}_\theta,\theta=\pi/2$]{\includegraphics[height=3.5 cm,width=8 cm]{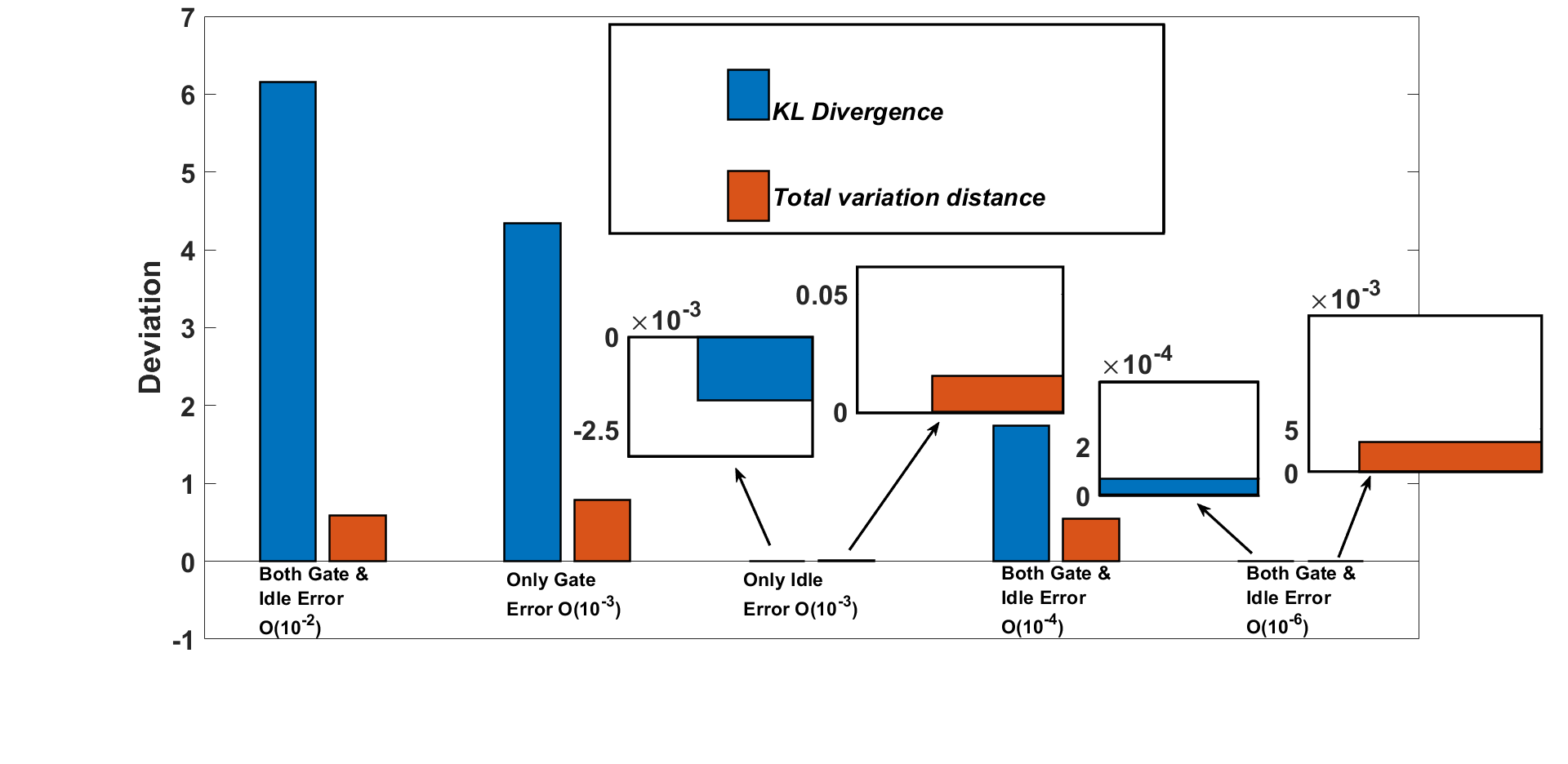}}
   \subfigure[$C\in \mathcal{Z}_\theta,\theta=\pi/3$]{\includegraphics[height=3.5 cm,width=8 cm]{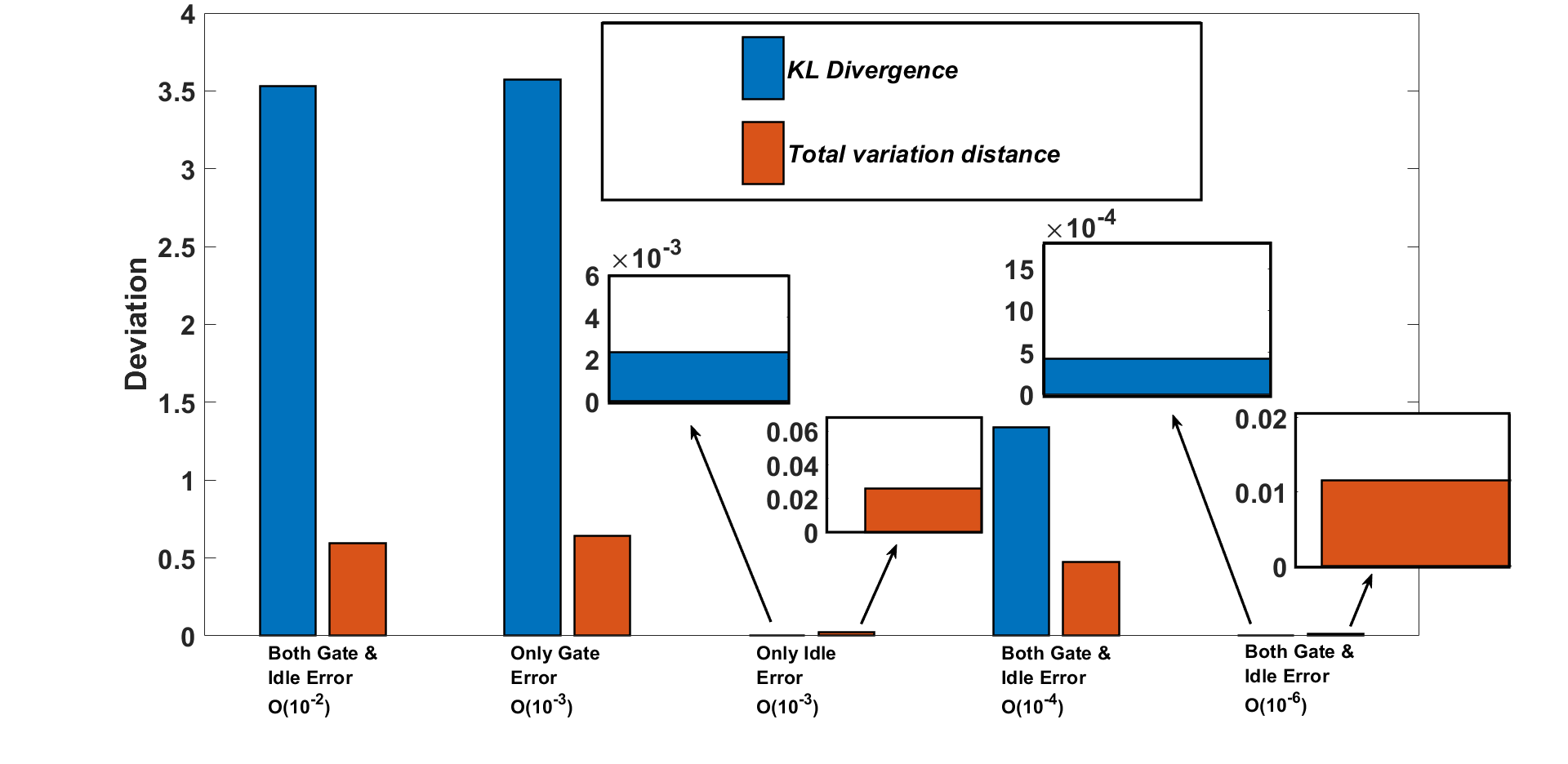}}
   \subfigure[$C\in \mathcal{W}_\theta,\theta=-\pi/4$]{\includegraphics[height=3.5 cm,width=8 cm]{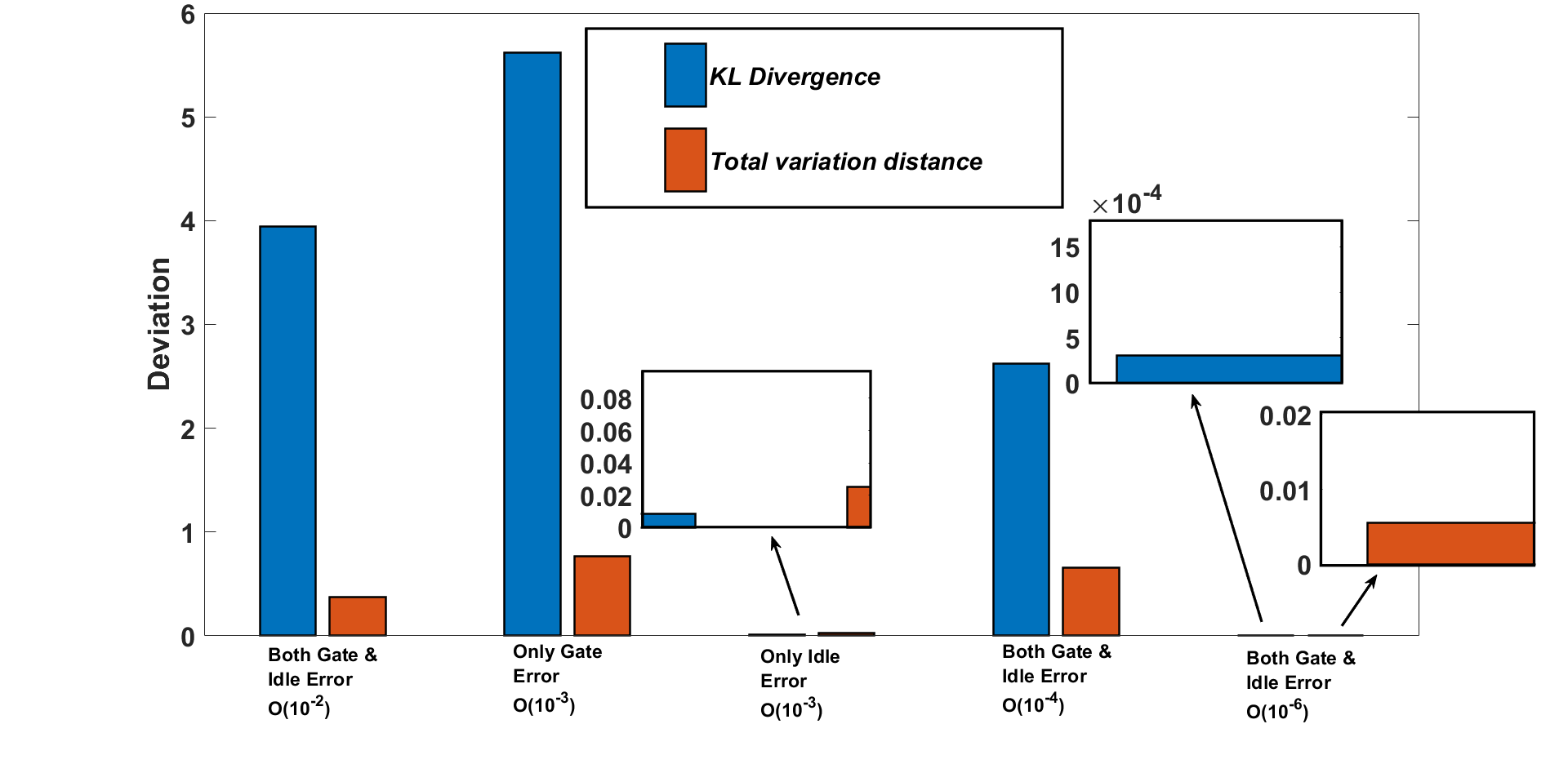}}
    \caption{Comparison $D_{KL}(P_{\mbox{Ideal}}\|P_{\mbox{Noisy}})$ and $TVD_{(P_{\mbox{Ideal}},P_{\mbox{Noisy}})}$ between the time-averaged probability distributions obtained through noiseless and noisy circuits for DTQWs in $\mathrm{Cay(D_{27},\{a,b\})}$ taking coins from the classes $,\mathcal{X}_\theta,\mathcal{Y}_\theta,\mathcal{Z}_\theta,\mathcal{W}_\theta$ with initial position $(1,0)$ and initial coin state $\ket{0}_3$. The time step is taken up to $300$. The generic depolarizer gate noise and phase damping idle noise is incorporated in the circuit and the error parameters are chosen from uniform distribution.}\label{fig:Timeavphcomp}
\end{figure}


Similar results comparing noisy and noiseless time-averaged probability distributions are obtained for DTQWs on $\mathrm{Cay}(\Z_{27},\{1,-1\})$. In Figures \ref{fig:Timeavampcompcy} and \ref{fig:Timeavphcompcy} we plot the KL divergence $D_{KL}(P_{\mbox{Ideal}}\|P_{\mbox{Noisy}})$ and the total variation distance $TVD_{(P_{\mbox{Ideal}},P_{\mbox{Noisy}})}$ between the noiseless and noisy time-average probability distributions for finding the walker over the vertices of $\mathrm{Cay(D_{27},\{a,b\})}$ for the initial coin state $\frac{1}{\sqrt{3}}(\ket{0}_3+\ket{1}_3+\ket{2}_3)$ and starting vertex $(0)$. Several coins from classes $\mathcal{X}_\theta,\mathcal{Y}_\theta,\mathcal{W}_\theta,\mathcal{Z}_\theta$ are chosen similar to Figure \ref{fig:Timeavampcomp} and Figure \ref{fig:Timeavphcomp}. The generic depolarizer gate noise is considered along with amplitude damping (Figure \ref{fig:Timeavampcompcy}) and  phase damping (Figure \ref{fig:Timeavphcompcy}) . In fact, for several instances in Figures \ref{fig:Timeavampcompcy} and \ref{fig:Timeavphcompcy}, it is observed more clearly how idle errors mitigate the effects of gate errors. 

\begin{figure}[H]
    \centering
     \subfigure[$C\in \mathcal{X}_\theta,\theta=\pi$]{\includegraphics[height=3.5 cm,width=8 cm]{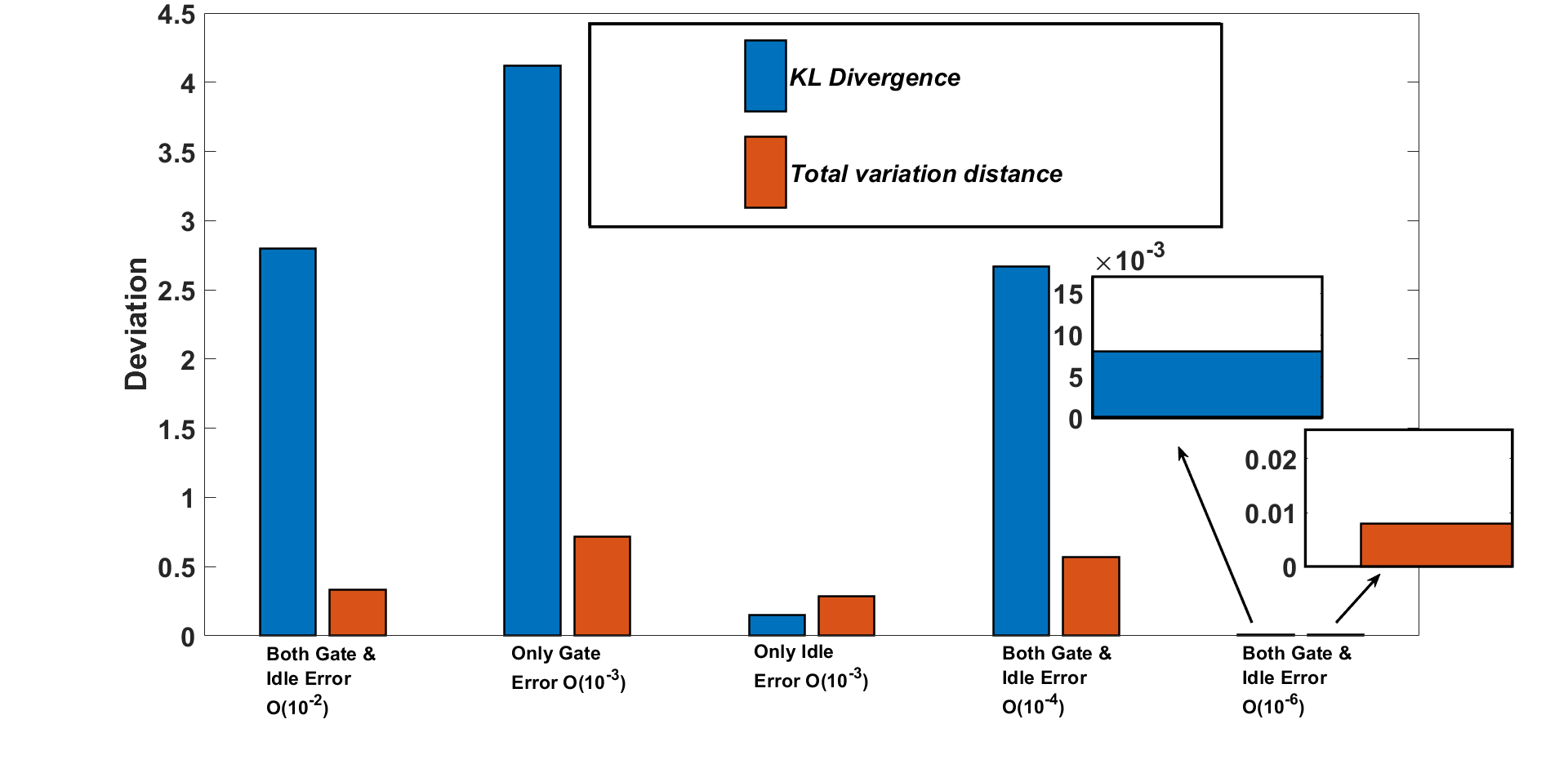}}
   \subfigure[$C\in \mathcal{Y}_\theta,\theta=\pi/2$]{\includegraphics[height=3.5 cm,width=8 cm]{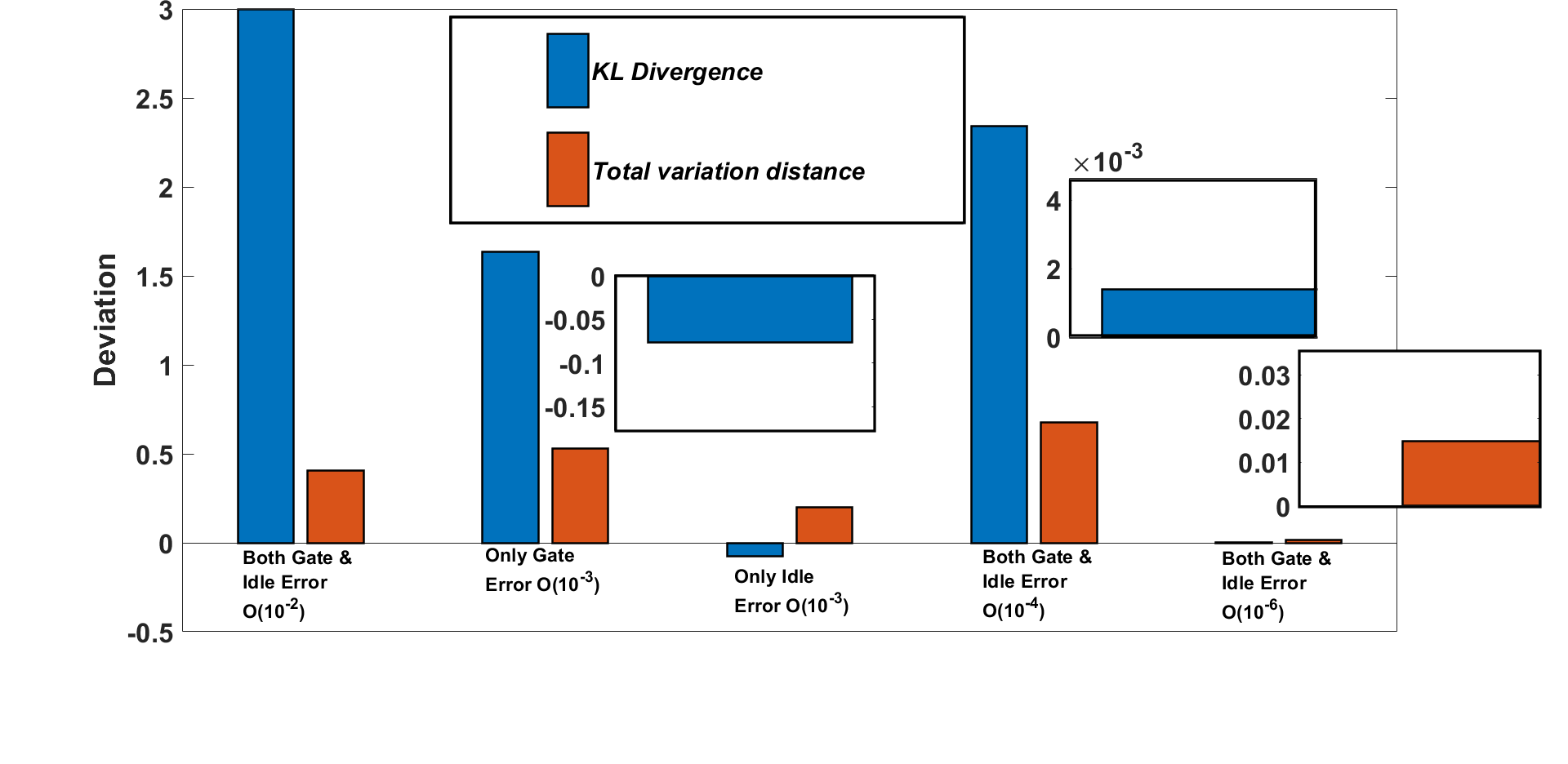}}
   \subfigure[$C\in \mathcal{Z}_\theta,\theta=\pi/3$]{\includegraphics[height=3.5 cm,width=8 cm]{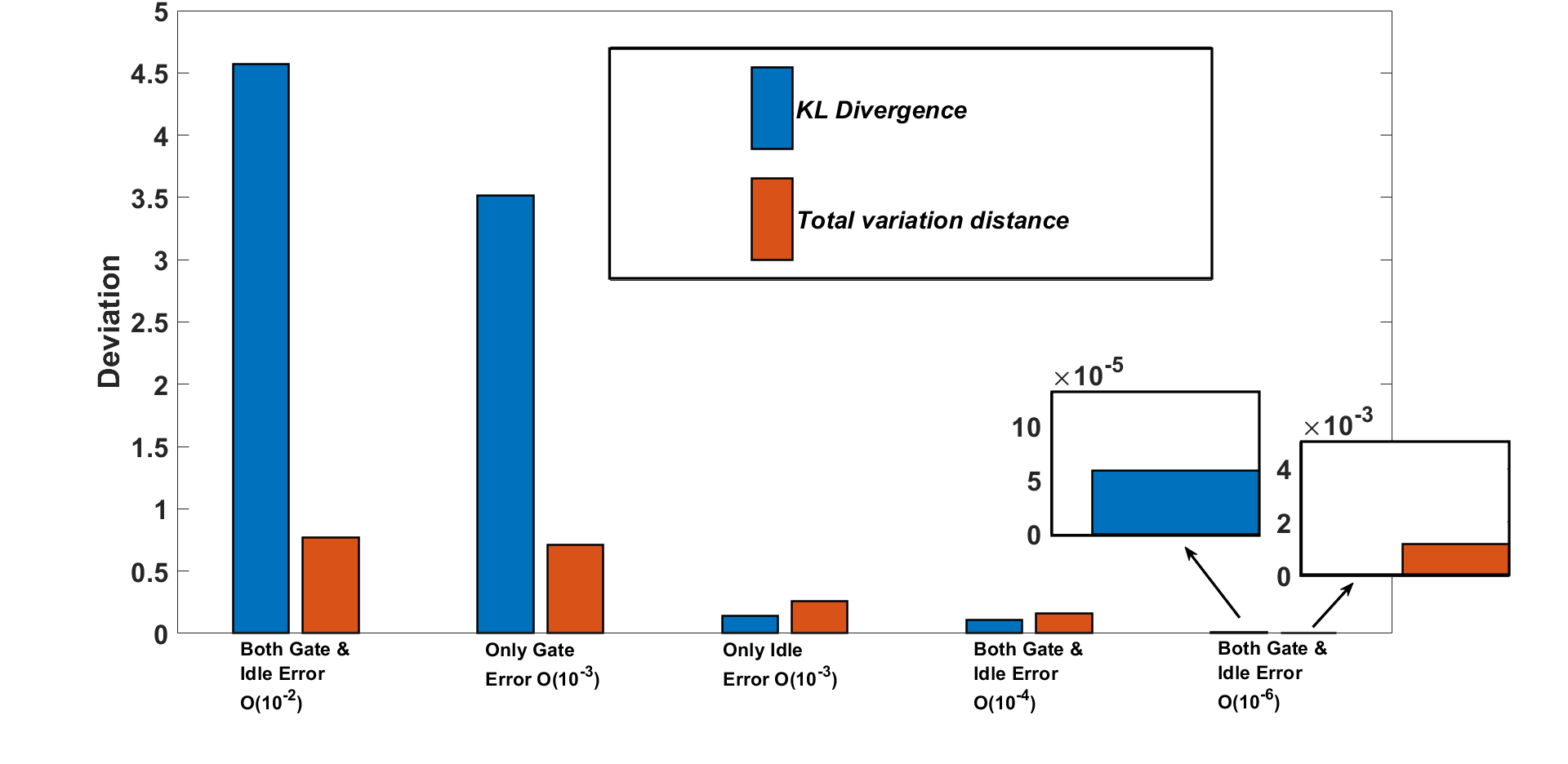}}
   \subfigure[$C\in \mathcal{W}_\theta,\theta=-\pi/4$]{\includegraphics[height=3.5 cm,width=8 cm]{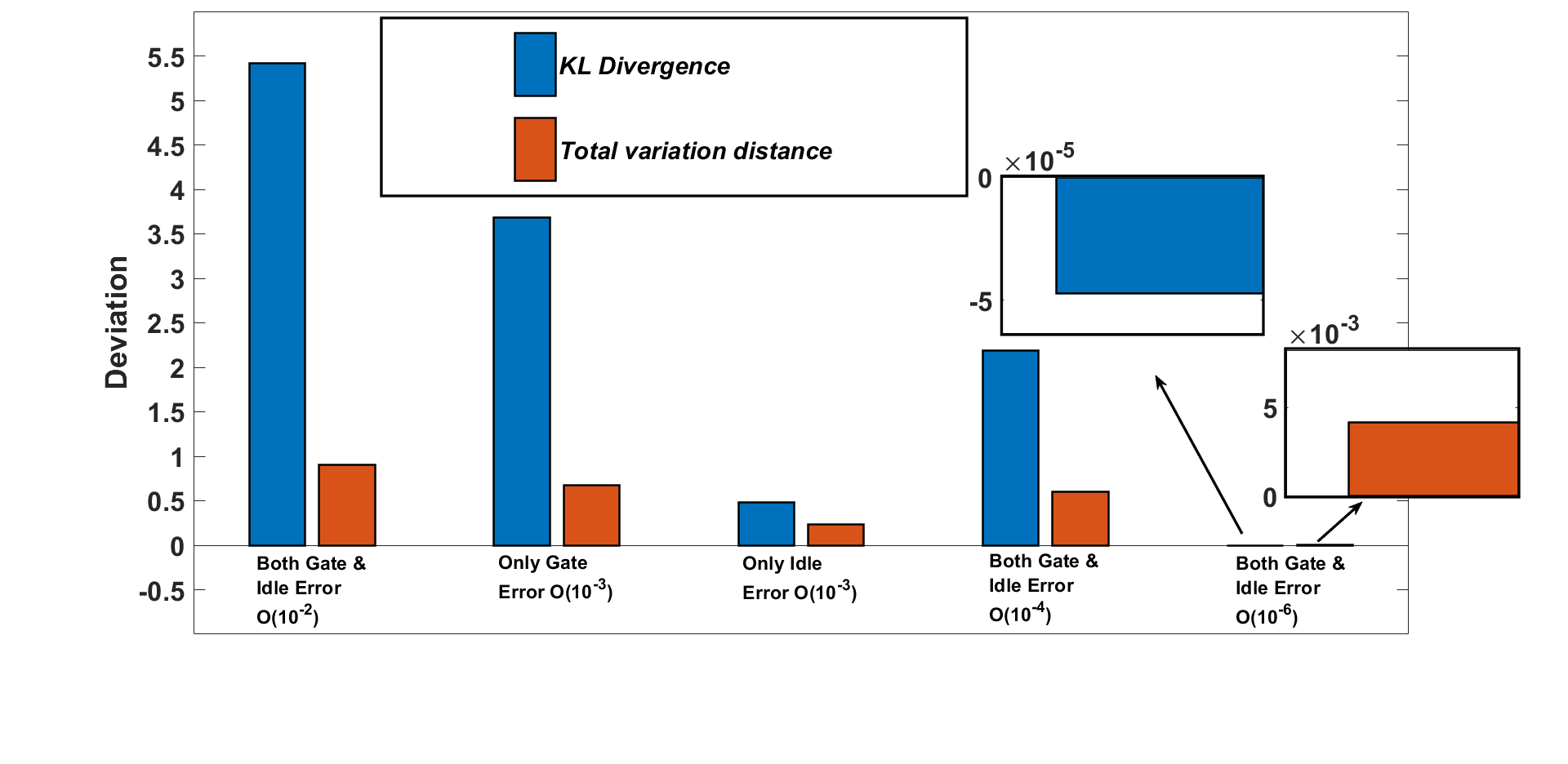}}
    \caption{Comparison $D_{KL}(P_{\mbox{Ideal}}\|P_{\mbox{Noisy}})$ and $TVD_{(P_{\mbox{Ideal}},P_{\mbox{Noisy}})}$ between the time-averaged probability distributions obtained through noiseless and noisy circuits for DTQWs in $\mathrm{Cay(\Z_{27},\{1,-1\})}$ taking coins from the classes $,\mathcal{X}_\theta,\mathcal{Y}_\theta,\mathcal{Z}_\theta,\mathcal{W}_\theta$ with initial position $0$ and initial coin state $\frac{1}{\sqrt{3}}(\ket{0}_3+\ket{1}_3+\ket{2}_3)$. The time step is taken up to $300$. The generic depolarizer gate noise and amplitude damping idle noise is incorporated in the circuit. The error parameters are chosen from uniform distribution.}\label{fig:Timeavampcompcy}
\end{figure}

\begin{figure}[H]
    \centering
     \subfigure[$C\in \mathcal{X}_\theta,\theta=\pi$]{\includegraphics[height=3.5 cm,width=8 cm]{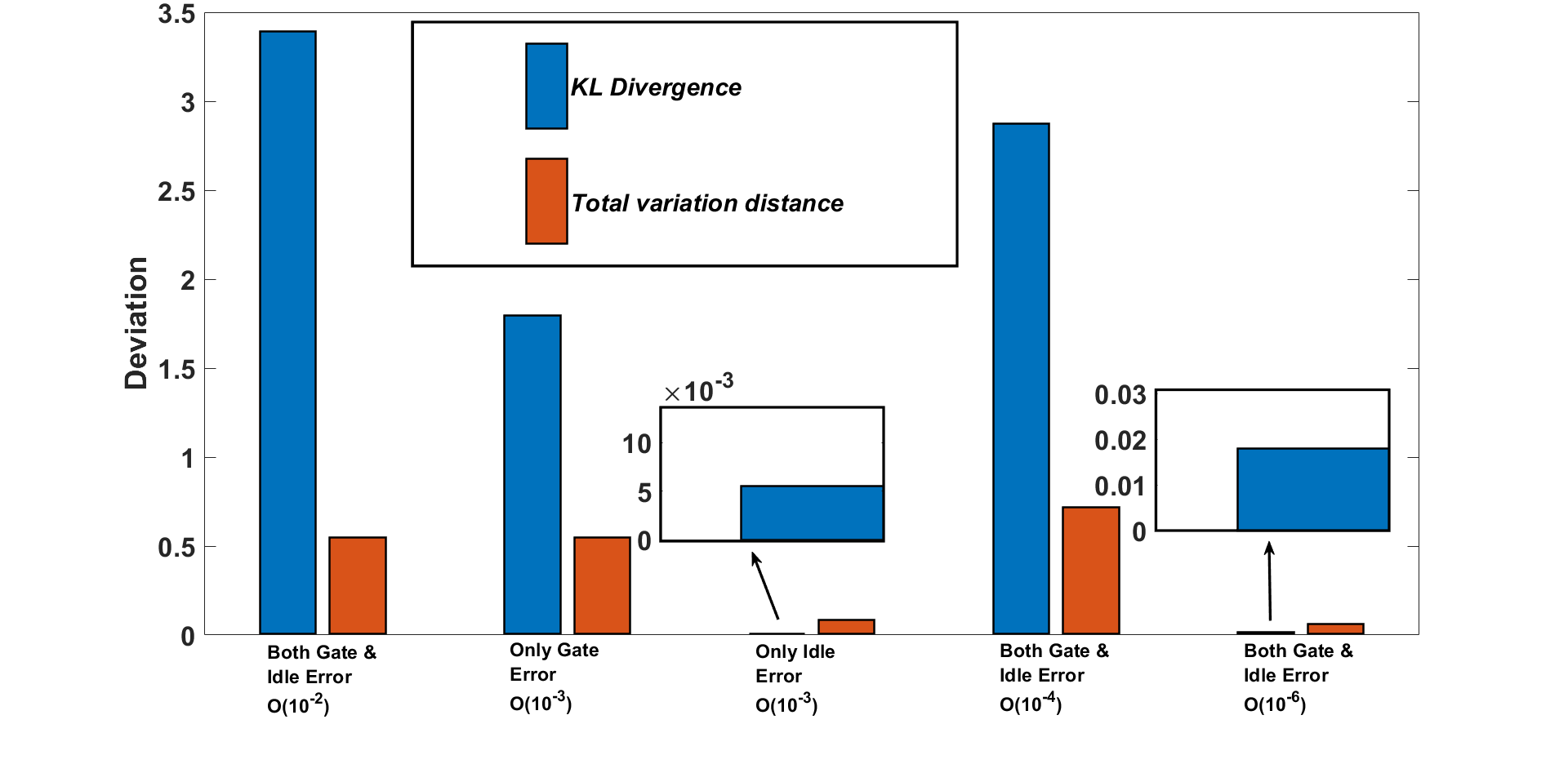}}
   \subfigure[$C\in \mathcal{Y}_\theta,\theta=\pi/2$]{\includegraphics[height=3.5 cm,width=8 cm]{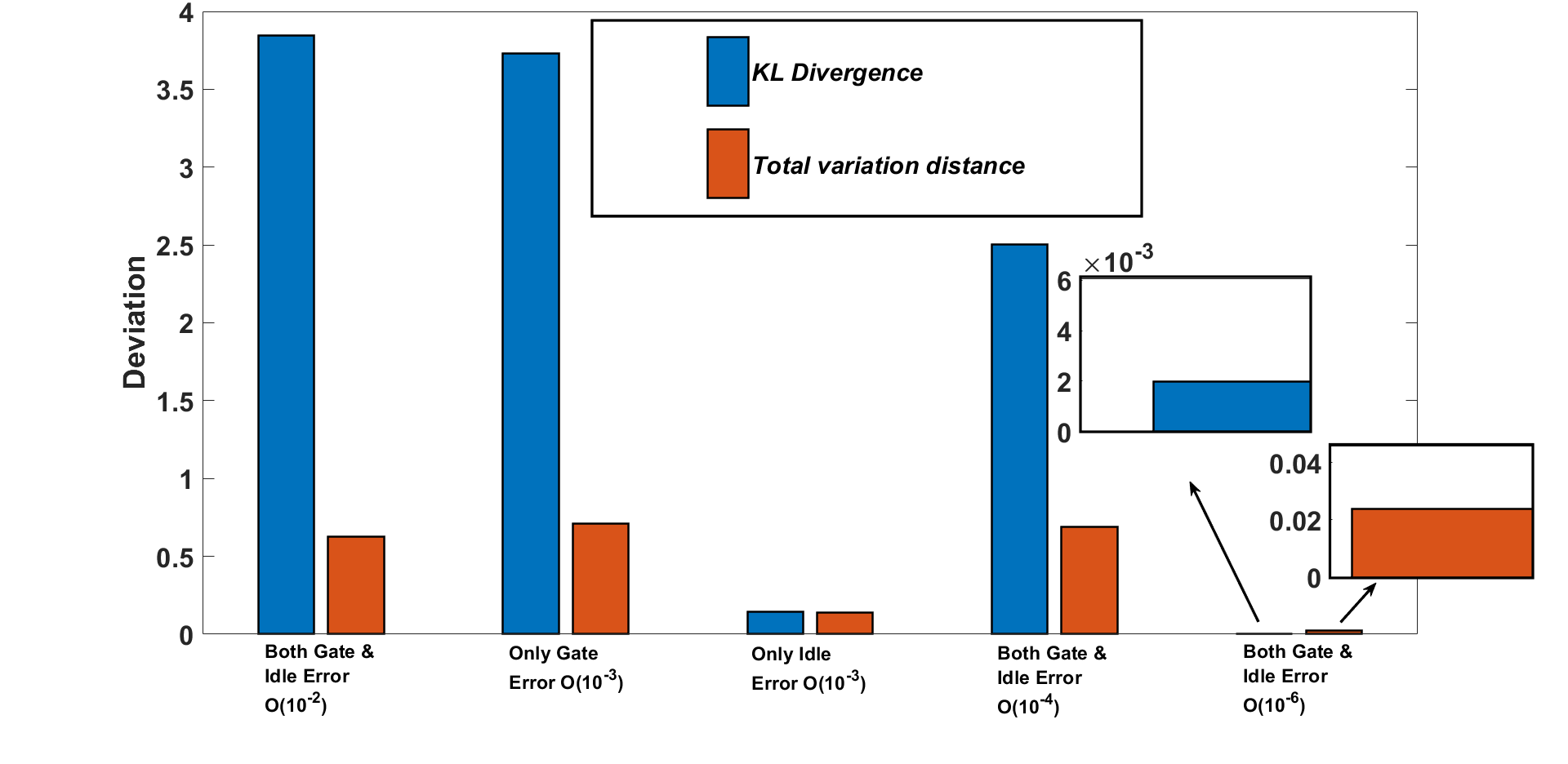}}
   \subfigure[$C\in \mathcal{Z}_\theta,\theta=\pi/3$]{\includegraphics[height=3.5 cm,width=8 cm]{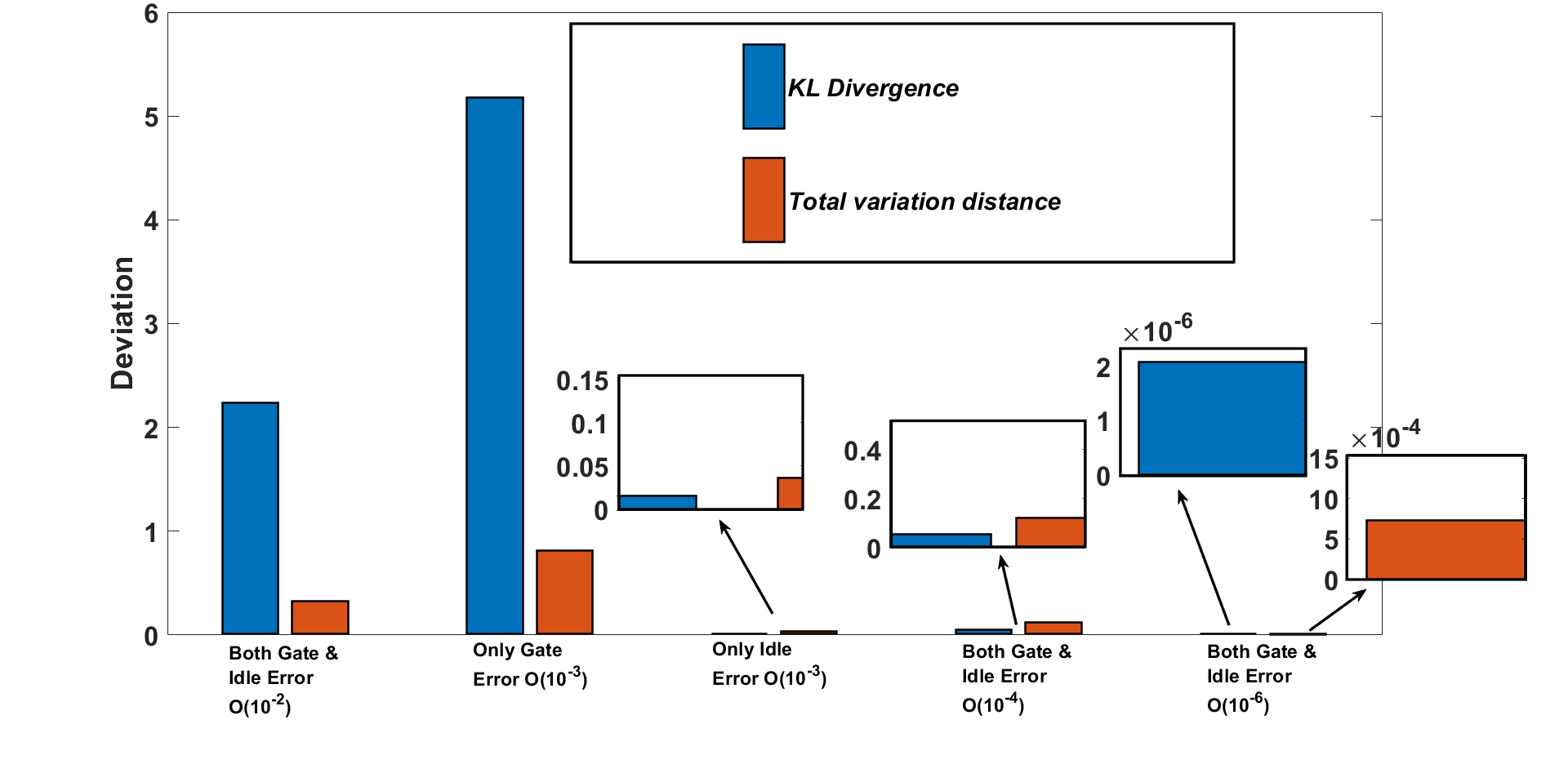}}
   \subfigure[$C\in \mathcal{W}_\theta,\theta=-\pi/4$]{\includegraphics[height=3.5 cm,width=8 cm]{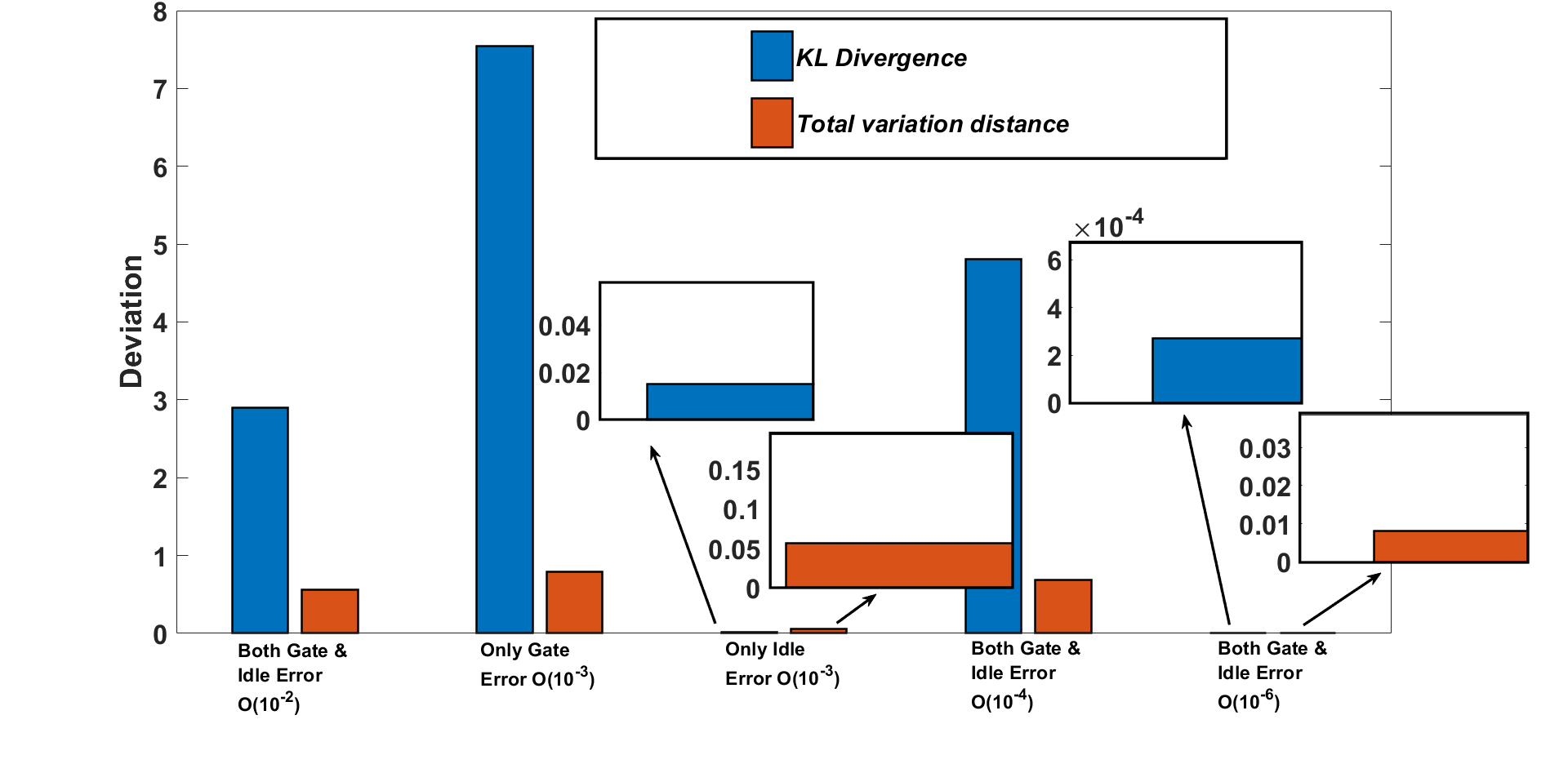}}
    \caption{Comparison $D_{KL}(P_{\mbox{Ideal}}\|P_{\mbox{Noisy}})$ and $TVD_{(P_{\mbox{Ideal}},P_{\mbox{Noisy}})}$ between the time-averaged probability distributions obtained through noiseless and noisy circuits for DTQWs in $\mathrm{Cay(\Z_{27},\{1,-1\})}$ taking coins from the classes $,\mathcal{X}_\theta,\mathcal{Y}_\theta,\mathcal{Z}_\theta,\mathcal{W}_\theta$ with initial position $0$ and initial coin state $\frac{1}{\sqrt{3}}(\ket{0}_3+\ket{1}_3+\ket{2}_3)$. The time step is taken up to $300$. The generic depolarizer gate noise and phase damping idle noise is incorporated in the circuit and the error parameters are chosen from uniform distribution.}\label{fig:Timeavphcompcy}
\end{figure}

\noindent{\bf Conclusion.} In this paper we develop qutrit quantum circuit models for quantum walks on Cayley graphs of Dihedral groups and additive group of integers modulo a positive integer. The circuits are based on elementary qutrit gates, and during the process we propose qutrit quantum circuit models for block diagonal special unitary matrices of order $3^n$ with diagonal blocks as special unitary matrices of order three. We derive the circuit complexity of the developed circuit models and we numerically simulate the time-averaged probability distributions of the walker employing the circuit models with various noise models. These results show that gate noises significantly impact the distributions of noiseless models than amplitude or phase damping errors.  
\\\\

\noindent{\bf Acknowledgement}
Rohit Sarma Sarkar acknowledges support through Prime Minister's Research Fellowship (PMRF), Government of India.

\end{document}